\documentclass[aps,pre,reprint,superscriptaddress,longbibliography]{revtex4-1}

\usepackage{color}
\usepackage{amsmath}
\usepackage{bm}
\usepackage{amssymb}
\usepackage{wrapfig}
\usepackage{braket}
\usepackage{amsthm}
\usepackage{natbib}
\usepackage{dsfont}
\usepackage{graphicx}
\usepackage[colorlinks = true, citecolor = blue, linkcolor = blue, urlcolor=blue]{hyperref}

\definecolor{green}{rgb}{0.0, 0.5, 0.0}

\definecolor{blue}{rgb}{0.0, 0.0, 0.5}

\DeclareFontFamily{OMX}{MnSymbolE}{}
\DeclareFontShape{OMX}{MnSymbolE}{m}{n}{
	<-6>  MnSymbolE5
	<6-7>  MnSymbolE6
	<7-8>  MnSymbolE7
	<8-9>  MnSymbolE8
	<9-10> MnSymbolE9
	<10-12> MnSymbolE10
	<12->   MnSymbolE12}{}
\DeclareSymbolFont{mnlargesymbols}{OMX}{MnSymbolE}{m}{n}
\SetSymbolFont{mnlargesymbols}{bold}{OMX}{MnSymbolE}{b}{n}
\DeclareMathDelimiter{\LL}{\mathopen}{mnlargesymbols}{'164}{mnlargesymbols}{'164}
\DeclareMathDelimiter{\rr}{\mathclose}{mnlargesymbols}{'171}{mnlargesymbols}{'171}

\usepackage{verbatim}
\usepackage[T1]{fontenc}
\usepackage[english]{babel}
\usepackage[utf8]{inputenc}

\newcommand{\Tr}{{\mathrm{Tr}}}

\renewcommand{\L}{{\mathcal{L}}}

\newcommand{\rhoss}{{\rho_\text{ss}}}
\newcommand{\trho}{{\tilde{\rho}}}

\newcommand{\C}{{\mathbf{C}}}

\newcommand{\pss}{\mathbf{p}_\text{ss}}
\newcommand{\ptss}{{\mathbf{\tilde{p}}_\text{ss}}}

\newcommand{\W}{{\mathbf{W}}}
\newcommand{\Wt}{{\mathbf{\tilde{W}}}}

\makeatletter
\def\l@subsubsection#1#2{}
\makeatother
\makeatletter
\def\@bibdataout@aps{%
	\immediate\write\@bibdataout{%
		@CONTROL{%
			apsrev41Control%
			\longbibliography@sw{%
				,author="08",editor="1",pages="1",title="0",year="1"%
			}{%
				,author="08",editor="1",pages="1",title="",year="1"%
			}%
		}%
	}%
	\if@filesw \immediate \write \@auxout {\string \citation {apsrev41Control}}\fi
}
\makeatother

\begin{document}
\title{Hierarchical classical metastability in an open quantum East model}

\author{Dominic C. Rose}
\affiliation{School of Physics and Astronomy, University of Nottingham, University Park,  Nottingham NG7 2RD, United Kingdom}
\affiliation{Centre for the Mathematics and Theoretical Physics of Quantum Non-Equilibrium Systems, University of Nottingham, University Park,  Nottingham NG7 2RD, United Kingdom}

\author{Katarzyna Macieszczak}
\affiliation{TCM Group, Cavendish  Laboratory,  University  of  Cambridge,	J.  J.  Thomson  Ave.,  Cambridge  CB3  0HE,  United Kingdom}

\author{Igor Lesanovsky}
\affiliation{School of Physics and Astronomy, University of Nottingham, University Park,  Nottingham NG7 2RD, United Kingdom}
\affiliation{Centre for the Mathematics and Theoretical Physics of Quantum Non-Equilibrium Systems, University of Nottingham, University Park,  Nottingham NG7 2RD, United Kingdom}
\affiliation{Institut f\"ur Theoretische Physik, Universit\"at Tübingen, Auf der Morgenstelle 14, 72076 T\"ubingen, Germany}

\author{Juan P. Garrahan}
\affiliation{School of Physics and Astronomy, University of Nottingham, University Park,  Nottingham NG7 2RD, United Kingdom}
\affiliation{Centre for the Mathematics and Theoretical Physics of Quantum Non-Equilibrium Systems, University of Nottingham, University Park,  Nottingham NG7 2RD, United Kingdom}

\begin{abstract}
We study in detail an open quantum generalisation of a classical kinetically constrained model --- the East model --- known to exhibit slow glassy dynamics stemming from a complex hierarchy of metastable states with distinct lifetimes. Using the recently introduced theory of classical metastability for open quantum systems, we show that the driven open quantum East model features a hierarchy of classical metastabilities at low temperature and weak driving field. We find that the effective long-time description of its dynamics is not only classical, but shares many properties with the classical East model, such as obeying an effective detailed balance condition, and lacking static interactions between excitations, but with this occurring within a modified set of metastable phases which are coherent, and with an effective temperature that is dependent on the coherent drive.
\end{abstract}


\maketitle

\tableofcontents

\section{Introduction}
With a strong focus of current research on non-equilibrium physics, open quantum systems have come to the fore as a natural platform for studying the associated phenomena: both through the natural occurrence of non-equilibrium behaviour, and through their use in quantum simulation based on, e.g., Rydberg atoms and optical lattices \cite{Pritchard2010,Blatt2012,Britton2012,Dudin2012,Peyronel2012,Guenter2013,Schmidt2013}.
This experimental prominence has been accompanied by the development of varied numerical approaches and analytical techniques, such as tensor networks \cite{Gangat2017}, Monte-Carlo methods \cite{Molmer92,Molmer93,Plenio1998,Daley2014}, field theoretical studies \cite{Torre2013,Sieberer2016,Maghrebi2016}, other variational approaches \cite{Weimer2015,Weimer2015a,Overbeck2017}, and machine learning \cite{Yoshioka2019,Hartmann2019,Nagy2019,Vicentini2019,BedollaMontiel2020}.

Despite the change implicitly present in the background of all non-equilibrium phenomena, most prior studies on open quantum systems have focused on their non-equilibrium steady states, with phase diagrams seeing a particular focus \cite{Mendoza-Arenas2016,Foss-Feig2017,Casteels2017,Letscher2017,Jin2018,Melo2016,Rodriguez2017}.
Classical phase transitions in the steady state, a distinctly time-independent phenomenon, are nevertheless accompanied by a critical slowing of the systems dynamics, with diverging timescales at the transition parameters.
For parameters near the transition, or finite system sizes, this slowing results in distinct timescales in the system dynamics, lending a rich structure to the time-evolution in such problems: this is commonly referred to as metastability.
With a deep theory for classical Markovian processes \cite{Gaveau1987,Gaveau1998,Gaveau1999,Schulman2001,Gaveau2006}, recent work has been done to extend this to open quantum systems \cite{Macieszczak2016a,Rose2016,Macieszczak2020}.

While metastability always arises as a consequence of proximity to phase transitions~\cite{Macieszczak2016a,Minganti2018,Landa2020}, it can occur without any significant change in the stationary state at all, through the presence of constraints in the dynamics.
In classical kinetically constrained models~\cite{Jaeckle1991,Sollich1999,Garrahan2002,Sollich2003,Binder2011,Biroli2013}, the evolution of system components is conditioned on the state of other components, which results in glassy dynamics with dynamical heterogeneity, i.e., excitations localised both in space and time, and a hierarchy of relaxation timescales in observable averages and correlations.
This complex dynamics corresponds to the occurrence of metastability despite the potential absence of phase transitions in the stationary state.
Quantum adaptions of these models have been developed through the concept of Rokhsar-Kivelson points~\cite{Rokhsar1988,Henley2004}, leading to quasi-many body localisation behaviour in closed quantum systems~\cite{Horssen2015,Hickey2016,Lan2018}.
Recent open quantum generalisations~\cite{Olmos2012,Lesanovsky2013,Olmos2014} are also known to display dynamical heterogeneity.
Here, we uncover its origin in the open quantum East model introduced in Ref.~\cite{Olmos2012} by utilising the non-Hermitian perturbation theory~\cite{Kato1995} and the recently formulated theory of classical metastability in Markovian open quantum systems~\cite{Macieszczak2016a,Rose2016,Macieszczak2020}.

\begin{figure}[t!]
	\includegraphics[width=1\linewidth]{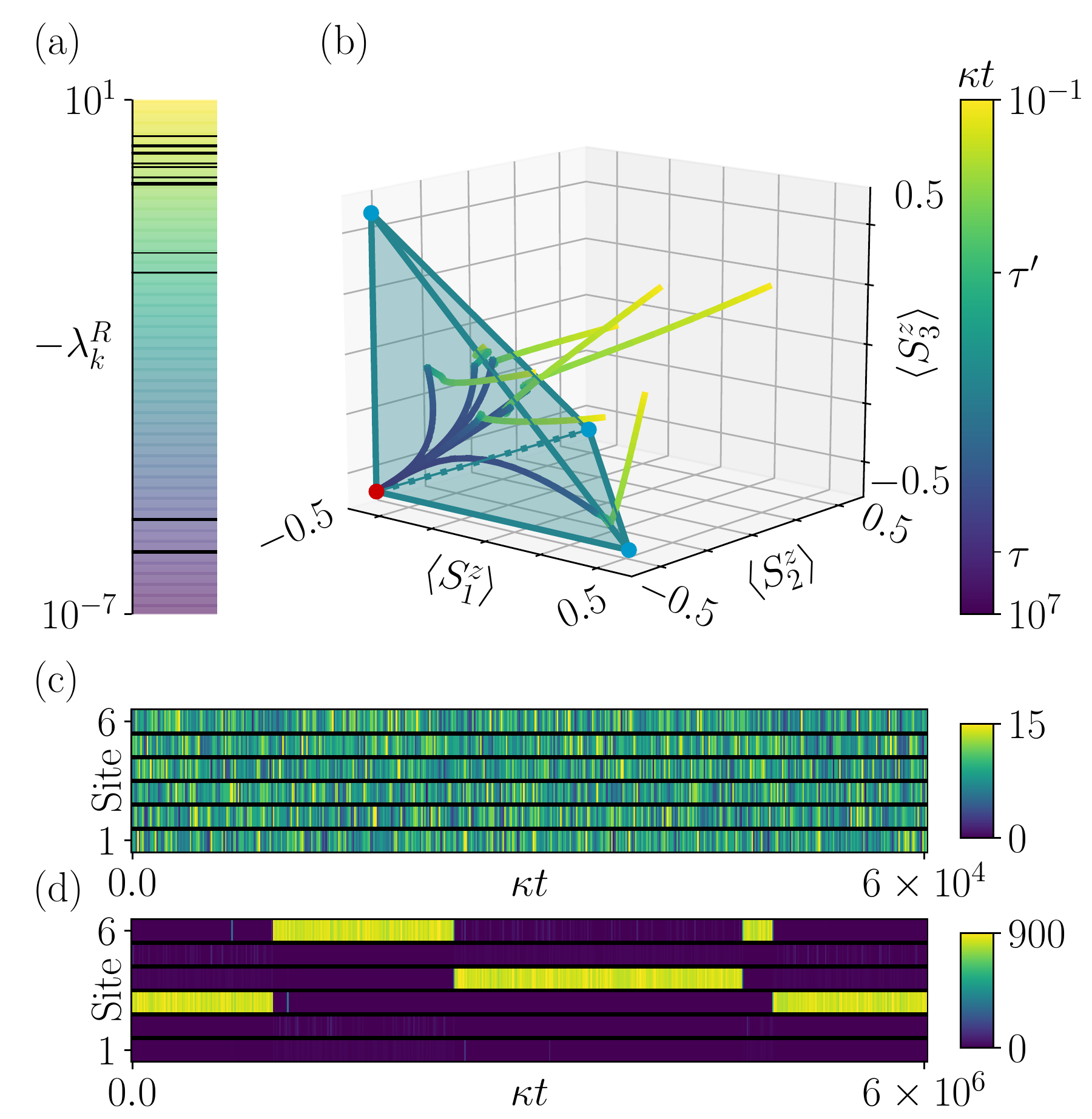}
	\caption{\label{fig:Spectrum}\textbf{Metastable dynamics in the open quantum East model}: (a) Spectrum of the open quantum East model of $N=3$ spins with a large separation $-\lambda^R_4\ll-\lambda^R_5$ [values shown in the log scale, $\lambda_1=0$ not visible; degeneracy $\lambda_2^R=\lambda_3^R$ due to the translation symmetry; cf.~Fig.~\ref{fig:3DMM}\textcolor{blue}{(a)}]. (b) Average site magnetisation for random initial states. Colour represents time in the log scale, shown on left. After the initial relaxation on the metastable timescale $\tau'$ (yellow-green) states approach a limited region of values captured by the blue simplex of the magnetisation with vertices corresponding to metastable phases (blue dots), before the final relaxation until the relaxation time $\tau$ (blue-purple) towards the unique stationary state (red dot); cf.~Fig.~\ref{fig:3DMM}. (c) and (d) Quantum jump Monte Carlo (QJMC) trajectories for $N=6$ spins, for the non-interacting unconstrained case (c), and for fully constrained case (d) with pronounced dynamical heterogeneity [colour indicates the number of jumps $J_j^-$,~\eqref{eq:J-}, for $j$th spin, grouped in $500\kappa t$ time bins]. Parameters: in panels (a) and (b) $\Omega/\kappa=0.1$ and $\gamma/\kappa=0.0001$, and $p=0.999$, in panels (c) and (d) $\Omega/\kappa=0.12$ and  $\gamma/\kappa=0.0096$ with $p=0$, $p=1$, respectively. See Appendixes~\ref{app:numerics_master} and~\ref{app:numerics_QJMC} for the numerical methods.
		\vspace*{-5mm}}
\end{figure}

The complex relaxation is a consequence of an emerging hierarchy of metastabilities: multiple timescales when average system states appear stationary, although different from the true, usually unique, stationary state.
This is visible in the spectrum of the master operator governing the dynamics as large separations between real parts of its eigenvalues; see Fig.~\ref{fig:Spectrum}\textcolor{blue}{(a)}.
We find that metastabilities are effectively classical, with any density matrix after sufficiently long evolution being a probabilistic mixture of distinct metastable phases; see Fig.~\ref{fig:Spectrum}\textcolor{blue}{(b)}. Analogously to the classical East model, these metastable phases  correspond to localised excitations but in a coherent basis and
 their number increases with system size, which is also corroborated  by the non-Hermitian perturbation theory analysis, which identifies the second-order dephasing as the mechanism beyond the emergence of fourth-order classical dynamics with respect to the driving field amplitude. Importantly, these phases arise not only in average dynamics but already in individual realisations of an experimental run or its numerical simulation: when coarse-grained in time over periods comparable to the metastable timescales, emission records jump sharply between the rates of the corresponding metastable phases  leading to dynamical heterogeneity; see Fig.~\ref{fig:Spectrum}\textcolor{blue}{(d)} [cf.~a trajectory without metastability in Fig.~\ref{fig:Spectrum}\textcolor{blue}{(c)}]. Furthermore, dynamics of coarse-grained emission records is determined
 by the effectively classical long-time dynamics of the average system state, which shares further properties with the classical East model:  detailed balance at an effective temperature dependent both on the temperature and the driving field, and the lack of interactions between excitation in metastable phases.
Additionally, we observe the emergence of an effective metastability for the emission activity, which is not  accompanied by a separation in the master operator spectrum, as it appears before metastable regimes for the system states.

This paper is organised as follows.
We begin by introducing open quantum East model~\cite{Olmos2012} in Sec.~\ref{sec:qEastMod}.
We verify the presence of a spectral gaps inducing a hierarchy of metastabilities in Sec.~\ref{sec:spectra-hierarchy} and discuss properties of the corresponding metastable phases in Sec.~\ref{sec:phases}.
We then investigate the structure of the classical long-time dynamics with focus both on the dynamics of the average system state in Secs. \ref{sec:effdyn_DB} and the dynamics of quantum trajectories in Secs. \ref{sec:dynhet} and \ref{sec:DPT}, as well as the emergence of effective metastability in Sec.~\ref{sec:dynobs}.


\section{Open quantum East model}\label{sec:qEastMod}

We now discuss the model we will consider in this paper, the open quantum East model~\cite{Olmos2012,Lesanovsky2013}, a generalisation of the classical kinetically constrained East model \cite{Jaeckle1991} studied in relation to glass physics~\cite{Garrahan2002,Ritort2003,Binder2011,Chandler2010,Biroli2013}.
Such classical systems often exhibit multiple stages of relaxation on different time scales, indicative of metastability, and we expect such behaviour to occur in their quantum counterparts.

\subsection{Model}\label{sec:model}
We consider dynamics of $N$ spins-$1/2$ governed by a Lindblad master operator as  (see Refs.~\cite{Lindblad1976,Gardiner2004,Breuer2002})
\begin{equation}\label{eq:rhodt}
\frac{\mathrm{d}}{\mathrm{d}t}\rho(t)=\L[\rho(t)],
\end{equation}
where $\rho(t)$ is the density matrix describing the system state at time $t$ and the Lindblad operator $\mathcal{L}$ is given by
\begin{eqnarray}\label{eq:lindblad-op}
\L(\rho) = \!\!\!\sum_{\substack{j=1,...,N \\ \alpha=-,+}}\!\!\left(\!-i[H_j,\rho]+ {J}^\alpha_{j}\rho{J}_{j}^{\alpha\dagger}-\frac{1}{2}\left\{{J}^{\alpha\dagger}_{j} {J}^{\alpha}_{j},\rho\right\}\!\right)\!,\quad\,\,
\end{eqnarray}
with $H_j$ being the Hamiltonian and ${J}_j^{-}$ and ${J}_j^{+}$ the jump operators that act locally on $j$th spin, constrained on the state of the preceding spin (see below).
These jump operators describe interactions between the system and its surrounding environment, which, if associated to emissions of energy quanta, can be detected via continuous measurements \cite{Gardiner2004}, e.g., by counting photons emitted by atoms coupled to the electromagnetic  vacuum  \cite{Ates2012,Olmos2012,Lesanovsky2013,Olmos2014}.

For $N=1$ spin, there are no constraints, and the dynamics is due to the interplay of the coherent field $\Omega$ and thermal fluctuations,
\begin{subequations}
\label{eq:HJJ1}
\begin{align}
\label{eq:H1}
H&=\Omega\,{S}^{x},\\
\label{eq:J-1}
{J}^{-}&=\sqrt{\kappa}\,{S}^{-},\\
\label{eq:J+1}
{J}^{+}&=\sqrt{\gamma}\,{S}^{+},
\end{align}
\end{subequations}
where ${S}^{x}$ and ${S}^{\mp}={S}^{x}\mp i{S}^{y}$ are the spin operators that can be associated with the photon emission and absorption, respectively.
This dynamics features a unique stationary state~\cite{Olmos2012},
\begin{equation}\label{eq:rho_ss1}
\rho_\text{ss,1} =\left[
\begin{array}{cc}
\frac{\Omega ^2+\kappa (\gamma +\kappa )}{(\gamma +\kappa )^2+2 \Omega ^2} & - i\frac{ (\gamma -\kappa ) \Omega }{(\gamma +\kappa )^2+2 \Omega ^2} \\ i\frac{ (\gamma -\kappa ) \Omega }{(\gamma +\kappa )^2+2 \Omega ^2} & \frac{\Omega ^2+\gamma  (\gamma +\kappa )}{(\gamma +\kappa )^2+2 \Omega ^2} \\
\end{array}
\right],
\end{equation}
expressed in the basis $|0\rangle$, $|1\rangle$. The eigenstates $|u\rangle$ and $|e\rangle$ of ${\rho}_{\text{ss},1}=\lambda_u\ket{u}\!\!\bra{u}+\lambda_e\ket{e}\!\!\bra{e}$, approach $\ket{0}$ and $\ket{1}$, as the coherent field tends to $0$, and we refer to them as the unexcited and the excited states, respectively (see Appendix~\ref{app:1spin}).

The many-body model~\cite{Olmos2012,Olmos2014} with $N\geq 2$, in analogy to the classical East model, is constructed
using a \emph{constraint} operator
\begin{equation}\label{eq:F}
F=(1-p)\mathds{1}+p\ket{e}\!\!\bra{e}.
\end{equation} The constraint, parametrised by $p$, is absent when $p=0$, for $p=1$  is referred as hard, and
for $0\leq p< 1$ as soft. The dynamics  in Eq.~\eqref{eq:lindblad-op} is then defined as [cf.~Eq.~\eqref{eq:HJJ1}]
\begin{subequations}
\label{eq:HJJ}
\begin{align}
\label{eq:H}
H_j&=\Omega\,{F}^{2}_{j-1}\, {S}^{x}_{j},\\
\label{eq:J-}
{J}^{-}_{j}&=\sqrt{\kappa}\,{F}_{j-1}\, {S}^{-}_{j},\\
\label{eq:J+}
{J}^{+}_{j}&=\sqrt{\gamma}\,{F}_{j-1}\,{S}^{+}_{j},
\end{align}
\end{subequations}
where the subscript $j$ denotes the operators acting on $j$th spin and we assume periodic boundary conditions, i.e., $0\mapsto N$ in operator indices.

For $p<1$, the stationary state of the dynamics is unique and given by a product state of the single-spin stationary state, ${\rho}_{\text{ss},1}^{\otimes{N}}$ [cf.~Eq.~\eqref{eq:rho_ss1}]. This follows directly from the construction of the dynamics, as the constraint commutes with the stationary state of a single spin, and as such, the state of a neighbouring spin is acted on as if the master operator were that of a non-interacting system, but with $\kappa$, $\gamma$ and $\Omega$ rescaled by $1-p\lambda_u$.
For the hard constraint, dynamics of the $j$th spin only occurs if the state of $(j-1)$th spin features some probability of being in the excited state $|e\rangle$. Therefore, the so called dark state ${(\ket{u}\!\!\bra{u})}^{\otimes{N}}$ is disconnected from the dynamics and thus stationary, as no constraint is fulfilled  [cf.~Fig.~\ref{fig:Spectrum}\textcolor{blue}{(d)} and see Appendix~\ref{app:1spin}].

As a consequence of its non-interacting structure, the stationary state features no static transitions, and cumulants of all system observables remain analytic. Nevertheless, at low temperatures  ($\gamma/\kappa\ll1$) and small values of coherent field ($|\Omega|/\kappa\ll1$), the dynamics manifests a significant change as $p$ tends to $1$, with jumps in trajectories becoming localised both spatially and temporally~\cite{Lesanovsky2013}, thus leading to dynamical heterogeneity [see Figs.~\ref{fig:Spectrum}\textcolor{blue}{(c)} and~\ref{fig:Spectrum}\textcolor{blue}{(d)}]. In this work, we unfold this dynamical phenomenon using the approach for classical metastability in open quantum systems  recently introduced in Ref.~\cite{Macieszczak2020}. In order to motivate the use of this new approach, we first discuss the approach via a mean-field approximation and results from the non-Hermitian perturbation theory.

\subsection{Mean-field theory} \label{sec:MF}
A common informative treatment of open many-body quantum systems is mean-field theory and its extensions~\cite{Biondi2017,Landa2020}, often resulting in the prediction of multiple stationary states~\cite{Ates2012,Maghrebi2016}.
These stationary states can often be identified with metastable phases in the finite-size system, with the mean-field describing short time evolution into the metastable manifold~\cite{Rose2016,Minganti2018,Landa2020}, and long-time dynamics neglected due to the lack of correlations acting as noise on this set of states~\cite{Landa2020}.
While the stationary state of the quantum (and classical) East model is homogeneous, the dynamical heterogeneity of trajectories suggests the long time dynamics takes place between states which are not translation-symmetric, and thus cannot be reproduced in the homogeneous mean-field ansatz $(\mathds{1}/2+x S^x+yS^y+zS^z)^{\otimes N}$.
Indeed, it is known that mean-field is ineffective in the classical case ($\Omega=0$) as the removal of spatial dependence in the state causes the constraint's directionality to be lost.
This will also be the case in the quantum regime ($\Omega\neq0$), unless we allow for the spacial dependence by considering a tensor product of different single-state density matrices, $\otimes_{j=1}^N (\mathds{1}/2+x_j S^x+y_j S^y +z_j S^z)$.
In this case, however, the number of parameters is reduced from $4^N-1$ merely to $3^N$, and at the price of solving non-linear (quadratic) differential equations.
As such, we forgo the mean-field treatment.

\subsection{Perturbation theory} \label{sec:PT}
The dynamical heterogeneity is present in the East model dynamics at low temperatures  ($\gamma/\kappa\ll1$), small values of coherent field ($|\Omega|/\kappa\ll1$), and constraint close to hard ($p=1-\epsilon\approx 1$, equivalent to $|\epsilon|\ll 1$).
Such separation of scale in the dynamical parameters, motivates the use of non-Hermitian perturbation theory~\cite{Kato1995} for the dynamics with jumps $J_j^-$ featuring the hard constraint, i.e.,
\begin{equation}
\label{eq:J0}
J_j^{(0)}=\sqrt{\kappa} |0\rangle\!\langle 0|_{j-1}\,|0\rangle\!\langle 1|_j,
\end{equation} perturbed with respect to the low temperature $\gamma$, the weak coherent field, $\Omega$,  and the soft constraint, $\epsilon$, to the dynamics with the Hamiltonian and jumps operators in  Eq.~\eqref{eq:HJJ} [cf.~Eq.~\eqref{eq:F}].
In Appendix~\ref{app:PT}, we derive the first-, second-, and third-order corrections to the dynamics of the system consisting of any number of spins, and also  discuss the finite-size effects.
Here, we summarize those results, with the further discussion in the context of findings of the approach from Ref.~\cite{Macieszczak2020} in the later sections.

The stationary states of jumps $J_j^{(0)}$ [Eq.~\eqref{eq:J0}] correspond to the states which feature no excitations, $|0...0\rangle$, or only isolated excitations, $|...010...\rangle$, as such states are dark to jump operators $J_j^{(0)}$, i.e., the action of the jump operators is $0$ on these states.
This further leads to all coherences between such states being stationary, so that they form a decoherence free subspace (DFS)~\cite{Zanardi1997,Zanardi1997a,Lidar1998} (see Appendix~\ref{app:dark}).
Upon perturbing, this DFS becomes a \emph{quantum metastable manifold}~\cite{Macieszczak2016a} and undergoes slow dynamics at the timescales we now discuss.\\

{\bf \em Low temperature}.
For $\gamma/\kappa\ll1 $, already in the first order, the perturbative dynamics proportional to $\gamma$ leads to the decay of the dark DFS towards states with excitations followed at least by two unexcited sites, i.e., $|0...0\rangle$, $|...00100...\rangle$, with no coherences being stationary any longer.
Therefore, the \emph{metastable manifold is classical} in the perturbative regime of low temperatures.
Furthermore, in higher orders of $\gamma$, non-decaying excitations are separated by a distance growing exponentially with the order of the corrections.
Ultimately, this leads to only two states being stationary: the state with no excitations, $|0...0\rangle$, and the uniform state with a single excitation, $N^{-1}\sum_{j=1}^N |0...01_j0...0\rangle$, which approximate, in the zero order of $\gamma$, the two stationary states of dynamics with hard constraint.
Higher-order corrections in the structure are also recovered by the perturbation theory (see Appendix~\ref{app:temp}).\\

{\bf \em Soft constraint}.
Softening the constraint in the regime $|\epsilon|\ll 1$, with $\epsilon=1-p$, leads to dynamics featuring removal of isolated excitations with the rate proportional to $\kappa \epsilon^2$  (see Appendix~\ref{app:soft_small}). This leads to decay of coherences and facilitates a unique stationary state, $|0...0\rangle$, which approximates in the zero order the unique stationary state at $p<1$. We note that even for finite values of temperature and coherent field, a perturbative dynamics between two disjoint stationary states takes place in the limit $|\epsilon|\ll 1$  (see Appendix~\ref{app:soft_finite}).\\

{\bf \em Weak coherent field}.
Here, the perturbation in $\Omega$ introduces both the Hamiltonian and the change of constraints [cf.~Eq.~\eqref{eq:HJJ}].
In Appendix~\ref{app:dynamicsOmega}, we show there are no odd-order corrections in $\Omega$ to the dynamics.
Furthermore, the second-order corrections correspond to dephasing of all coherences in the DFS with rates proportional to $\Omega^2/\kappa$, while probabilistic mixtures of the states of none, $|0...0\rangle$, or only isolated excitations $|...010...\rangle$, remain stationary. Meanwhile, the first order-corrections to the state structure introduce the rotation of $|0\rangle$ and $|1\rangle$ towards coherent states $|u\rangle$ and $|e\rangle$, with coefficients proportional to $\Omega/\kappa$.
Therefore, also in this case we conclude that the \emph{metastable manifold is classical}, but with respect to a now-coherent basis, which we expect to coincide with $|u\rangle$ and $|e\rangle$.
Although the classical metastable manifold is analogous to the case  of the perturbative dynamics due to $\gamma$,  we have that the dynamics of excitations, e.g., the removal of one of a pair of excitations separated just by a single neighbour, can take place at earliest in the fourth-order, with rates proportional to $\Omega^4/\kappa^3$.
Indeed, the stationary state [Eq.~\eqref{eq:rho_ss1}] with probabilities $\lambda_u=1-\lambda_e=1-\gamma/\kappa-16\Omega^4/\kappa^4+...$ (see Appendix~\ref{app:1spin}), suggests that the coherent field in the fourth order may play an analogous role to the temperature in the coherent basis $|u\rangle$, $|e\rangle$.
Because of the complexity of the fourth-order non-Hermitian perturbation theory, we investigate those hypotheses using instead the approach of Ref.~\cite{Macieszczak2020}.\\

{\bf \em Open boundary conditions}. Interestingly, in the case of hard constraint and open boundary conditions, the dark state $(|u\rangle\!\langle u|)^{\otimes N}$ and $N$ states with isolated excitations $(|u\rangle\!\langle u|)^{\otimes (j-1)}\otimes |e\rangle\!\langle e| \otimes \rho_\text{ss,1}^{\otimes (N-j)}$ are stationary. Furthermore, in the limit of small temperature and weak coherent field, the soft constraint connects single excitations mostly to the dark state, but not to one other, and the dark state has a significantly longer lifetime (see Appendix~\ref{app:obc}), so that the facilitated dynamics features excitations localised both in time and space, which is characteristic of the dynamical heterogeneity.\\

Finally, we note that the perturbations in the temperature, the coherent field, or softness of the constraint, are local [cf.~Eq.~\eqref{eq:HJJ}].
Therefore, the timescales of the resulting perturbative dynamics may be proportional to the system size, in which case the validity of the perturbation theory is limited to $\gamma N\ll \kappa$, $|\Omega| N\ll \kappa$, and $|1-p| N\ll \kappa$ (as $\kappa/2$ is the slowest eigenvalue of the dynamics with $J^-$ at the hard constraint); see, e.g.,  Appendix~\ref{app:soft_finite}.
This size-dependent regime is not an issue for the numerical methods of Ref.~\cite{Macieszczak2020}, which we exploit in the rest of the paper.

\section{Classical metastable manifold}\label{sec:approx}

We now investigate the presence and character of metastability in the open quantum East model using the theory of metastability in open quantum systems introduced in Refs.~\cite{Macieszczak2016a,Macieszczak2020}. For metastability in classical stochastic dynamics, see Refs.~\cite{Gaveau1987,Gaveau1998,Gaveau1999,Schulman2001,Gaveau2006}. \\

\subsection{Hierarchy of metastabilities}\label{sec:spectra-hierarchy}
Since the operator in Eq.~\eqref{eq:lindblad-op} defining the time evolution of the average  state $\rho(t)$ is linear, the timescales of the dynamics are determined by its eigenvalues through the expansion
\begin{equation}\label{eq:Expansion0}
\rho(t)={e}^{t\L}[\rho(0)]={\rho}_{\rm ss}+\sum_{k\geq 2}{e}^{t{\lambda}_{k}}{c}_{k}{R}_{k},
\end{equation}
where ${R}_{k}$ is the eigenmode corresponding to the eigenvalue ${\lambda}_{k}$, and the coefficient ${c}_{k}=\Tr[{L}_{k}\rho(0)]$, with $L_k$ being the eigenmode of $\L^\dagger$ with the same eigenvalue, normalised such that $\Tr({L}_{k}^\dagger{R}_{l})={\delta}_{kl}$.
The real parts of eigenvalues must satisfy ${\lambda}_{k}^{R}\leq 0$,  where zero eigenvalues correspond to stationary states~\cite{Baumgartner2008,Albert2016}, and we order the eigenvalues by decreasing real part, so that $\lambda_1=0$.
For a unique stationary state (i.e., $p<1$; see Sec.~\ref{sec:model}), we have $R_1=\rho_{\rm ss}$ and $L_1=\mathds{1}$ (from trace preservation), while  ${\tau}=-1/{\lambda}_{2}^{R}$ is the time scale of the final relaxation.

\begin{figure}[t!]
	\includegraphics[width=1\linewidth]{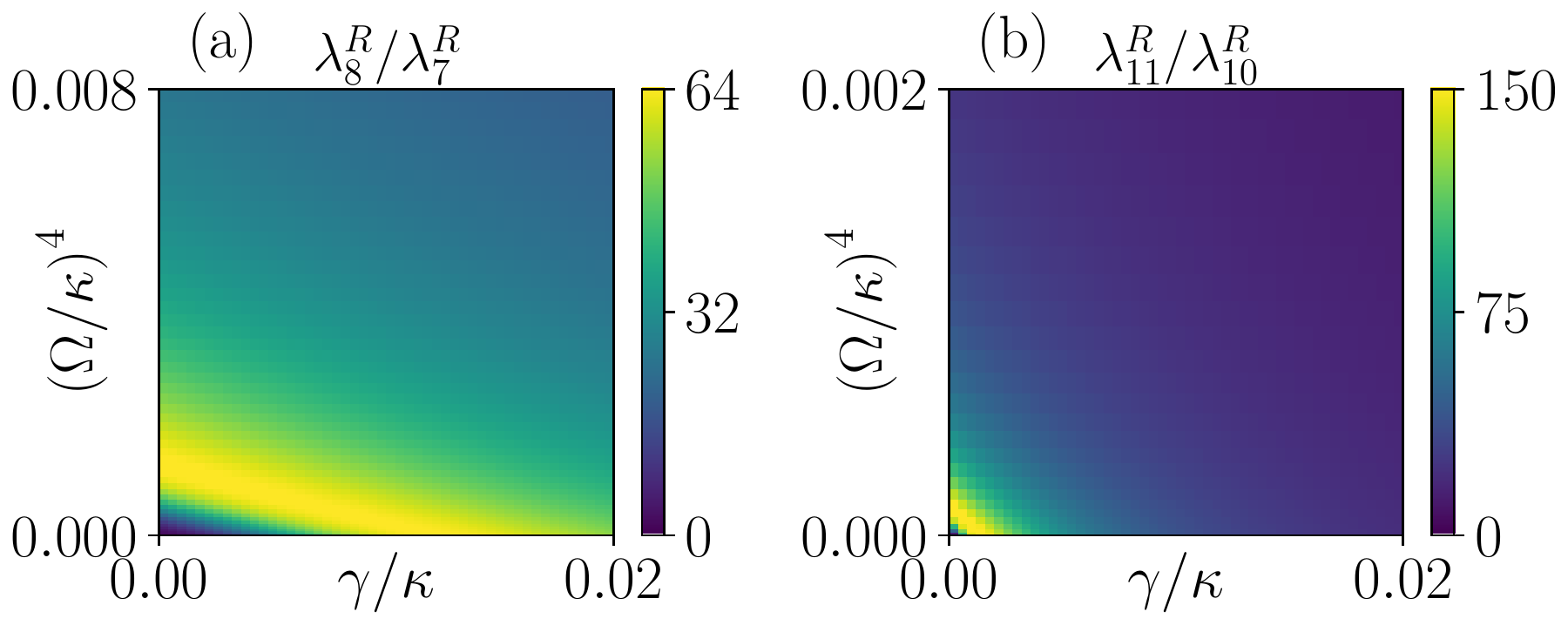}
	\caption{\label{fig:spectral}\textbf{Separation in master operator spectrum}: The ratios of the master operator eigenvalues (the real parts) demonstrate two gaps in the spectrum at (a) $m=7$ and (b) $m=10$ for $N=6$ spins, which shows a hierarchy of metastabilities in the open quantum East model (note the difference in scale of vertical axes). Softness $p=0.99$. Sampling: in panel (a) (51x51) points and in panel (b) (51x26) points, linearly spaced for $\gamma/\kappa$ and $(\Omega/\kappa)^2$.
	}	
\end{figure}

In the rest of this work, we focus on  the dynamics of the open quantum East model with $N=6$ spins and  softness $p=0.99$. Here, in the presence of small temperature and weak coherent field, we observe a large separation in the spectrum between $\lambda_{m}^R$ and $\lambda_{m+1}^R$ for $m=7$ [see Fig.~\ref{fig:spectral}\textcolor{blue}{(a)}]; and, at smaller values of the temperature and the field, another separation for $m=10$ [see Fig.~\ref{fig:spectral}\textcolor{blue}{(b)}].
A large enough separation in the real part of the spectrum is known to correspond to the occurrence of metastability~\cite{Macieszczak2016a}, since for time $ -1/\lambda_{m+1}^R\ll t\ll -1/\lambda_{m}^R$ any system state can be approximated as stationary,
\begin{equation}\label{eq:Expansion}
	\rho(t)={\rho}_{\rm ss}+\sum_{k=2}^m {c}_{k} {R}_{k}+...,
\end{equation}
by neglecting the presence of the fast modes, $k>m$, and the decay of slow modes, $2\leq k\leq m$ [cf.~Eq.~\eqref{eq:Expansion0}]. Such states are called metastable and  the corresponding time regime referred to as  the metastable regime with the relaxation time scale $-1/{\lambda}_{m+1}^{R}$.
In particular, at intermediate values of the field and temperature, we have a hierarchy of two metastable regimes in the open quantum East model, and a hierarchy of the relaxation timescales given by $-1/{\lambda}_{11}^{R}$, $-1/{\lambda}_{8}^{R}$ and $-1/{\lambda}_{2}^{R}$.
A similar hierarchy of metastabilities  can be observed at other system sizes, which, as we will see, is a consequence of the classical and local structure of the manifold of metastable states and of the dynamics within.

\subsection{Hierarchy of metastable phases}\label{sec:phases}

The manifold of metastable states is fully characterised by linear combinations of the stationary state $\rho_\text{ss}$ and the low-lying modes $R_2,...,R_m$ with coefficients $(c_2,...,c_m)$ [cf.~Eq.~\eqref{eq:Expansion}]. However, the modes do not represent physical states of the system [as $\text{Tr}({R}_{k})=0$ for $k>2$ from orthogonality of the modes]. Nevertheless, we will show that the structure of the metastable manifolds in the open quantum East model is classical, with metastable states  approximated  as probabilistic mixtures of $m$ distinct metastable phases with localised excitations.

\subsubsection{Classicality}

\begin{figure}[t!]
	\includegraphics[width=1\linewidth]{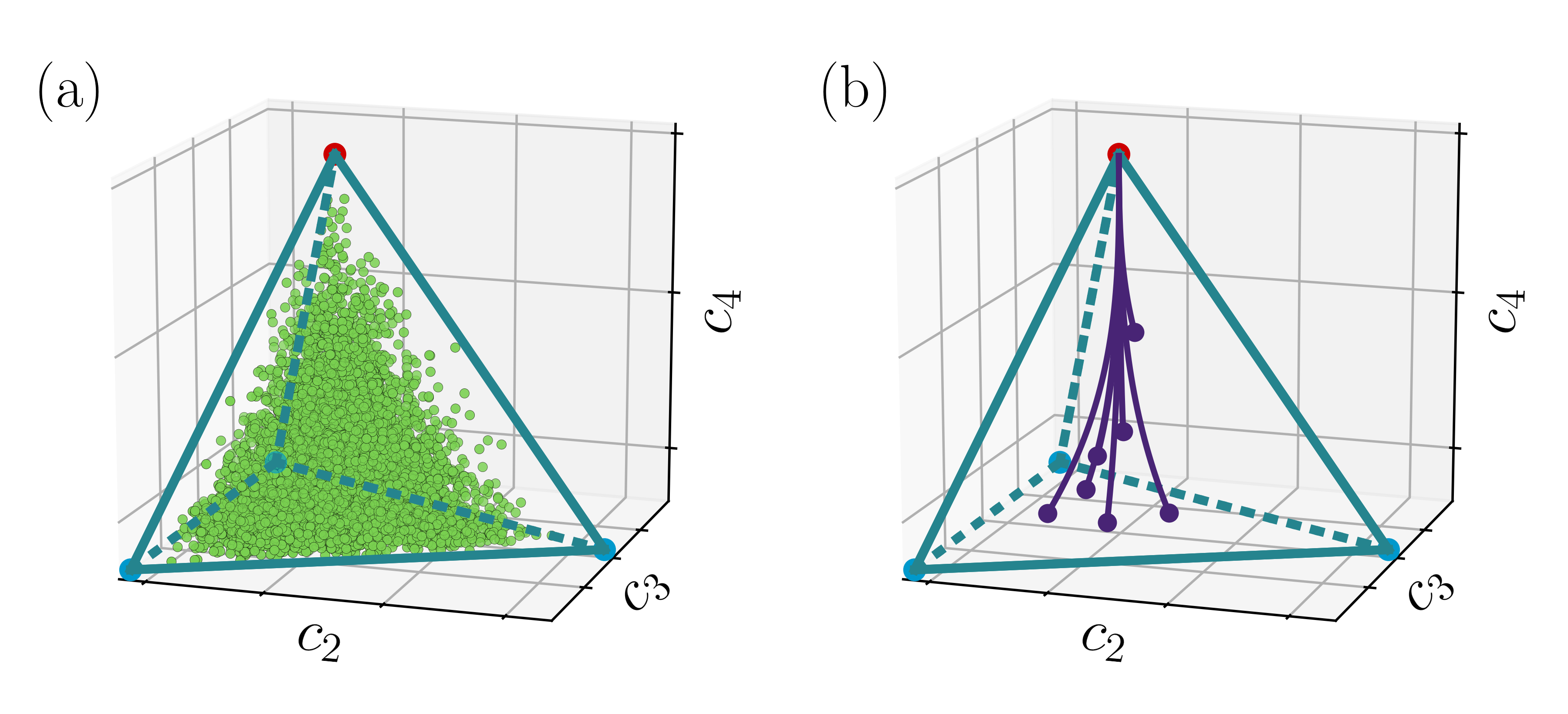}
	\vspace*{-6.75mm}
	\caption{\label{fig:3DMM}\textbf{Example of classical metastable manifold}: (a)~The metastable manifold in the open quantum East model of $N=3$ spins with periodic boundary conditions, illustrated by the coefficients $(c_2,c_3,c_4)$ [cf.~Eq.~\eqref{eq:Expansion}] of uniformly sampled pure initial states (small green dots). The blue lines show the simplex of $m=N+1=4$ metastable phases (found at the vertices). (b) The long-time evolution inside the metastable manifold towards the stationary state [red circle corresponding to (0,0,0) coefficients] for initial states shown in Fig.~\ref{fig:Spectrum}\textcolor{blue}{(b)}. 
		Softness $p=0.999$ and other parameters as in Fig.~\ref{fig:Spectrum}\textcolor{blue}{(a)}.
		\vspace*{+1mm}}
\end{figure}

For the system with  $N=3$ spins, a single gap at $m=4$ is present in the spectrum, so that the metastable manifold can be sampled by plotting the coefficients for random pure initial states as in Fig.~\ref{fig:3DMM}\textcolor{blue}{(a)}. We observe that the metastable manifold is classical, that is,  approximated by a simplex, with coefficients of any metastable state approximated by a probabilistic mixture of the coefficients corresponding to the simplex vertices, which describe states with a single or no excitation [cf.~Fig.~\ref{fig:Spectrum}\textcolor{blue}{(b)}]. Since for $m>4$, as relevant for larger system sizes, such a visual verification of metastable manifold classicality  is not possible,  we instead turn to the recently proposed approach from Ref.~\cite{Macieszczak2020}, which we sketch now.

For a set of $m$ candidate states $\rho_1,...., \rho_m$, the corresponding metastable states $\rhoss+\sum_{k=2}^m c_k^{(l)} R_k=\trho_l$, $l=1,...,m$, can be considered as the new physical basis replacing the low-lying modes, so that  [cf.~Eq.~\eqref{eq:Expansion}]
\begin{equation}\label{eq:Expansion2}
	\rho(t)= \sum_{l=1}^m \tilde{p}_l \, \trho_l+....
\end{equation}
Here, $\tilde{p}_l=\sum_{k=1}^m(\C^{-1})_{lk} c_k$ with $(\C)_{kl}= c_k^{(l)}$, are the barycentric coordinates with respect to the simplex of $\trho_1$, ..., $\trho_m$ in the coefficient space. When the distance of barycentric coordinates from probability distributions is negligible, the metastable state can be approximated as a probabilistic mixture of $\trho_1$, ..., $\trho_m$. If this is true for any metastable state, the metastable manifold is classical and we refer to $\trho_1$, ..., $\trho_m$ as \emph{metastable phases}.

\begin{figure}[t!]
	\vspace*{+2mm}
	\includegraphics[width=1\linewidth]{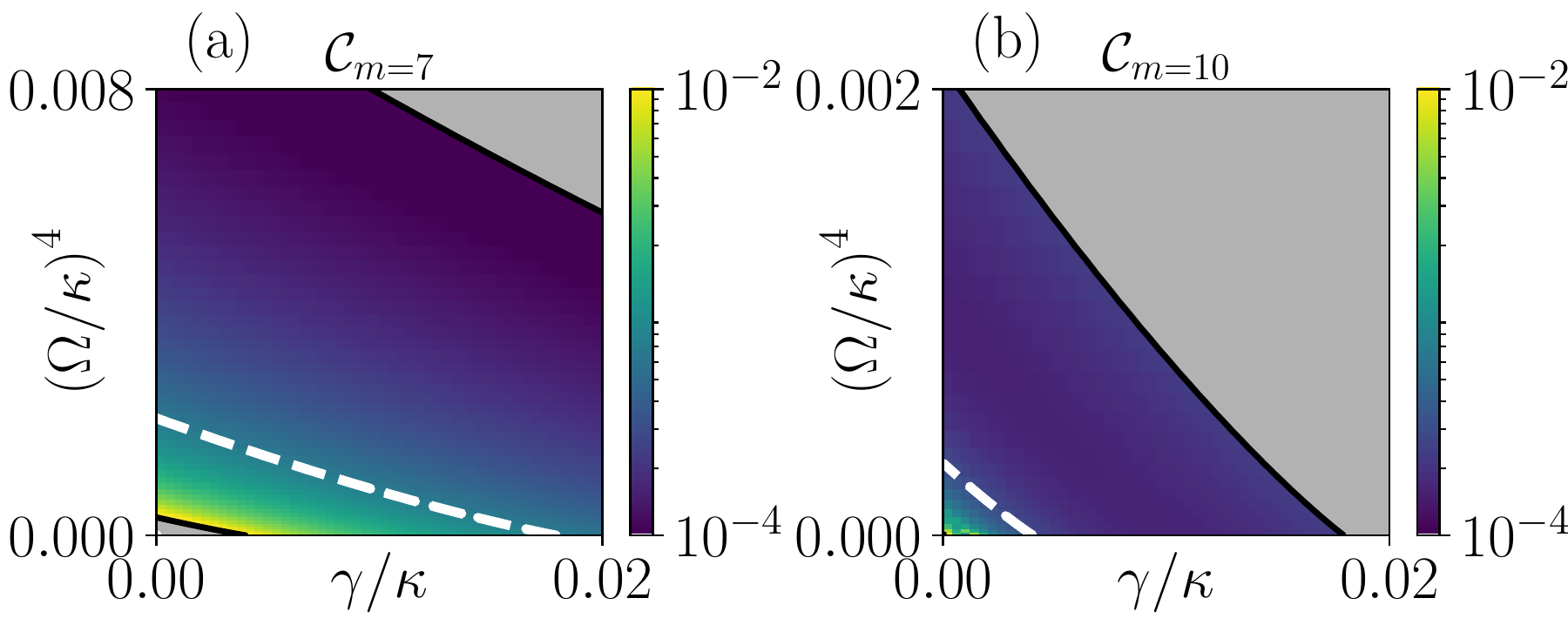}
	\caption{\label{fig:classicality}\textbf{Accuracy of classical approximation}: An upper bound on the average distance (on the maximal distance when multiplied by $2^{N-1}$) of barycentric coordinates to probability distributions for: (a) $m=7$, and (b) $m=10$. We consider $L1$ norm, in which probability distributions are normalised;  see Appendix~\ref{app:numerics_phases} for the definition of the bound. Softness $p=0.99$ and size $N=6$. Data is greyed out for parameters where  the relevant gaps are not present in the spectrum of master operator, while white dashed lines indicate that for smaller parameters in panel (a) the gap at $m=10$ is present, while in panel (b) the gap at $m=7$ is absent (cf.~Fig.~\ref{fig:spectral}).
	}	
\end{figure}

 For the range of temperatures and field amplitudes corresponding to presence of  gaps in the spectrum of the master operators (cf.~Fig.~\ref{fig:spectral}),  using a version of the algorithm from Ref.~\cite{Macieszczak2020} (see Appendix~\ref{app:numerics_phases} for details),  we found sets of $m$ states for which both the average distance and the maximal distance of barycentric coordinates to probability distributions are negligible; see Fig.~\ref{fig:classicality}. In particular,  Figure~\ref{fig:classicality}\textcolor{blue}{(a)} shows that the metastable manifold with $m=7$ is well approximated by $7$ metastable phases for broad regime of low temperatures and weak coherent field
exactly corresponding to the large separation at $m=7$ in the master operator spectrum -  with the parameter values above a certain threshold [shown as black line with grey region below; cf.~Fig.~\ref{fig:spectral}\textcolor{blue}{(a)}].
Below the threshold, the metastable manifold instead consists of $m=10$ metastable phases  [see~Fig.~\ref{fig:classicality}\textcolor{blue}{(b)} with the discussed threshold now shown as white dashed line], except for negligibly small values where  the separation at $m=10$ in the spectrum also disappears (cf.~Fig.~\ref{fig:spectral}). These phases remain metastable also above the threshold, but for a smaller range of values of the field and temperature than in the case of $m=7$, which correspond to the hierarchy of metastabilities, i.e., two  gaps in the spectrum of master operator  at $m=7$  and $m=10$ [cf.~Fig.~\ref{fig:spectral}\textcolor{blue}{(b)}].
These results are in agreement with the perturbation theory results derived in Appendix~\ref{app:PT}, which predict emergence of a classical manifold from a quantum metastable manifold at small $\gamma$ and $\Omega$, but at zero temperature and in the absence of the field indicate that softening the constraint leads to quantum decay of excitations within the quantum metastable manifold (cf.~Sec.~\ref{sec:PT}).

We conclude that, at the chosen soft constrain, the  \emph{metastable manifolds are classical} for low temperatures and weak coherent fields (except for their negligibly small values) and there exists an intermediate parameter region with a hierarchy of metastabilities corresponding to two classical metastable manifolds. We will understand the emergence of hierarchy by studying properties of the metastable phases and their long-time dynamics.

\begin{figure*}[t!]
	\includegraphics[width=1\linewidth]{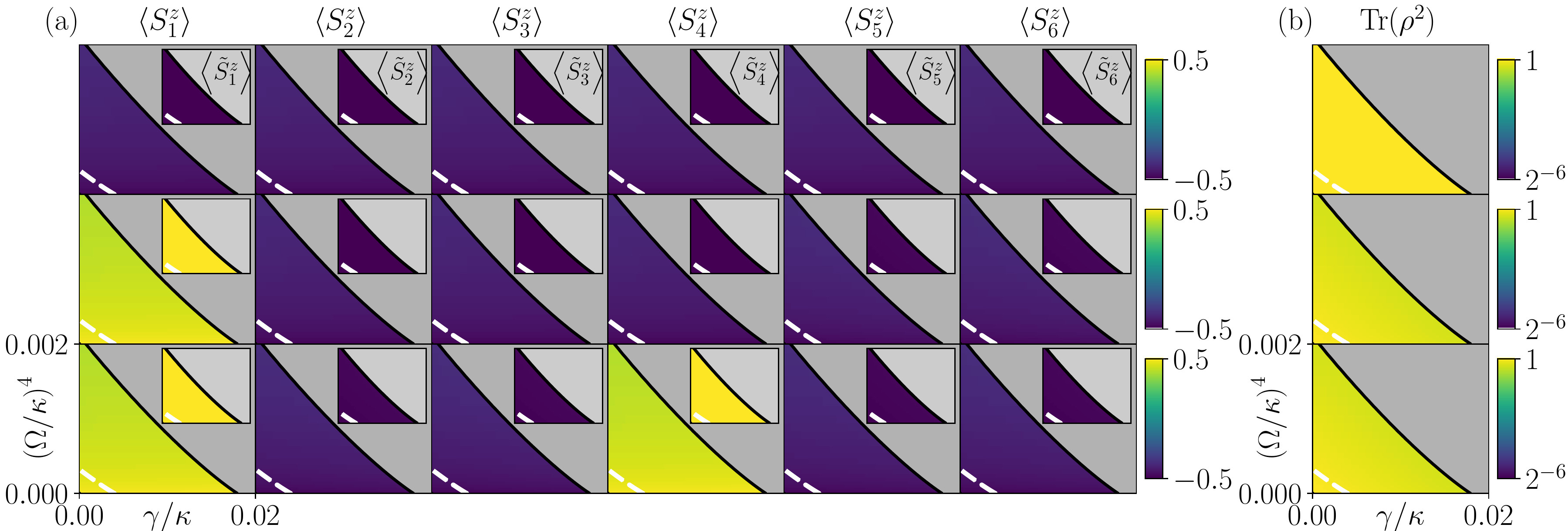}
	\caption{\label{fig:eMSProp}\textbf{Classical metastable manifold}: (a) The average site-wise $z$-magnetisation for the metastable phases: dark state, and examples of the single- and double-excitation states, respectively. The set of all metastable phases is formed by adding all translations of these states with single (6 states) and double (3 states) excitations. The insets show the site-wise $z$-magnetisation in the $\ket{u}$, $\ket{e}$ basis,  $\tilde{S}^{z}=(\ket{e}\!\!\bra{e}-\ket{u}\!\!\bra{u})/2$. (b) The purities of the metastable phases, chosen as in panel (a). Data is greyed out for parameters where the gap in the spectrum is not present at $m=10$; cf.~Fig.~\ref{fig:spectral}\textcolor{blue}{(b)}.
	}
\end{figure*}

\subsubsection{Metastable phases}

We now discuss the properties of the metastable phases whose probabilistic mixtures approximate the classical manifold of metastable states present in the open quantum East model at low temperatures and weak field [cf.~Eq.~\eqref{eq:Expansion2}]. We focus on the parameter regime where there exists a gap in the master operators at $m=10$ [below white dashed line in Fig.~\ref{fig:classicality}\textcolor{blue}{(a)}], which captures all values for which a hierarchy of metastabilities is present, but also a region where a gap at $m=7$ is absent [below white dashed line in Fig.~\ref{fig:classicality}\textcolor{blue}{(b)}].

In Fig.~\ref{fig:eMSProp}\textcolor{blue}{(a)}, we show the spin magnetisation along $z$-axis for the metastable phases.
For $m=7$ (first two rows, above the white dashed line), the metastable manifold consist of the state with all spins down (no excitation), and six states with a single spin up (a single excitation).
For $m=10$, the manifold additionally contains three states with two excitations at maximally separated sites, i.e., followed by two empty sites [see third row in  Fig.~\ref{fig:eMSProp}\textcolor{blue}{(a)}].

As the probability of a spin up or down seems to decrease with the stronger coherent field, we also confirm (see the insets), that the spins in metastable phases are actually aligned with the rotated eigenbasis, $|u\rangle$ and $|e\rangle$, of the stationary state [see Eq.~\eqref{eq:rho_ss1} and Appendix~\ref{app:1spin}].
Therefore, the metastable phases with no excitations, single excitation and two excitations can be approximately viewed as $|uuuuuu\rangle$, $|euuuuu\rangle$, $|euueuu\rangle$, respectively, and their translations. We obtained such a structure in the first-order perturbation theory with respect to temperature ($|000000\rangle$, $|100000\rangle$, $|100100\rangle$ with translations; see Appendix~\ref{app:temp}), and now we confirm it is the case in the presence of the coherent field.

These pure states, however, are not stable since the presence of an excited spin facilitates dynamics on the spin to its right, in turn facilitating dynamics further along the chain: the metastable states thus feature excitations as much separated as possible, so that the relaxation is as slow as possible.
Furthermore, the dynamics facilitated by these excitations cause photon emissions from their right neighbour, resulting in a mixed rather than pure metastable state, i.e., $|e\rangle\!\langle e|\otimes |u\rangle\!\langle u|$ replaced by $|e\rangle\!\langle e|\otimes\rho_\text{ss,1}$ (cf.~Appendix~\ref{app:obc} and Sec.~\ref{sec:dynhet}).
This is confirmed by the purity of the metastable phases in Fig.~\ref{fig:eMSProp}\textcolor{blue}{(b)}, where
the phases with a single or double excitation feature a purity slightly below $1$, with a lower purity for the state with more excitations.
Furthermore, in the first-order corrections due to temperature, purity is lowered proportionally to $\gamma/\kappa$, and Figure~\ref{fig:eMSProp}\textcolor{blue}{(b)} suggests it is also the case for the coherent field, with the lowest order contribution scaling with $\Omega^4/\kappa^4$.

Finally, we note that the pure states are exactly orthogonal, and thus the metastable phases are approximately disjoint, as expected from the general theory~\cite{Macieszczak2020}.  Furthermore, the set metastable phases is invariant under the translation symmetry, which is a consequence of the metastable manifold inheriting the symmetry of the dynamics in Eq.~\eqref{eq:lindblad-op} with periodic boundary conditions~\cite{Minganti2018,Macieszczak2020}.

\section{Classical long-time dynamics}\label{sec:effdyn}

After a metastable regime, $t\gtrsim -1/\lambda_{m}^R$, the decay of low-lying modes can no longer be neglected [cf.~Eqs.~\eqref{eq:Expansion0} and~\eqref{eq:Expansion}],
\begin{equation}\label{eq:truncated-evolution}
\rho(t)={\rho}_{\rm ss}+\sum_{k=2}^m {c}_{k} e^{t\lambda_k}{R}_{k}+....
\end{equation}
Nevertheless, since the contribution from the fast modes can be neglected, the long-time dynamics takes place essentially inside the metastable manifold [see Figs.~\ref{fig:Spectrum}\textcolor{blue}{(b)} and~\ref{fig:3DMM}\textcolor{blue}{(b)}].

In the basis of metastable phases, the long-time dynamics corresponds to the dynamics of barycentric coordinates [cf.~Eq.~\eqref{eq:Expansion2}]
\begin{equation}\label{eq:phase-evolution}
	\rho(t)= \sum_{l=1}^m \tilde{p}_l(t) \, \trho_l+...,
\end{equation}
where $\tilde{p}_l(t)=\sum_{k=1}^m (\mathbf{C}^{-1})_{lk} e^{t\lambda_k} c_k$.  The dynamics is linear,
 \begin{equation}\label{eq:pdt}
	\frac{\mathrm{d}}{\mathrm{d}t}\, \tilde{p}_l(t) =\sum_{k=1}^m (\Wt)_{lk} \,\tilde{p}_k(t),
\end{equation}
with   $(\Wt)_{lk}= \sum_{n=1}^m (\mathbf{C}^{-1})_{ln}  \lambda_{n} (\mathbf{C})_{nk} $.
This generator corresponds to the master operator in Eq.~\eqref{eq:lindblad-op} expressed in the metastable phase basis (when it is restricted to low-lying modes) and  we will use it to understand the physical properties of the long-time dynamics in the open quantum East model from a classical perspective.

\begin{figure}[t!]
	\includegraphics[width=1\linewidth]{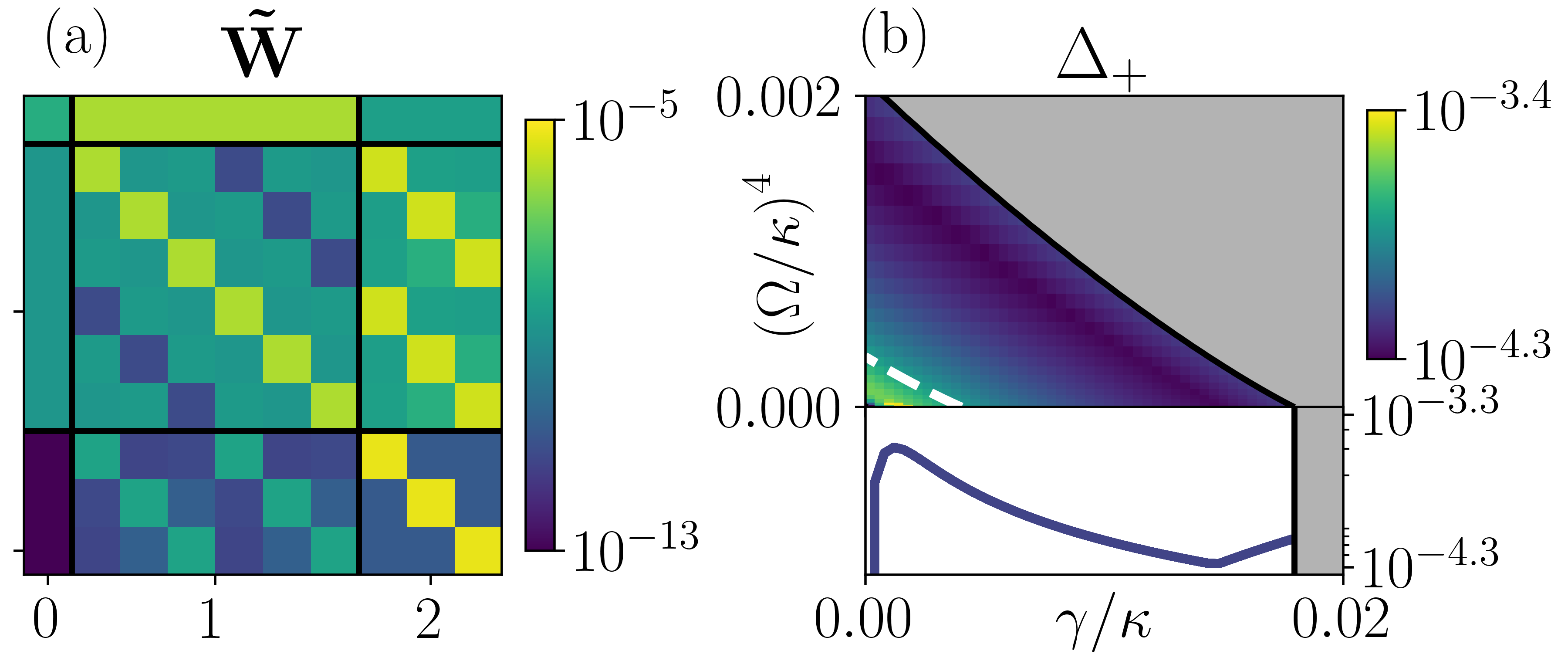}
	\caption{\label{fig:EffDyn} \textbf{Effective classical generator}: (a) Absolute values of the effective master operator entries in the basis of metastable phases for $\Omega/\kappa=0.024$, $\gamma/\kappa=0.0016$, and $m=10$. 
		The horizontal labels indicate the number of excitations in a metastable phase, delineated by the black lines. (b) The normalised distance $\Delta_+$ to the closest classical stochastic generator  over (top) the metastable region of the parameter space and (bottom) the classical $\Omega/\kappa=0$ cross section. We consider the operator norm induced by $L1$ norm; see Appendix~\ref{app:Wdistance}. Data is greyed out in panel (b) for parameters where the gap at $m=10$ in the spectrum is not present; cf.~Fig.~\ref{fig:spectral}\textcolor{blue}{(b)}.}
	\vspace*{-3mm}
\end{figure}

\subsection{Properties of long-time dynamics} \label{sec:effdyn_DB}

We now verify that the dynamics within the metastable manifold is classical. This enables us to investigate classical features in the dynamics characteristic of the classical East model: the presence of the detailed balance and  the absence of interactions in the stationary state.

\subsubsection{Classicality}

The effective generator $\Wt$, pictured in Fig.~\ref{fig:EffDyn}\textcolor{blue}{(a)} encodes all information needed to predict the evolution of the average system state at long times. It conserves the sum of barycentric coordinates, i.e., $\sum_{l=1}^m (\Wt)_{lk}=0$,
which is a consequence of the master operator in Eq.~\eqref{eq:lindblad-op} being trace-preserving~\cite{Macieszczak2020}. Although it does not generate positive dynamics (cf.~Appendix~\ref{app:W}), its diagonal elements are negative, while its off-diagonal approximately positive, so that it can be  approximated by a classical stochastic generator (cf.~Appendix~\ref{app:Wdistance}). Importantly, Figure~\ref{fig:EffDyn}\textcolor{blue}{(b)} confirms that the \emph{effective dynamics can be approximated by classical stochastic dynamics} across the entire metastable region of the parameter space, with the normalised distance of $\Wt$ to the set of classical stochastic generators much smaller than $1$. In fact, this is a  consequence of the classicality of the metastable manifold~\cite{Macieszczak2020} we discussed in~Sec.~\ref{sec:phases}.

We also note that in Fig.~\ref{fig:EffDyn}\textcolor{blue}{(a)}, the dynamics features the translation symmetry, i.e., $(\Wt)_{\pi(l)\pi(k)}=(\Wt)_{lk}$, where $\pi$ is the permutation that  the metastable phases undergo under the translation of spins [cf.~Fig.~\ref{fig:eMSProp}\textcolor{blue}{(a)}]. This symmetry is inherited from the translation symmetry of the open quantum East model with periodic boundary conditions~\cite{Macieszczak2020}. While it reduces the free parameters of the effective dynamics [to $10$ for $N=6$  and $m=10$], it does not guarantee the presence of detailed balance we demonstrate next.

\subsubsection{Detailed balance}

\begin{figure}[t!]
	\includegraphics[width=1\linewidth]{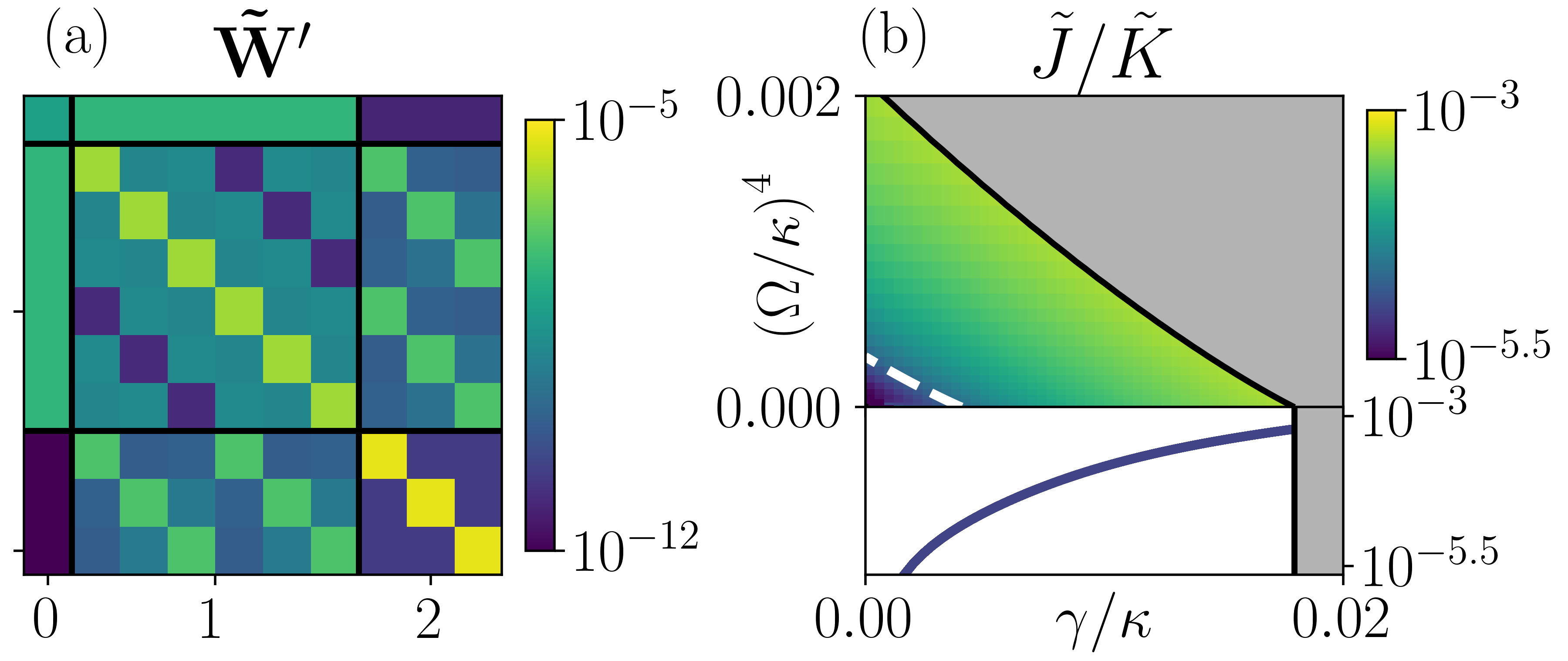}
	\caption{\label{fig:EffDyn1} \textbf{Detailed balance}: (a) Absolute values of entries in the similarity-transformed effective master operator in Fig~\ref{fig:EffDyn}\textcolor{blue}{(a)} [cf.~Eq.~\eqref{Wprime}] display the transposition symmetry associated with the detailed balance. Furthermore, the diagonal elements are negative while all the off-diagonal are positive. (b) The ratio of the total current and total activity in the stationary state [Eqs.~\eqref{eq:Jtilde} and~\eqref{eq:Ktilde}] over (top) the metastable region of the parameter space and (bottom) the classical $\Omega/\kappa=0$ cross section  confirms the approximate detailed balance. Data is greyed out in panel (b) for parameters where the gap at $m=10$ in the spectrum is not present; cf.~Fig.~\ref{fig:spectral}\textcolor{blue}{(b)}.}
	\vspace*{-3mm}
\end{figure}

In the effective dynamics, the stationary probability current between the $k$th and $l$th metastable phase is given by $(\Wt)_{kl} (\ptss)_l -(\Wt)_{lk} (\ptss)_k$,
where $\ptss$ is the stationary distribution of $\Wt$ (or equivalently the barycentric coordinates for $\rhoss$).
Detailed balance is then defined to be when a systems stationary state exhibits no currents (see Appendix~\ref{app:detailed-balance}).

As a first check of detailed balance in the effective dynamics, we consider the similarity transformation which renders classical detailed-balance generators symmetric,
\begin{equation}\label{Wprime}
 (\Wt')_{lk}=(\ptss)_l^{-\frac{1}{2}} (\Wt)_{lk}(\ptss)_k^{\frac{1}{2}}.
\end{equation}
For the effective generator in Fig.~\ref{fig:EffDyn}\textcolor{blue}{(a)}, we indeed obtain an approximately symmetric matrix in Fig.~\ref{fig:EffDyn1}\textcolor{blue}{(a)}.

 To verify detailed balance across the range of parameters for which metastability occurs, we consider in Fig.~\ref{fig:EffDyn2}\textcolor{blue}{(b)}  the ratio of the total stationary current
\begin{equation}
\tilde{J}=\frac{1}{2}\sum_{k,l=1}^m\left|(\Wt)_{kl}(\ptss)_l-(\Wt)_{lk}(\ptss)_k\right|\label{eq:Jtilde}
\end{equation}
to the total activity
\begin{equation}
\tilde{K}=-\sum_{l=1}^m (\Wt)_{ll}(\ptss)_l,\label{eq:Ktilde}
\end{equation}
which ratio bounds the normalised distance to the closest detailed balance dynamics (see~Appendix~\ref{app:detailed-balance}). We observe that the current across all metastable parameters is small compared to the system's activity, and thus the long-time dynamics can be well approximated by dynamics with detailed balance. This is also the case for the classical East model ($\Omega=0$), which features only approximate detailed balance when restricted to the metastable manifold [see the bottom panel in Fig.~\ref{fig:EffDyn1}\textcolor{blue}{(b)}], although
 there are no stationary currents between the $2^N$ configurations of up and down spins in the classical system.

For the classical model, these results can be traced back to the perturbative dynamics between configurations.
The perturbation effect of the soft constraint removes one excitation at a time with rates proportional to $(1-p)^2\kappa$, or reintroduces, removes or shifts a single excitation, at rates proportional to $(1-p)^2\gamma$.
For the small temperature, phases with double excitations are reduced to a single excitation at rates proportional to $\gamma^2/\kappa$, while at rates proportional to $\gamma^3/\kappa^2$ a second excitation can be introduced or removed, or a single excitation can be shifted.
The result is a ladder structure of the dynamics with respect to the number of excitations which necessarily implies detailed balance, though the approximation worsens for larger $\gamma$ or $(1-p)$ due to higher order corrections; see Appendices~\ref{app:temp} and~\ref{app:soft} for details.

Approximate detailed balance observed also in the presence of a weak coherent field in Fig.~\ref{fig:EffDyn1}\textcolor{blue}{(b)}, suggests that a similar mechanism may be responsible for the long-time dynamics in the open quantum East model.
This is indeed confirmed for the parameters chosen in Fig.~\ref{fig:EffDyn}\textcolor{blue}{(a)}: the most probable transitions (yellow-light green) are associated with the removal of the second excitation, or removal of a single excitation towards the unexcited state; the less likely transitions (green) correspond to a shift of a single excitation, the introduction of one excitation, or removal of two excitations; while the least likely transitions (blue) correspond to the introduction of two excitations simultaneously, or shift of two excitations [see also Figs.~\ref{fig:EffDyn2}\textcolor{blue}{(b)} and~\ref{fig:EffDyn2}\textcolor{blue}{(c)}].

\subsubsection{Non-interacting stationary state}
\begin{figure}[t!]
	\includegraphics[width=1\linewidth]{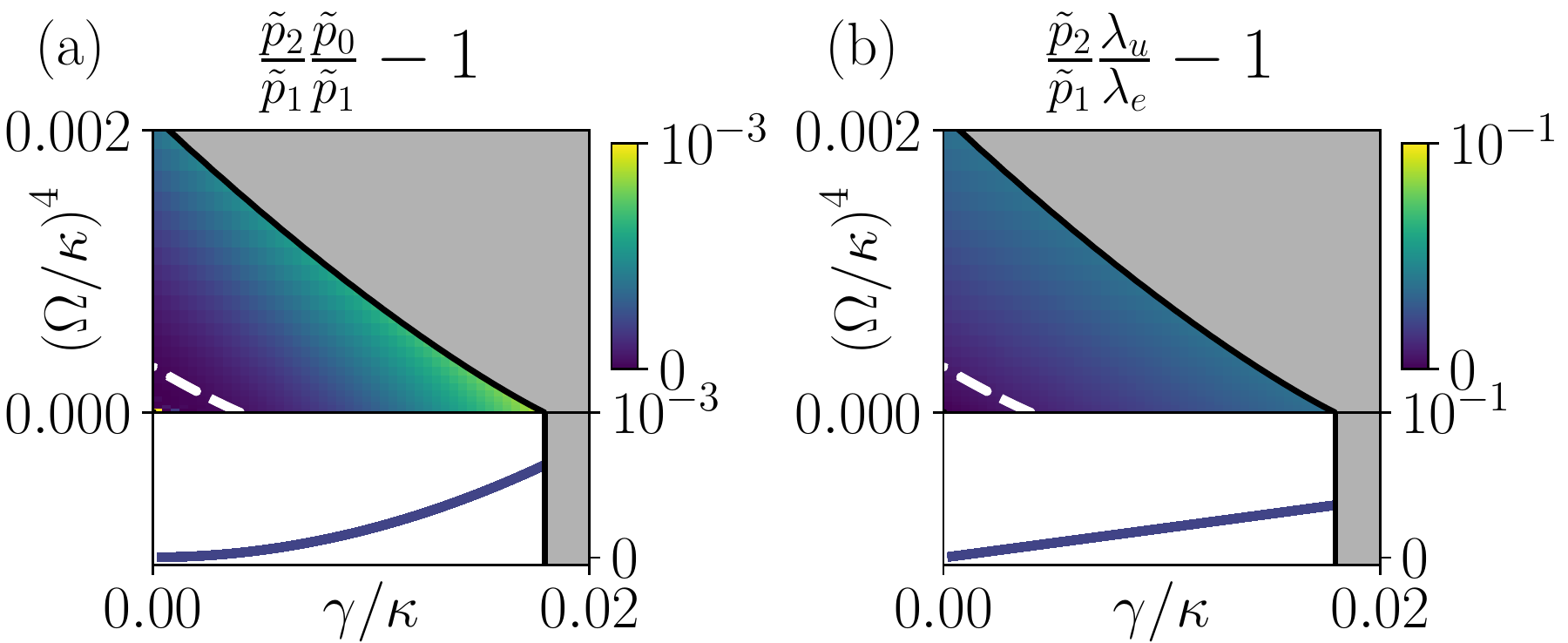}
	\caption{\label{fig:SteadyState} \textbf{Stationary state properties}: (a) The free energy of the metastable phases does not depend on the number of interactions, as the ratio of the stationary probabilities between the state with a single excitation and the unexcited state, $\tilde{p}_1/\tilde{p}_0$ approximately equals the ratio of the probabilities for a double excitation and a single excitation $\tilde{p}_2/\tilde{p}_1$. (b) The ratio $\tilde{p}_1/\tilde{p}_0$ actually corresponds to the ratio of the stationary probabilities in as single state dynamics, Eq.~\eqref{eq:rho_ss1}, of the excited state, $\lambda_e$, and unexcited state, $\lambda_u$. Both plots are shown over (top) the metastable region of the parameter space and (bottom) the classical $\Omega/\kappa=0$ cross section. Data is greyed out for parameters where the gap at $m=10$ in the spectrum is not present; cf.~Fig.~\ref{fig:spectral}\textcolor{blue}{(b)}.}
		\vspace*{-3mm}
\end{figure}

Since the long-time dynamics possesses approximate detailed balance, the stationary state of the effective master operator is effectively thermal. We discuss now its effective free energy function.

In the full classical East model, whose stationary state is a product state, the free energy is simply a function of the number of excitations, but not their relative distance (i.e., it is not dependent on any type of interactions).
Thanks to the exponential form of a thermal distribution, this can be tested by considering ratios of state probabilities: the exponents only state dependence will be a linear function of the number of excitations.
We test this property in Fig.~\ref{fig:SteadyState}\textcolor{blue}{(a)} for  the distribution over the metastable phases of the stationary state of the the open quantum East model, by comparing the ratio of probabilities of finding the stationary state in one of the single or double excited states, $\tilde{p}_2/\tilde{p}_1$, to the ratio of finding it in the unexcited state or one of the single excitation states, $\tilde{p}_1/\tilde{p}_0$.
We would expect these ratios to differ when interactions contribute to the effective energy of the phase, due to the presence of multiple excitations in the phase with probability $\tilde{p}_2$.
However, we find that the ratio of these ratios is close to $1$ at all metastable parameters, suggesting that in the quantum model interactions do not play a role in the free energy of the metastable phases, as in the classical East model.

Furthermore, the free energy per excitation is directly determined by the ratio of probabilities for excited and unexcited states in the single-site dynamics; see Fig.~\ref{fig:SteadyState}\textcolor{blue}{(b)}.
This is again the consequence of the product structure of the stationary state, and the metastable states being approximated simply as pure states with a single or double excitations [cf.~Fig.~\ref{fig:eMSProp}\textcolor{blue}{(a)}], which appear as the leading order corrections to the stationary state above the no-excitation contribution
\footnote{In the case of the classical model ($\Omega=0$), in the perturbative regime, the metastable manifold is known to consist of isolated excitations at any system size (see Appendix~\ref{app:PT}), and thus the free energy over metastable phases is again only the function of number of excitations, and so we can expect, also in the open quantum East model for larger system sizes}.
Therefore,  the error of this non-interacting approximation of the free energy will increase as temperatures and coherent field values grow [cf.~Fig.~\ref{fig:SteadyState}\textcolor{blue}{(b)}].

\subsection{Dynamical heterogeneity}\label{sec:dynhet}

\begin{figure*}[t!]
	\includegraphics[width=1\linewidth]{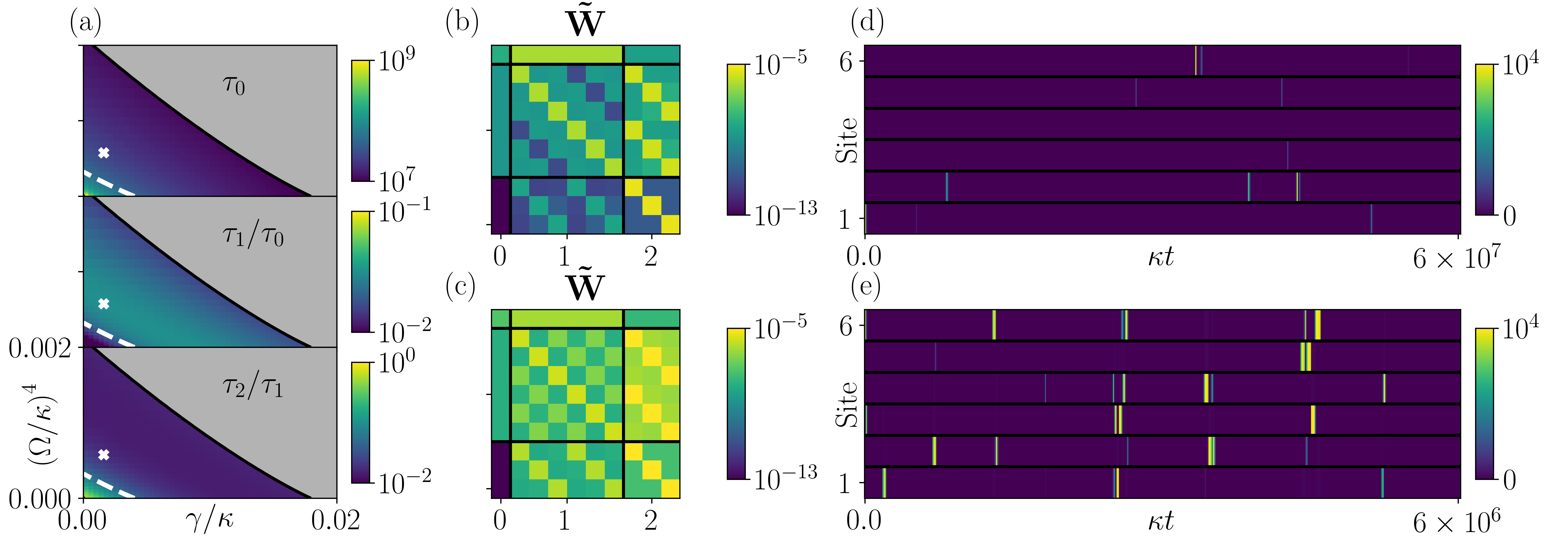}
	\caption{\label{fig:EffDyn2} \textbf{Effective stochastic dynamics and quantum trajectories}: (a) Effective lifetimes of the metastable phases as a function of the parameters, plotted relative to the next longest lifetime. Data is greyed out for parameters where the gaps at $m=10$ in the spectrum is not present; cf.~Fig.~\ref{fig:spectral}\textcolor{blue}{(b)}.
		(b) and (c) The component magnitudes of the effective master operators for (b) $\Omega/\kappa=0.024$, $\gamma/\kappa=0.0016$ and (c) $\Omega/\kappa=0.12$, $\gamma/\kappa=0.0096$. (d) and (e) Sample QJMC trajectories corresponding to panels (b) and (c), respectively. Trajectories are split into 500 time-bins for which the total activity of jumps $J_j^-$ (photon emissions from $j$th spin) within each bin is plotted [see Eq.~\eqref{eq:J-}]. Note the presence of simultaneous excitations from two sites.}
\end{figure*}

The long-time dynamics between metastable phases is directly related to dynamics of quantum trajectories: periods of higher or lower activity in trajectories are identifiable with metastable phases featuring different numbers of excitations, with transition rates between these distinct periods described by the effective generator~\cite{Rose2016,Macieszczak2020}.
We now discuss this correspondence in terms of the dynamical heterogeneity observed in the open quantum East model~\cite{Olmos2012,Lesanovsky2013}. We also demonstrate the resulting proximity to dynamical phase transitions~\cite{Garrahan2010}.

\subsubsection{Lifetimes of metastable phases}
The effective lifetimes of the metastable phases, i.e., the distinct periods in quantum trajectories, are given by the inverse magnitude of the diagonal elements of the effective generator.
From the translation symmetry of the metastable manifold, the lifetimes of metastable phases connected under the symmetry must be the same, i.e., we have: the lifetime $\tau_0$ of the homogeneous unexcited phase, the lifetime $\tau_1$ of six phases with a single excitation, and the lifetime $\tau_2$ of three phases with double excitations [see Fig.~\ref{fig:EffDyn2}\textcolor{blue}{(a)}].
	
The unexcited state is the longest lived metastable phase, as
it can only be excited via to the softness of constraint in $H_j$ or $J_j^+$ (for any $j=1,...,N$) [cf.~Eq.~\eqref{eq:HJJ}], which takes place at the rate proportional to $N (1-p)^2(\Omega^2/\kappa +\gamma)+...$ in the perturbative regime $|1-p|\ll 1$  (see Appendix~\ref{app:soft_finite}).
The simple dependence of the rate on the system size $N$ is due to the translation symmetry: the excitation can be introduced at any of the $N$ spins.
In contrast, the removal of a single excitation of $j$th spin, again requires the softness of constraint in $J_j^-$ (or in $H_j$, which however is of lower order), and thus takes place at the faster rate $\kappa (1-p)^2+...$, leading to the separation of timescales $\tau_1/\tau_0\approx N (\Omega^2/\kappa^2 +\gamma/\kappa)$.
Furthermore, when two gaps are present in the master operator spectrum [above the white dashed line in Fig.~\ref{fig:EffDyn2}\textcolor{blue}{(a)}], the hierarchy of metastabilities  ($m=7,10$; cf.~Sec.~\ref{sec:phases}) is manifested in the distinct values of the lifetimes $\tau_1$ and $\tau_2$, while for a single gap ($m=10$; below the white dashed line), these lifetimes are necessarily comparable, which we now explain.

The softness of constraint causes the removal rate of an excitation from metastable phases with a single excitation and double excitation to be the same (except from the fact that two possible sites to decay in the double excited state), while for hard constraint only the second excitation can be removed due to the temperature of the coherent field (by flipping the unexcited spins between two excitations which allows for the hard constraint to be fulfilled).
When the constraint is soft, the absence or presence of separation between  $\tau_1$ and $\tau_2$ is determined by the softness-induced and temperature-induced dynamics being faster, respectively.
In Fig.~\ref{fig:EffDyn2}\textcolor{blue}{(a)}, the regime of smaller (greater) $\gamma$ and $\Omega$ below (above) the threshold corresponds to the former (latter) process being the fundamental mechanism in relaxation of the metastable phase with double excitation towards the stationary state.

\subsubsection{Structure of effective dynamics}
In Figs.~\ref{fig:EffDyn2}\textcolor{blue}{(b)} and~\ref{fig:EffDyn2}\textcolor{blue}{(c)}, we show examples of the effective master operator for two sets of parameters, indicated by the crosses in Fig.~\ref{fig:EffDyn2}\textcolor{blue}{(a)}, corresponding to the cases with a single metastability and a hierarchy of two metastabilities.
In both cases, a double-excitation phase is most likely transformed into one of two single-excitation phases (equally likely due to the translation symmetry by $3$ sites of the double-excitation phase), with one excitation inducing relaxation of the other.
A single excitation phase is most likely transformed into no-excitation phase, which in turn gets excited most likely with only a single excitation (into one of six single-excitation phases).
This ladder structure of the effective classical dynamics supports detailed balance in the dynamics discussed in Sec.~\ref{sec:effdyn_DB}.
	
Beyond those leading order transformations, however, shifts of a single excitations are possible due to two different mechanisms.
In Fig.~\ref{fig:EffDyn2}\textcolor{blue}{(b)}, we observe that the shift of a single excitation is possible to all sites except that corresponding to the possible position of a second excitation, in which case the second excitation is introduced instead.
This indicates that the shift is actually facilitated by the introduction of excitations and their subsequent decay, which can be facilitated either by several excitations by the temperature/coherent field, or the softness of constraint allowing for introduction of excitations directly in unexcited sites.
The uniform probability of shifts to different sites in Fig.~\ref{fig:EffDyn2}\textcolor{blue}{(c)}, suggest that the two processes contribute equally, while, in the case of hierarchy in Fig.~\ref{fig:EffDyn2}\textcolor{blue}{(b)}, the larger values of the coherent field and temperature dominate the latter process and only some shifts are possible.
This is directly supported by the perturbation results in the classical model; see Appendix~\ref{app:PT}. We note however, that for considered system size of $N=6$ and the chosen constraint with $p=0.99$, we do not yet capture the hallmark behaviour of the classical East model, where required order of temperature contributing to the dynamics of excitations scales logarithmically with their distance (cf.~Appendix~\ref{app:dynamicsT}). In particular, we cannot verify whether  the  necessarily (quadratically) higher orders in which the local coherent field contributes to the dynamics of in the open quantum East model  (cf.~Appendix~\ref{app:dynamicsOmega2}) alter this characteristic.

Although there is no apparent directionality in the dynamics for both cases, which is likely due to the small system size $N=6$ (cf.~Appendix~\ref{sec:finitesize}), we would in general expect this to follow from the presence of the constraint to the left [cf.~Eq.~\eqref{eq:HJJ}], and such directionality is present in perturbation theory with respect to temperature for larger system sizes. Nevertheless, we observe in both cases that the unexcited metastable lifetime is much longer than the metastable phases with a single or double excitations; in trajectories of the classical effective dynamics most time is spent in the unexcited state, and excitations are present at isolated moments in time. These periods are also isolated in space, due to the symmetry structure of the metastable manifold (see also Appendix~\ref{app:obc}).

\subsubsection{Dynamical heterogeneity}
We now discuss how metastability and the structure of long-time dynamics manifests itself in the emission patterns in individual experimental realisations of the system dynamics~\cite{Olmos2012,Lesanovsky2013}.
	
Consider first dynamics in the case of the state being (on average) in one of the metastable phases featuring a single or double excitations.
An excitation of site $(j-1)$ fulfils the hard constraint of the single spin dynamics of the $j$th spin [cf.~Eqs.~\eqref{eq:HJJ1} and~\eqref{eq:HJJ}], enabling dynamics on this site and thus its relaxation towards the single-spin stationary state in Eq.~\eqref{eq:rho_ss1}. Thus, for times shorter than relaxation of the considered metastable phase, $t\lesssim \tau_1,\tau_2$, in an individual realisation of an experiment (or quantum trajectories obtained in QJMC simulations) photon emissions occur from $j$th spin corresponding to the jump $J_j^-$ [Eq.~\eqref{eq:J-}], so that the metastable phase with an excitations appears \emph{locally bright}.
These emissions occur at the rate $\kappa \Tr(S^+ S^-\rho_\text{ss,1}) \approx\Omega^2/\kappa+\gamma+...$,
so that the site next to the excitation appears brighter for higher temperature or coherent field values [see Figs.~\ref{fig:EffDyn2}\textcolor{blue}{(d)} and~\ref{fig:EffDyn2}\textcolor{blue}{(e)}].
In contrast, for the unexcited metastable phase, the hard constraint in the dynamics is not fulfilled, and therefore, this phase  appears \emph{dark} in quantum trajectories, before it relaxes due to the soft constraint introducing of a single excitation at  $t\gtrsim \tau_0$. 
	
At longer times, higher order processes introducing several excitations or exploiting softness of constraint become non-negligible on average, contributing to the long-time dynamics of the metastable phases by connecting disjoint parts of state space.
In individual quantum trajectories these processes take place separately and at fluctuating times, but are typically followed by the fast decay of excitations due to satisfied hard constraints (on timescales $t\lesssim  -1/\lambda_{m+1}^R$; $m=10$) towards another metastable phase.
Therefore, a time coarse-graining of quantum trajectories leads to the system transitioning only between metastable phases.
As averaging over trajectories returns the evolution with the master operator [Eq.~\eqref{eq:rhodt}], transitions in coarse-grained quantum trajectories must be governed by the effective long-time generator [Eq.~\eqref{eq:pdt}].
This is corroborated in Fig.~\ref{fig:EffDyn2}\textcolor{blue}{(d)} where, correspondingly with the transition rates of the effective stochastic generator, in the case of a single metastable regime, mostly transitions between the excited states and dark states are observed. Meanwhile, in Fig.~\ref{fig:EffDyn2}\textcolor{blue}{(e)},  for the hierarchy of two metastabilities, there is also a significant presence of transitions between states with a single excitation, shifting the location of emissions.
		
We conclude that  the dynamical heterogeneity in the quantum trajectories is the microscopic counterpart to the classical stochastic jumps between phases with different numbers of excitations at different sites, which arise as a result of \emph{temporal coarse-graining of quantum trajectories}. This is a general phenomenon, detailed theoretically in Ref. \cite{Macieszczak2020}. In particular, this relation could be used to explore possible differences in the contributions to the dynamics from the temperature and the coherent field at larger system sizes accessible in QJMC simulations.

\subsubsection{Proximity to dynamical phase transitions}\label{sec:DPT}

Systems with intermittent dynamics are commonly found to exist near a dynamical phase transition in the statistics of the activity, i.e., the number of jumps per unit time~\cite{Garrahan2010,Garrahan2011,Ates2012,Lesanovsky2013}.
Here, we demonstrate this for the global activity for  jumps related to loss of excitations.
Since our system exhibits dynamical heterogeneity, we also  find the system in proximity to  transitions in the statistics of the local activity.\\

{\bf \em Dynamical phase transitions in global jump activity}.
The intermittent emissions in trajectories have a direct effect on the time integrated statistics of their corresponding jumps.
The statistics of a trajectory-observable chosen as the number $K^-(t)$ of $J_j^-$ jumps up to time $t$ summed across all sites,  is encoded by the cumulant generating function
\begin{equation}\label{eq:cgf}
\Theta(s,t)=\ln [Z(s,t)],
\end{equation}
where
\begin{eqnarray}\label{eq:mgf}
	Z(s,t)=\sum_{K^-}p(K^-,t)e^{-sK^-}=\Tr(e^{t\mathcal{L}_s\rho}),
\end{eqnarray}
can be obtained using the biased master operator
\begin{equation}
	\label{eq:Ls}
	\L_s(\rho)=\L(\rho)+(e^{-s}-1)\sum_{j=1}^N J_j^- \rho \,J_j^{-\dagger}
\end{equation}
 [cf.~Eq.~\eqref{eq:lindblad-op}].
Furthermore, the statistics of the activity $k^-(t)=K^-(t)/t$ is encoded in the long time limit by the
scaled cumulant generating function (SCGF)
\begin{eqnarray}\label{eq:scgf}
	\theta(s) =\lim_{t\rightarrow\infty}\frac{\Theta(s,t)}{t},
\end{eqnarray}
  given by the leading eigenvalue of $\mathcal{L}_s$ [cf.~Eqs.~\eqref{eq:cgf} and~\eqref{eq:mgf}].
The corresponding eigenmode $\rhoss(s)$ of $\mathcal{L}_s$  is the average asymptotic state in the biased trajectory ensemble, where each trajectory is weighted by $e^{-s K^-(t)}$, before the overall ensemble is then renormalised [cf.~Eq.~\eqref{eq:mgf}].
The SCGF plays the role of free energy in non-equilibrium statistical mechanics~\cite{Touchette2009}, with its non-analyticities corresponding to \emph{dynamical phase transitions}~\cite{Garrahan2010,Garrahan2011,Ates2012,Lesanovsky2013}.

\begin{figure}[t!]
	\includegraphics[width=1\linewidth]{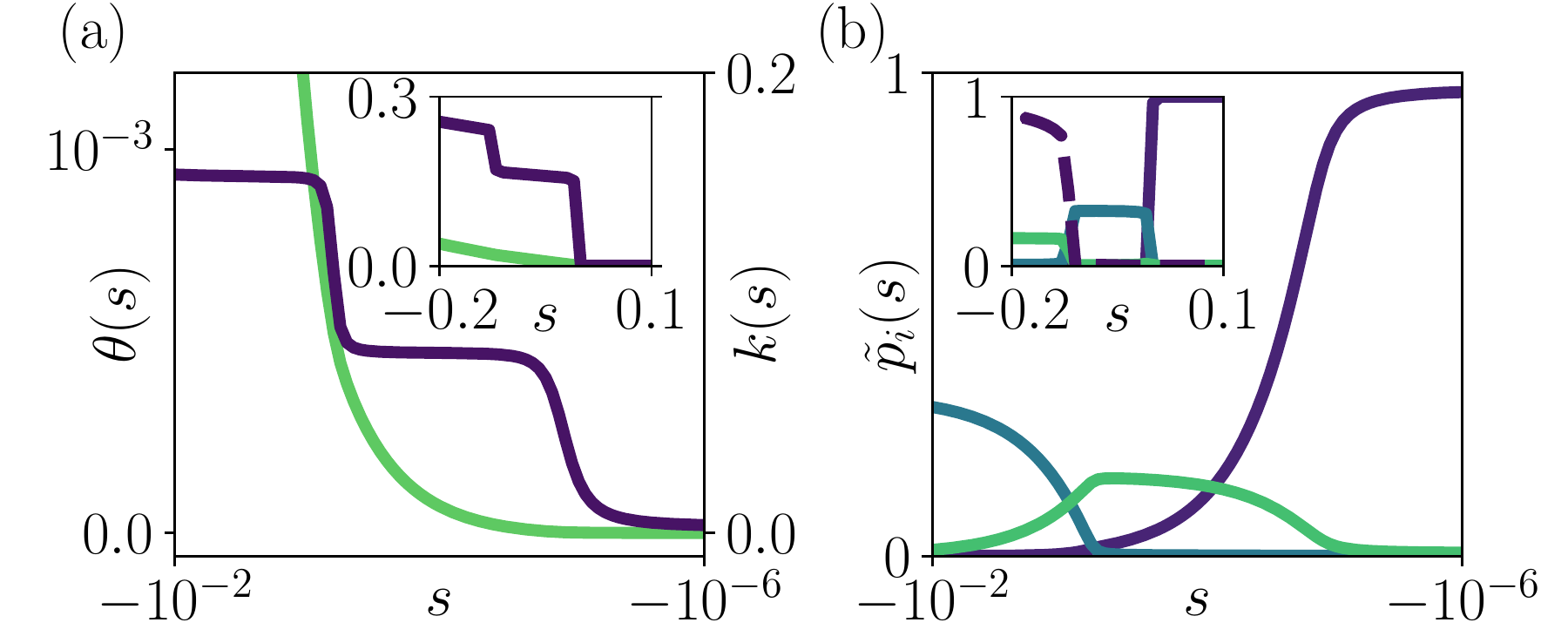}
	\caption{\label{fig:FullSEnsemble} \textbf{Global jump statistics}: (a) The leading eigenvalue $\theta(s)$ (light) of the biased master operator, Eq.~\eqref{eq:Ls}, and the corresponding activity $k(s)$ (dark), plotted on a log / linear scale respectively for negative values of $s$. The inset shows a larger range of parameters on a linear scale with an additional transition corresponding to the effective metastability (see Sec.~\ref{sec:dynobs}). (b) The barycentric coordinates of $\rhoss(s)$ with respect to: the dark state (dark), a single excited phase (light) and a two-excitation phase (intermediate). The inset shows a larger range of parameters, along with the distance between $\rhoss(s)$ and the corresponding metastable state (dashed) [cf.~Eq.~\eqref{eq:Expansion}]. Parameters are chosen as $\Omega/\kappa=0.15$, $\gamma/\kappa=0.0004$, and $p=0.999$, while $\theta(s)$ and $\rhoss(s)$ are obtained by numerical diagonalisation of $\L_s$.}
	\vspace*{-3mm}
\end{figure}

In  Fig.~\ref{fig:FullSEnsemble}\textcolor{blue}{(a)}, two sharp changes are found in the first derivative of $\theta(s)$,  i.e., the average activity $k(s)=-\mathrm{d}\theta(s)/\mathrm{d}s$, at negative bias $s$ close to $0$, between the values equal zero, one or twice the average single spin activity proportional to $\Omega^2/\kappa+2\gamma+...$  [see Eq.~\eqref{eq:rho_ss1}].
This indicates that the proximity of the physical dynamics $s=0$ to two first-order dynamical phase transitions.
Furthermore, these changes occur as the perturbation due to the bias becomes larger than $\lambda_m$ for $m=7$ and $m=10$, which indicates their relation to the presence of the hierarchy of metastabilities.
This is also supported by Fig.~\ref{fig:FullSEnsemble}\textcolor{blue}{(b)}, where in a decomposition of $\rhoss(s)$ between metastable phases (in its barycentric coordinates), at $s=0$ it corresponds mostly to the dark metastable phase, while for a large enough negative bias $s$ (towards more active trajectories) can be characterised as the equal mixture of six single-excitation metastable phases ($1/6$th probability each) or the equal mixture of three double-excitation metastable phases ($1/3$rd probability each). This homogeneity follows from the translation symmetry of $\L_s$.

These results actually follow from the correspondence of $\theta(s)$ to an SCGF for integrated metastable phase activity in classical trajectories of the long-times dynamics~\cite{Macieszczak2020}  (cf.~Ref.~\cite{Macieszczak2016}), which holds for small to moderate values of $s$ [when contributions from fast modes are negligible; cf.~the inset in Fig.~\ref{fig:FullSEnsemble}\textcolor{blue}{(b)}] and metastable phases distinguished by the average jump activities dominating rates of the long-time dynamics.
This is exactly the situation in the open quantum East model due to the constraint in $J_j^-$ fulfilled by excitations present in metastable phases [cf.~Fig.~\ref{fig:eMSProp}\textcolor{blue}{(a)}].
Indeed, biasing trajectories with negative $s$ effectuates post-selection towards more active trajectories, which in this case correspond to metastable phases with a single or double activity for smaller or larger $|s|$ (respectively, in order to make up for the shorter lifetime $\tau_2\ll \tau_1$ due to the hierarchy of metastabilities present for the chosen parameters; probability of trajectories with even higher activity remains negligible).
In contrast, for positive $s$ inactive trajectories are preferred, corresponding to the dark metastable phase with no excitations.\\

\begin{figure}[t!]
	\includegraphics[width=1\linewidth]{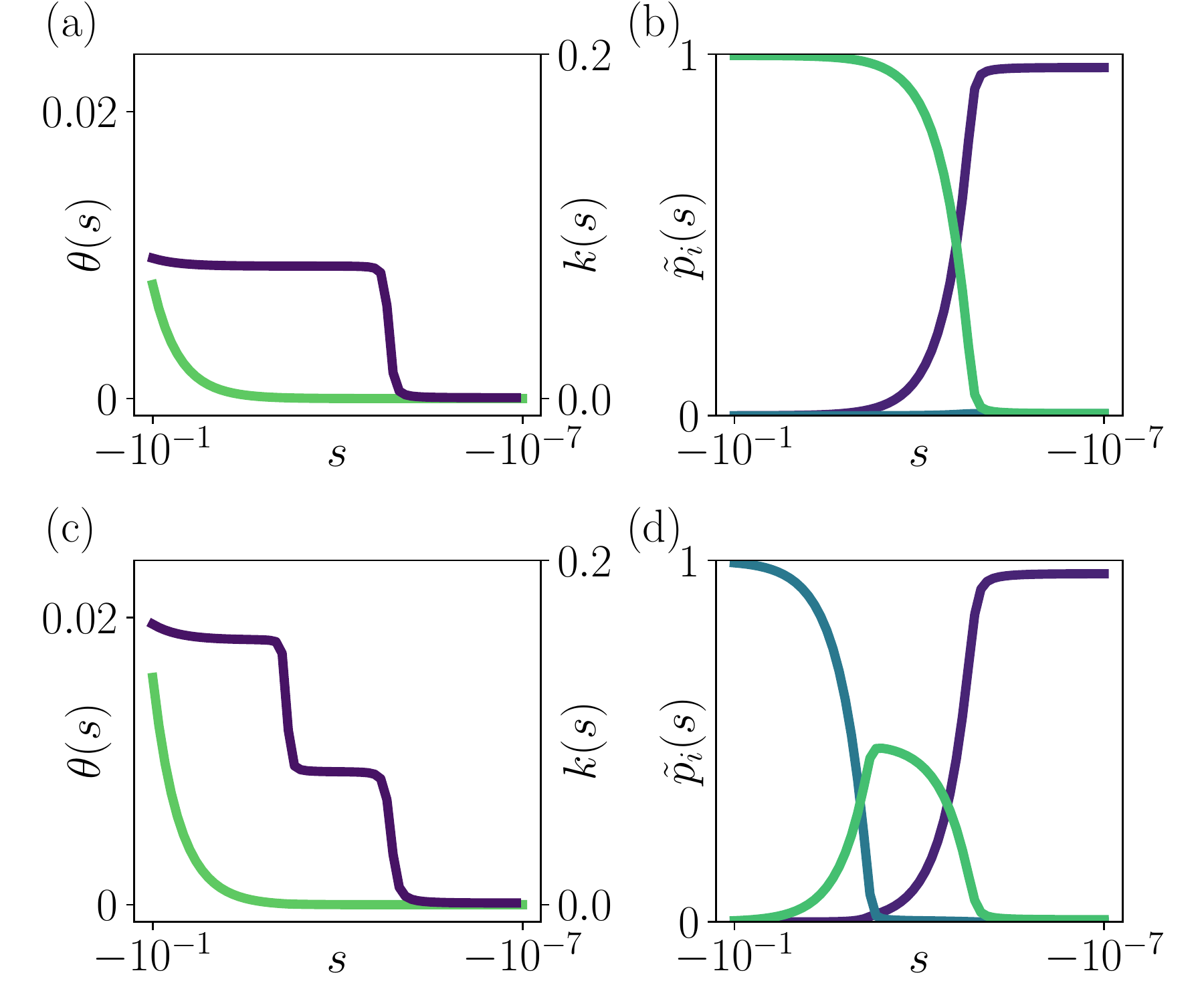}
	\caption{\textbf{Local jump statistics}: \label{fig:TwoSiteSEnsemble} (a,c) The leading eigenvalue of the biased master operator in Eq.~\eqref{eq:Lsj} (left axis) and the corresponding activity (right axis) as a function of the local bias, (a) on a single site $s_1=s$, (c) on two sites $s_1=s_4=s$. (b,d) The barycentric coordinates of $\rhoss(s)$ with respect to: the dark phase (dark), the phase with a single excitation on the third or sixth site (light) and the phase with  two excitations on the third and sixth site (intermediate). Parameters are chosen as $\Omega/\kappa=0.15$, $\gamma/\kappa=0.0004$, and $p=0.999$, while $\theta(s)$ and $\rhoss(s)$ are obtained by numerical diagonalisation of $\L_{s_1,...,s_6}$.}
	\vspace*{-5mm}
\end{figure}

{\bf \em Dynamical phase transitions in local jump activity}.
Rather than the global activity of jumps across the system, we can consider local jump activity, with the locally biased master operator [cf.~Eq.~\eqref{eq:Ls}]
\begin{equation}
\label{eq:Lsj}
\L_{s_1,...,s_N}(\rho)=\L(\rho)+\sum_{j=1}^N (e^{-s_j}-1) J_j^- \rho \,J_j^{-\dagger}
\end{equation}
encoding the joint statistics of number $K_j^-$ of jumps $J_j^-$ at sites $j=1,...,N$ observed up to time $t$.
Similarly to Fig.~\ref{fig:FullSEnsemble} for the full jump statistics,  in Fig.~\ref{fig:TwoSiteSEnsemble}, we observe sharp changes in the first derivative of the corresponding maximal eigenvalue of $\L_{s_1,...,s_N}$.

In Figs.~\ref{fig:TwoSiteSEnsemble}\textcolor{blue}{(a)} and~\ref{fig:TwoSiteSEnsemble}\textcolor{blue}{(b)}, we look at a cross-section with $s_1=s$ and $s_j=0$ for $j\neq1$.
This biases towards trajectories containing significant periods of the single excitation metastable phases, and ignores the double excitation phases: indeed, in Fig.~\ref{fig:TwoSiteSEnsemble}\textcolor{blue}{(a)} there is only a single jump in the activity to a value corresponding to the activity of single excitation phases, while the overlap with the phase featuring an excitation at site $6$ that induces emissions on site $1$, turns out to be dominant at negative values of $s$ in Fig.~\ref{fig:TwoSiteSEnsemble}\textcolor{blue}{(b)}.

To target a double excitation state, we look  in Figs.~\ref{fig:TwoSiteSEnsemble}\textcolor{blue}{(c)} and~\ref{fig:TwoSiteSEnsemble}\textcolor{blue}{(d)} at a cross-section with $s_1=s_4=s$ and $s_j=0$ for $j\neq1,4$, targeting the phase with excitations on sites $3$ and $6$.
As expected, there is a pair of jumps in the activity in Fig.~\ref{fig:TwoSiteSEnsemble}\textcolor{blue}{(c)},  corresponding to the activity of single excitation phases and double excitation phases respectively.
For smaller negative values of $s$, the overlap with the metastable phases in Fig.~\ref{fig:TwoSiteSEnsemble}\textcolor{blue}{(b)} is split evenly across the single excitation phases on sites $3$ and $6$, as expected in comparison with Fig.~\ref{fig:FullSEnsemble}; for large values the only relevant overlap becomes the double excitation phase that was targeted with this choice of bias. \\

{\bf \em Metastable phases from biased trajectories}. Beyond clarifying the relation of first-order dynamical phase transitions to metastability, Figs.~\ref{fig:FullSEnsemble} and~\ref{fig:TwoSiteSEnsemble} demonstrate that metastable phases differing in activity can be obtained as the asymptotic average states in the biased ensemble of trajectories. This result indicates an alternative method to uncover the structure of metastable manifold, which does not require the diagonalisation of  the master operator (cf.~Appendix~\ref{app:numerics_phases}). While methods for efficient sampling of biased classical trajectory ensembles are common \cite{Crooks2001,Bolhuis2002,Lecomte2007,Giardina2011,Ferre2018,Rose2020}, with some work in this direction for quantum systems \cite{Schile2018,Carollo2019}, more development is needed to achieve the speed needed for many-body quantum systems. A possible direction could be the use of tensor network techniques, as done in recent classical large deviation studies \cite{Banuls2019,Causer2020}.

\subsection{Effective metastability of observables}\label{sec:dynobs}
Metastability can be observed experimentally in the behaviour of statistical quantities such as expectation values or autocorrelations of system observables~\cite{Macieszczak2016a,Sciolla2015,Rose2016,Macieszczak2020}, with each metastable regime manifesting as a plateau in the observable dynamics.
In particular, for the average of an observable $M$ we have [cf.~Eq.~\eqref{eq:Expansion0}]
\begin{eqnarray}\label{eq:full-expansion_obs}
	\langle M(t)\rangle&=&\Tr[M \rho(t)]=\langle M\rangle_\text{ss}+\sum_{k\geq 2}{e}^{t{\lambda}_{k}}\,{c}_{k} {d}_{k},
\end{eqnarray}
where $\langle M\rangle_\text{ss}$ is the average in the stationary state $\rhoss$ and ${d}_{k}=\Tr(M R_k)$ are coefficients of the decomposition of $M$ into the eigenmodes $L_k$.
After the relaxation towards a metastable regime, $t\gg -1/\lambda_{m+1}^R$, evolution of the average is determined only by the slow modes [cf.~Eq.~\eqref{eq:truncated-evolution}]
\begin{equation}\label{eq:Expansion1_obs}
	\langle M(t)\rangle=\langle M\rangle_\text{ss}+\sum_{k= 2}^m {e}^{t{\lambda}_{k}}\,{c}_{k} {d}_{k}+...,
\end{equation}
so that during the metastable regime, $-1/\lambda_{m+1}^R \ll t\ll -1/\lambda_{m}^R$,  the average is approximately stationary  manifesting as a visible plateau on a log-timescale.

\begin{figure}[t!]
	\vspace*{-3mm}
	\includegraphics[width=1\linewidth]{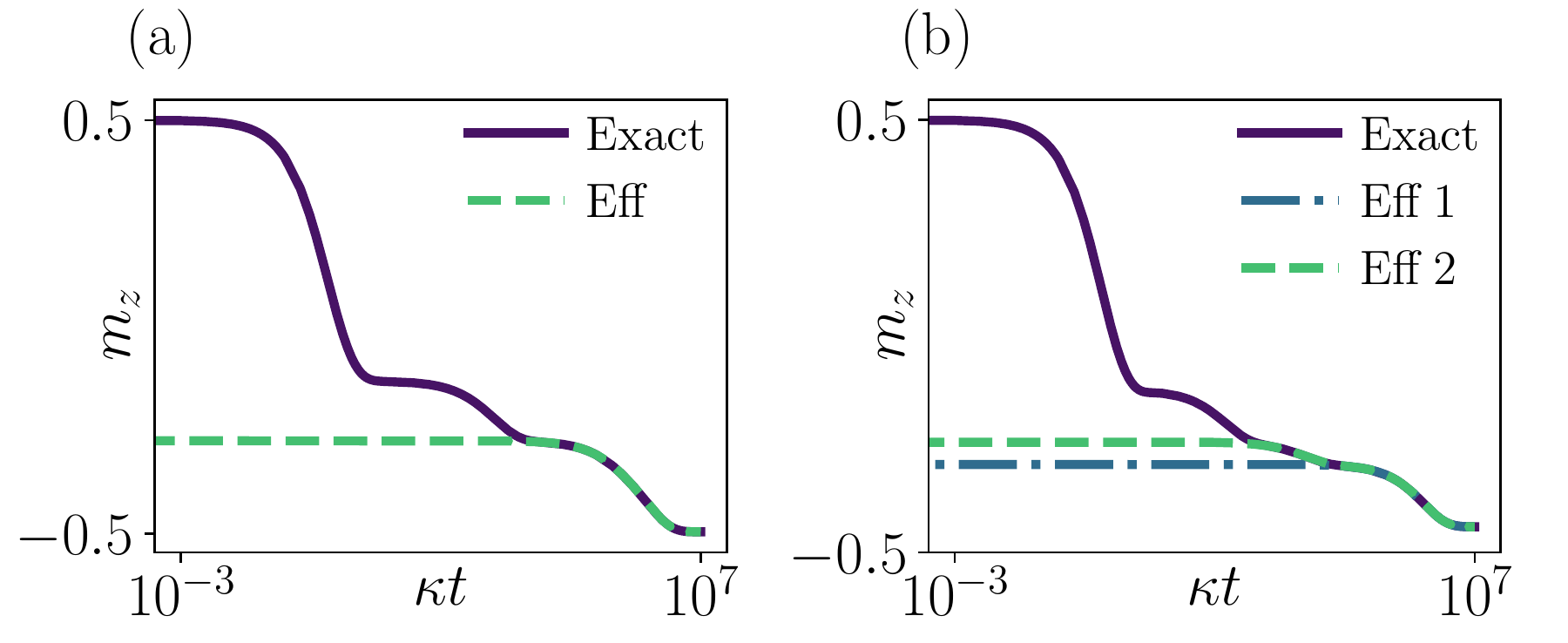}
	\caption{\label{fig:eff_stat} \textbf{Effective metastability of magnetisation}: (a) and (b) Exact (solid) and effective (dashed, dash-dotted) evolution of the total $z$ magnetisation from an initial all up state, corresponding to master operators with one and two metastable timescales shown in Figs.~\ref{fig:EffDyn2}\textcolor{blue}{(b)} and~\ref{fig:EffDyn2}\textcolor{blue}{(c)}, respectively. The two effective curves in panel (b) correspond to restricting the effective master operator to only the $m=7$ (slower) modes (Eff~1) or all $m=10$ low-lying modes (Eff~2). Parameters are chosen as $\Omega/\kappa=0.15$, $\gamma/\kappa=0.0004$, and $p=0.999$.   }
	\vspace*{-3mm}
\end{figure}

 In Fig.~\ref{fig:eff_stat},  for $N=6$ spins of the open quantum East model initialised from the all up state, the observable is chosen as $z$-magnetisation per spin, $m_z(t)=\sum_{j=1}^N \langle S^z_j(t)\rangle/N$, which corresponds to the number of excitations in the system [cf.~Fig.~\ref{fig:eMSProp} and see also Fig.~\ref{fig:Spectrum}\textcolor{blue}{(b)}]. We indeed observe plateaus due to the presence of metastable regimes, as indicated by the agreement with the long-time dynamics [Eq.~\eqref{eq:Expansion1_obs}]. These are preceded by the necessary decay of excitations, as metastable phases feature at most two excitations, while final relaxation removes all excitations to reach the unexcited stationary state.

Interestingly, the exact dynamics features an additional (anomalous) plateau at earlier times, which is not due to any further gap present in the spectrum of the master operator, but results instead from the zero overlap of either the initial state ($c_k=0$)  or the observable ($d_k=0$) with many eigenmodes in Eq.~\eqref{eq:full-expansion_obs}, creating an effective gap in the eigenvalues of the master operator that do contribute to the dynamics, and thus an \emph{effective metastability}. We have verified that this gap does not simply arise due to the choice of a symmetric observable, i.e., is not present in the eigenvalues of the symmetric modes.

Furthermore, the average magnetisation is related to  instant activity of jumps $J_j^-$ [Eq.~\eqref{eq:J-}]  per spin $\sum_{j=1}^N\langle J_j^{-\dagger} J_j^-\rangle/N=m_z+1/2$. This links the existence of metastable phases differing in magnetisation to sharp changes in the activity of quantum trajectories (cf.~Fig.~\ref{fig:FullSEnsemble}). When the effective metastability is present, also another jump in the activity occurs corresponding to a higher number of excitations than in the metastable phases and at a more negative bias [see the inset in Fig.~\ref{fig:FullSEnsemble}\textcolor{blue}{(a)} and the inset in Fig.~\ref{fig:FullSEnsemble}\textcolor{blue}{(b)} where the dashed line represents the distance between the average state in trajectories with a given activity and its projection onto the metastable manifold]. This suggests the effective metastability results from the magnetisation overlaps with the modes, rather than the specific choice of the initial state.

\section{Conclusions}\label{sec:conclusion}
In this work, we have investigated a quantum generalisation of the classical East model, uncovering a hierarchy of classical metastable manifolds characterised by the number of excited sites, similar to the case of the classical East model.
The long time effective dynamics of the model was shown not only to be classical and featuring a hierarchy of timescales, but also to possess detailed balance, with an effective free energy depending only on the number of excitations and not their distance: both properties also found in the classical East model, but here persisting even in the presence of a coherent driving that is comparable in strength to the temperature.
The dynamics thus mimics the classical model, with an effective temperature taking into account both the coherent driving field and temperature, and effective classical configurations given by a modified basis of quantum states.
This demonstrates for the first time the usefulness the methods for metastability in open quantum systems introduced in Ref.~\cite{Macieszczak2016a,Macieszczak2020} for uncovering complex relaxation dynamics in many-body quantum systems without static phase transitions.\\

\acknowledgments
K.M.\ gratefully acknowledges support from a Henslow Research Fellowship. J.P.G.\ is grateful to All Souls College, Oxford, for support through a Visiting Fellowship during the latter stages of this work. I.L. acknowledges support from the ”Wissenschaftler-R\"uckkehrprogramm GSO/CZS of the Carl-Zeiss-Stiftung and the German Scholars Organization e.V., as well as through the  Deutsche Forschungsgemeinsschaft (DFG, German Research Foundation) under Project No.43569660. We acknowledge financial support from EPSRC Grant no.\ EP/R04421X/1, and by University of Nottingham, Grant No.\ FiF1/3. We are also grateful for access to the University of Nottingham High Performance Computing Facility.


%

\vspace*{5cm}
\begin{appendix}

\section*{Appendix}

\section{Numerics}\label{app:numerics}
Here, we summarise key points of numerical methods used in this work.

\subsection{Diagonalisation of master operator}\label{app:numerics_master}
To diagonalise the master operator in Eq.~\eqref{eq:lindblad-op}, with the spectrum shown in Figs.~\ref{fig:Spectrum}\textcolor{blue}{(a)}, ~\ref{fig:Spectrum}\textcolor{blue}{(b)},  and~\ref{fig:spectral}, we use the Liouville representation for a chosen basis $\{|\psi_k\rangle\}_{i=1}^{2^N}$ of the system space.  The density matrix is represented as a vector,
\begin{equation}\label{eq:rho-vector}
	\overline{\rho}=\sum_{k,l=1}^{2^N} \langle \psi_k|\rho|\psi_l\rangle \, |\psi_k\rangle\otimes|\psi_l\rangle,
\end{equation}
and the master operator as matrix,
\begin{align}\label{eq:lindblad-matrix}
	\overline{\mathcal{L}}=&-i\left(H\otimes I - I\otimes H^T\right)\\\nonumber
	&+\sum_{j}\left[{J}_{j}\otimes{J}_{j}^{*} - \frac{1}{2}\left({J}_{j}^{\dagger}{J}_{j}\otimes I +I\otimes{J}_{j}^{T}{J}_{j}^*\right)\right],
\end{align}
where the transposition and complex conjugation take place in the chosen basis. $\overline{\mathcal{L}}$ shares the same spectrum with $\mathcal{L}$, and the eigenmodes of $\L$ can be recovered from eigenvectors of $\overline{\mathcal{L}}$ by the inverse transformation to Eq.~\eqref{eq:rho-vector}. This approach can also be used for diagonalisation of the biased operators in Eqs.~\eqref{eq:Ls} and~\eqref{eq:Lsj}.

The translation symmetry of the master operator  with periodic boundary conditions(and of the biased operator for the total activity) is exploited by considering a basis of states invariant under the translation of spins (up to a phase);  cf.~Refs.~\cite{Buvca2012,Albert2014,Nigro2020}.	The construction of the matrix in Eq.~\eqref{eq:lindblad-matrix} is further simplified by considering plane-wave jump operators instead of local jump operators~\cite{Macieszczak2020b}.

\subsection{QJMC simulations}  \label{app:numerics_QJMC}
The QJMC algorithm,
which is used to obtain trajectories shown in Figs.~\ref{fig:Spectrum}\textcolor{blue}{(c)},~\ref{fig:Spectrum}\textcolor{blue}{(d)},~\ref{fig:EffDyn2}\textcolor{blue}{(d)}, and~\ref{fig:EffDyn2}\textcolor{blue}{(e)},
is implemented largely following the standard procedure (see e.g., Ref.~\cite{Daley2014}) with one key difference: the time of a jump is found utilizing a binary search as proposed in Chapter 2 of Ref.~\cite{Everest2017a} (see also the implementation in Ref.~\cite{Everest2017b}). We sketch it below.

While it is standard to discretise the time evolution for efficiency, this leads to a competition between accuracy of jump times, requiring a fine-grained discretisation, and efficiency of time evolution, desiring larger time steps.
To meet both these aims, rather than restricting to a single time step for evolution we use a set of $N_U$ non-unitary evolution operators $U_k$ for $k=1,...,N_U$, related by $U_k=U^2_{k+1}$.
With $U_{N_U}$ inducing a time evolution of $\Delta t$, $U_k$ thus induces a time evolution of $2^{N_U-k}\Delta t$.
Evolution between jumps is then initially done using $U_{1}$, allowing for large steps in time of $2^{N_U-1}\Delta t$.
Once the norm of the state drops below the random number drawn to determine when a jump occurs, the sequence of unitaries is then used to perform a binary search, identifying the time of the jump with the chosen accuracy at much higher speed.

\subsection{Generation of metastable phases} \label{app:numerics_phases}
	We now sketch a version of  the computational approach in Ref.~\cite{Macieszczak2020}, which we use in this work  to verify the classicality of the metastable manifold in the open quantum East model (cf.~Fig.~\ref{fig:classicality}) and to find its metastable phases (cf.~Fig.~\ref{fig:eMSProp}).

\subsubsection{Upper bound on distance to probability distributions}

We first explain how to estimate from above the distance of barycentric coordinates in Eq.~\eqref{eq:Expansion2} from probability distribution for any metastable state. This allows us verify how well the metastable manifold is approximated by probabilistic mixtures of the chosen metastable states.

For a set of candidate states $\rho_1$, ..., $\rho_m$, the corresponding metastable states
	\begin{equation}\label{eq:rhotilde}
	\tilde\rho_l\equiv\rhoss+\sum_{k=2}^m c_k^{(l)} R_k, \quad l=1,...,m,
\end{equation}
can be used as a new basis replacing the stationary state $\rhoss$ an the low-lying modes $R_2,...,R_k$,  provided that they are linearly independent. The decomposition of a metastable state in terms of barycentric coordinates for Eq.~\eqref{eq:rhotilde} is given in Eq.~\eqref{eq:Expansion2}.  The barycentric coordinates can be obtained
as $\tilde{p}_l=\Tr (	\tilde{P}_l\rho)$, where
\begin{equation}\label{eq:Ptilde}
	\tilde{P}_l=\sum_{k=1}^m(\mathbf C^{-1})_{kl} \,L_k, \quad l=1,...,m,
\end{equation}
is the dual basis to Eq.~\eqref{eq:rhotilde}, analogously to the coefficients defined as $c_k=\Tr(L_k \rho)$. Note that Eq.~\eqref{eq:Ptilde}  is well defined only when the coefficient matrix for candidate states, $(\mathbf C)_{kl}=c_k^{(l)}$, is invertible, i.e., for linearly independent $\trho_1$, ..., $\trho_m$. We use the dual basis to test the accuracy of the approximation of the metastable manifold in terms of probabilistic mixtures of Eq.~\eqref{eq:rhotilde}, as follows.

While $\sum_{k=1}^m\tilde p_k=1$ is guaranteed to hold for all states by definition of barycentric coordinates, individual coordinates are not restricted to be positive in contrast to probability distributions. In particular, whenever a coordinate takes a value below $0$ or above $1$, the corresponding metastable state is found beyond the simplex of states in Eq.~\eqref{eq:rhotilde}.
Since the barycentric coordinates correspond to the expected values of the dual basis elements in Eq.~\eqref{eq:Ptilde}, their average distance in $L1$ norm from probability distributions can be bounded from above by  (cf.~Appendix C 1 in Ref.~\cite{Macieszczak2020})
\begin{eqnarray}\label{eq:test-average}
	\mathcal{C}=\frac{1}{2^{N-1}}\sum_{l=1}^m \sum_{k=1}^{2^N}\!\left\{\max\!\left[-\lambda_{k}^{(l)}\!,0\right]\!+\!\max\!\left[\lambda_{k}^{(l)}\!\!-\!1,0\right]\!\right\}\!,\qquad
\end{eqnarray}
where $\lambda_{k}^{(l)}$ is $k$th eigenvalue of $\tilde{P}_l$ and we consider uniformly sampled pure initial states.	Note that $\mathcal{C}$ is simply proportional the sum over $l$ of distances of $\tilde{P}_l$ spectrum to $[0,1]$ interval. Furthermore, $2^{N-1} \mathcal{C}$ is an upper bound on the distance of barycentric coordinates for any initial state to probability distribution.  This bound is shown in Fig.~\ref{fig:classicality} for the metastable candidate states in Fig.~\ref{fig:eMSProp}.

We conclude that when $\mathcal{C}$ of Eq.~\eqref{eq:test-average} is small in comparison to $1$ (which is the $L1$ norm of probability distributions), the metastable manifold is well approximated by probabilistic mixtures of the candidate metastable states in Eq.~\eqref{eq:rhotilde}.
Therefore, by checking the spectrum of the dual basis in Eq.~\eqref{eq:Ptilde}, we can investigate the classicality of the metastable manifold, as long as candidate states can be generated efficiently, which we discuss next.

	\subsubsection{Generation of candidate states}
	
	We now explain how sets of candidate states can be generated efficiently. We work with the assumption that the metastable manifold is classical, i.e.,  has an approximate simplex structure in the coefficient space [cf.~Fig.~\ref{fig:3DMM}\textcolor{blue}{(a)}] and, thus, we attempt to find a set of $m$ metastable states which define the largest volume simplex contained within the subset of the coefficient space corresponding to the MM. We also make use of the translation symmetry of the open quantum East model with periodic boundary conditions. The approach used here is a simplified version of the algorithm introduced in Ref.~\cite{Macieszczak2020}.\\
	
	{\bf \em Algorithm}.  Below, we outline the steps to efficiently generate sets of candidate states.
	\begin{enumerate}
		\item Diagonalise $\L$ to find the left low-lying  eigenmodes,  $L_k$ with $k=2,...,m$.
		\item Construct candidate metastable states:
		\begin{enumerate}
			\item Diagonalise the (randomly rotated) eigenmatrices $L_k$.
			\item Add the eigenstates associated to their extreme eigenvalues as initial states for candidate metastable states.
			\item Repeat Steps 2~i and 2~ii for $r$ random rotations.
		\item Apply spin translations to the candidate metastable states to construct their cycles.
		\item Cluster cycles according to their relative distance in the space of coefficients.
			\end{enumerate}
		\item Find best candidate metastable states:
			\begin{enumerate}
		\item Choose sets of cycles providing the simplex with the largest volume, i.e., the largest $|\text{det}\C|$.
		\item Calculate the corresponding corrections $\mathcal{C}$.
			\end{enumerate}
	\end{enumerate}

	{\bf \em Discussion}.  In the above approach, we assume that the eigenmodes $L_k$ found in Step~1 are Hermitian. Such a choice is always possible due to the system dynamics being Hermiticity preserving, $\mathcal{L}(\rho^\dagger)=\mathcal{L}(\rho)^\dagger$,  as follows. First, for a real eigenvalue $\lambda_k$, both $L_k$ and $R_k$ can be chosen Hermitian. Second, for a complex $\lambda_k$, there exists another eigenvalue equal $\lambda_k^*$ with the corresponding left and right eigenmodes $L_k^\dagger$ and $R_k^\dagger$.
In this case, instead of  $L_k$ and $L_k^\dagger$, we consider their Hermitian and anti-Hermitian part, $(L_k+L_k^\dagger)/2$ and $(L_k-L_k^\dagger)/2i$, respectively [while the right eigenmodes are replaced with $(R_k+R_k^\dagger)$ and $i(R_k-R_k^\dagger)$]. Furthermore, to consider all coefficients on equal footing, we normalise $L_k$ so that the difference between its extreme eigenvalues equals $1$.

	In Step~2, we construct metastable states which achieve extreme values of coefficients in order to find vertices of the maximal simplex within the metastable manifold.  For $(m-1)$ left eigenmodes, we obtain $2(m-1)$ candidate states in Steps~2~i and~2~ii. A metastable state corresponding to an extreme value of a coefficient necessary corresponds to a metastable phase (or their mixture, in the degenerate case of many vertices of the maximal simplex featuring the same value of the coefficient). Although it is not guaranteed that all vertices achieve an extreme value in at least one coefficients, this is remedied by additionally considering random rotations of  $L_2$, ...., $L_m$ in Step~2~iii (which also removes degeneracy of coefficients for the maximal simplex vertices). Furthermore, as the set of metastable phases is known to be invariant under any symmetry of the dynamics~\cite{Minganti2018,Macieszczak2020},  candidate metastable states should form cycles under the symmetry, which motivates Step~2~iv. Actually, for $N=6$ spins in our model, we find that this step removes the need for considering random rotations (this is due to the presence of both the hierarchy and translation symmetry). Finally, any repetitions in candidate metastable states are removed in  Step~2~v.
	
	In general, there are more than $m$ candidate metastable states obtained in Step~2, because beyond metastable phases, we also obtain their mixtures as a result of degeneracies of extreme values of coefficients. Therefore, we next consider the volumes of sets of  $m$ candidate metastable states, and choose those corresponding to the simplex with the largest volume in the coefficient space [where the volume equals $|\text{det}\C|/(m-1)!$]. Importantly, thanks to the translation symmetry of the model, in Step~3.~i, it is enough to consider only sets of cycles with lengths summing up to $m$, i.e., the required number  of phases. Finally, in Step~3.~ii,  the quality of the corresponding approximation of the metastable manifold is assessed using Eq.~\eqref{eq:test-average}, and, if required, can be further improved by increasing the number $r$ of rotations in Step~2~iii.

	We note that the presence of symmetry can be exploited even further in the algorithm; see Ref.~\cite{Macieszczak2020}.  Nevertheless, in this work, we successfully  identify the metastable states corresponding to the hierarchy of two classical metastable manifolds (cf.~Figs.~\ref{fig:classicality} and~\ref{fig:eMSProp}).

\section{Stationary states of open quantum East model}\label{app:1spin}
	
	\subsection{Stationary state of single spin}
	The unique stationary state of a single spin dynamics, with the Hamiltonian $H_1$ and jumps $J_1^-$ and $J_1^+$ in Eq.~\eqref{eq:HJJ1},
	is given by  Eq.~\eqref{eq:rho_ss1}, which diagonalises~\cite{Olmos2012},
	\begin{eqnarray}
	\rho_\text{ss,1}=\lambda_u\,|u\rangle\! \langle u|+\lambda_e\,|e\rangle\! \langle e|,
	\end{eqnarray}
	with the probabilities
	\begin{eqnarray}\label{eq:p_ue}
	\lambda_{u,e}=\frac{1}{2}\pm\frac{(\kappa -\gamma ) \Delta}{(\kappa +\gamma )^2+\Delta ^2},
	\end{eqnarray}
	and the eigenstates
\begin{eqnarray}\label{eq:u_matrix}
	|u\rangle \!\langle u|&=&\left(
	\begin{array}{cc}
		\frac{1}{2}+\frac{\kappa +\gamma }{ 2\Delta} & \frac{i \Omega }{\Delta} \\ -\frac{i \Omega }{\Delta} &\frac{1}{2}-\frac{\kappa +\gamma }{ 2\Delta}  \\
	\end{array}
	\right),\\\label{eq:e_matrix}
	|e\rangle\! \langle e|&=&\left(
	\begin{array}{cc}
			\frac{1}{2}-\frac{\kappa +\gamma }{ 2\Delta} & -\frac{i \Omega }{\Delta} \\ \frac{i \Omega }{\Delta} &\frac{1}{2}+\frac{\kappa +\gamma }{ 2\Delta}  \\
	\end{array}
	\right),
\end{eqnarray}
	where $\Delta=\sqrt{(\kappa+\gamma  )^2+4 \Omega ^2}$ and we considered the basis $|0\rangle$, $|1\rangle$.

	\subsection{Stationary states of constrained dynamics}
	 In the presence of hard constraint ($\epsilon=1-p=0$), there are two stationary states of the open quantum East model with periodic boundary conditions (for open boundary conditions, see Appendix~\ref{app:obc}),
	\begin{eqnarray}\label{eq:rhoss_pbc}
	\rho_\text{ss}^{(0)}&=&\lVert u\rr ^{\otimes N},\\\nonumber
	\rho_\text{ss}^{(1)}&=& [ (\lambda_u\lVert u\rr+\lambda_e\lVert e\rr )^{\otimes N} - \lambda_u^N \lVert u\rr ^{\otimes N} ]/(1-\lambda_u^N),\quad
	\end{eqnarray}
	where $j=1,...,N$, and we introduced $\lVert...\rr=|...\rangle \langle...|$ to denote a density matrix. Note that $\rho_\text{ss}^{(0)}$ is disconnected from the dynamics, as it is orthogonal to the constrain $|e\rangle \langle e|$, and in the second stationary state $\rho_\text{ss}^{(1)}$, we subtracted the contribution without excitations to make the two stationary states disjoint (orthogonal).
	
	In the presence of soft constrain ($\epsilon\neq0$), the stationary state is unique,
	\begin{eqnarray}
\rhoss=\lambda_u^N	\rho_\text{ss}^{(0)}+\rho_\text{ss}^{(1)}(1-\lambda_u^N)&=& (\lambda_u\lVert u\rr+\lambda_e\lVert e\rr )^{\otimes N}\nonumber\\
&=&\rho_\text{ss,1}^{\otimes N},\label{eq:rhoss_pbc_soft}
	\end{eqnarray}
	which features no correlations as a product state of single-spin stationary states [Eq.~\eqref{eq:rho_ss1}]. For the dynamics leading from Eq.~\eqref{eq:rhoss_pbc} to Eq.~\eqref{eq:rhoss_pbc_soft}, see Appendix~\ref{app:soft_finite}.

	\section{Perturbation theory  for open quantum East model with periodic boundary conditions}\label{app:PT}
	
	We consider non-Hermitian perturbation theory~\cite{Kato1995,Zanardi2014,Zanardi2015,Zanardi2016,Macieszczak2016} in the following parameters: the coherent field $\Omega$, the temperature, $\gamma$, and the constrain softness with $\epsilon=1-p$.  We first consider independent contributions from the temperature and the field, and discuss influence of a soft constrain on the dynamics. We discuss the mixed contributions at the end.  Periodic boundary conditions are assumed throughout. The case of open boundary conditions and its relation to dynamical heterogeneity are discussed in Appendix~\ref{app:obc}.
	
	\subsection{Dark stationary states at zero temperature and without coherent field}  \label{app:dark}
	We consider stationary states of dissipative dynamics with the jump operators
	\begin{equation}\label{eq:Jkappa}
	J^-=\sqrt{\kappa}  |1\rangle\!\langle 1|\otimes |0\rangle\!\langle 1|,
	\end{equation}
	which remove an excitation provided that the neighbouring spin to the left is in the excited state. The stationary states are formed by \emph{dark states} with, if present, isolated excitations, i.e., composed of empty sites, $|0\rangle$, and single excitations followed by an empty site, $|\mathbf{1}_2\rangle=|10\rangle$,
	\begin{equation}\label{eq:dark}
	|...0...0...\rangle, \qquad |...\mathbf{1}_2...0...\rangle,...\qquad |...\mathbf{1}_2...\mathbf{1}_2...\rangle.
	\end{equation}
	As these stationary states are dark, i.e., $J^-|\mathbf{1}_2\rangle=0=J^-|00\rangle$, also coherences between them are stationary, forming a \emph{decoherence free subspace}~\cite{Zanardi1997,Zanardi1997a,Lidar1998}.

	\subsection{Classical dynamics due to temperature} \label{app:temp}
	
	We first consider influence of classical dynamics due to non-zero temperature, i.e., jumps
	\begin{equation}\label{eq:Jgamma}
	J^+=\sqrt{\gamma}  |1\rangle\!\langle 1|\otimes |1\rangle\!\langle 0|,
	\end{equation}
	in order to see how a classical metastable manifold arises from the quantum DFS in Eq.~\eqref{eq:dark}.
	
	\subsubsection{Stationary states}
	The stationary states for $\kappa,\gamma\neq 0$ (without coherent field, $\Omega=0$, and the hard constraint, $\epsilon=0$) are known to be [cf.~Eqs.~\eqref{eq:u_matrix} and~\eqref{eq:e_matrix}],
	\begin{eqnarray}\label{eq:rho_ssT}
	\rho_\text{ss}^{(0)}&=&\lVert 0\rr ^{\otimes N},\\\nonumber
	\rho_\text{ss}^{(1)}&=&\left[\left(\lambda_0\lVert 0\rr+\lambda_1\lVert 1\rr\right)^{\otimes N} - \lambda_0^N \lVert 0\rr ^{\otimes N}\right]/(1-\lambda_0^N),\quad
	\end{eqnarray}
	where
	\begin{equation}
	\lambda_0=\frac{\kappa}{\kappa+\gamma}, \quad \lambda_1=\frac{\gamma}{\kappa+\gamma}
	\end{equation}
	[cf.~Eq.~\eqref{eq:p_ue}].
	In particular, due to a hard constrain ($\epsilon=0$), $\rho_\text{ss}^{(1)}$ is disconnected from the dynamics, and the final contribution to it in the asymptotic state equals the initial contribution, $\lim_{t\rightarrow\infty}\rho_t= \lambda\rho_\text{ss}^{(0)}+(1-\lambda) \rho_\text{ss}^{(1)}$, where $\lambda= \langle 0|^{\otimes N} \rho_0\,|0\rangle^{\otimes N} $. \\
	
	In particular, considering $\gamma/\kappa$ as a small parameter, i.e., the low-temperature limit, we recover from Eq.~\eqref{eq:rho_ssT}
	\begin{eqnarray}\label{eq:rhoss_Gamma}
	\rho_\text{ss}^{(0)}&=&\lVert 0\rr ^{\otimes N},\\
	\rho_\text{ss}^{(1)}&=& \left[\!1-\frac{\gamma}{\kappa}(N-1)\right]\frac{1}{N}\sum_{j=1}^N  \lVert ...1_j...\rr\nonumber\\&&\quad+ \frac{\gamma}{\kappa}\frac{1}{N}\sum_{j=1}^N\sum_{j>k}^N  \lVert ...1_j...1_k...\rr+...,\quad\,
	\end{eqnarray}
	where we introduced the notation $\lVert ...1_j...\rr=\lVert 0\rr^{\otimes(j-1)} \otimes\lVert 1\rr\otimes\lVert 0\rr^{\otimes(N-j)} $ and $\lVert ...1_j...1_k...\rr=\lVert 0\rr^{\otimes(j-1)} \otimes\lVert 1\rr\otimes\lVert 0\rr^{\otimes(k-j-1)}\otimes\lVert 1\rr\otimes \lVert 0\rr^{\otimes(N-k)}$.

	\subsubsection{Perturbative dynamics}\label{app:dynamicsT}
	
	Before the discussion of the perturbative dynamics, let us note that the state $\lVert 0\rr^{\otimes N}=\rho_\text{ss}^{(0)}$, in agreement with Eq.~\eqref{eq:rho_ssT}, is decoupled from the dynamics to all orders, as the hard constraint in the no-zero temperature dynamics [Eq.~\eqref{eq:Jgamma}] cannot be satisfied.\\

	{\bf \em First-order dynamics}. In the first order, we obtain following transformation, which corresponds to the decay of closest isolated excitations,
	\begin{eqnarray*}
		|...\mathbf{1}_2\mathbf{1}_2...\rangle\!\langle...\mathbf{1}_2\mathbf{1}_2...|  &\longmapsto& \frac{\gamma}{2} \big(|...\mathbf{1}_2 \mathbf{0}_2...\rangle\!\langle...\mathbf{1}_2\mathbf{0}_2...|\\
		&&\quad\,\, -|...\mathbf{1}_2\mathbf{1}_2...\rangle\!\langle...\mathbf{1}_2\mathbf{1}_2...|\big),\\
		|...\mathbf{1}_2 \mathbf{0}_2...\rangle\!\langle...\mathbf{1}_2 \mathbf{0}_2...|  &\longmapsto& 0,\\
		|...\mathbf{1}_2 \bar{\mathbf{1}}_2...\rangle\!\langle...\mathbf{1}_2 \bar{\mathbf{1}}_2...|  &\longmapsto& 0,
		\\
		|... \bar{\mathbf{1}}_2...\rangle\!\langle...\bar{\mathbf{1}}_2...|  &\longmapsto& 0,
		\\
		|...{\mathbf{0}}_2...\rangle\!\langle...{\mathbf{0}}_2...|  &\longmapsto& 0,
	\end{eqnarray*}
	where we introduced $|\bar{\mathbf{1}}_2\rangle=|01\rangle$ and $|\mathbf{0}_2\rangle=|00\rangle$, and
	$...$ in $ |...\mathbf{1}_2\mathbf{1}_2...\rangle\!\langle...\mathbf{1}_2\mathbf{1}_2...|$ denote any configuration allowed by Eq.~\eqref{eq:dark}, also configurations corresponding to coherences, i.e., different in the ket and bra [this is in direct analogy to the tensor product structure in $J^+$ which acts on the state of a pair of spins, independently from the state of the rest of spins]. The notation used hereafter describes the dynamics due to the (first) perturbation acting on the leftmost spin. In order to recover full dynamics of a given state, it is necessary to consider the above dynamics with respect to each of the spins in the system (i.e., all translations).
	
	The coherences are affected by non-zero temperature as follows,
	\begin{eqnarray*}
		|...\mathbf{1}_2 \mathbf{0}_2...\rangle\!\langle...\mathbf{1}_2\mathbf{1}_2...|  &\longmapsto& -\frac{\gamma}{3} |...\mathbf{1}_2 \mathbf{0}_2...\rangle\!\langle...\mathbf{1}_2\mathbf{1}_2...|, \\
		|...\mathbf{1}_2 \bar{\mathbf{1}}_2...\rangle\!\langle...\mathbf{1}_2\mathbf{1}_2...|  &\longmapsto& -\frac{\gamma}{3} |...\mathbf{1}_2 \bar{\mathbf{1}}_2...\rangle\!\langle...\mathbf{1}_2\mathbf{1}_2...|, \\
			|...\mathbf{1}_2 \mathbf{0}_2...\rangle\!\langle...\mathbf{1}_2\bar{\mathbf{1}}_2...|  &\longmapsto& 0, \\
		|...\mathbf{1}_2...\rangle\!\langle...\bar{\mathbf{1}}_2...|  &\longmapsto& -\frac{\gamma}{2} |...\mathbf{1}_2...\rangle\!\langle...\bar{\mathbf{1}}_2...|, \\
		|...\mathbf{1}_2...\rangle\!\langle...{\mathbf{0}}_2...|  &\longmapsto& -\frac{\gamma}{2} |...\mathbf{1}_2...\rangle\!\langle...{\mathbf{0}}_2...|, \\
		|...\bar{\mathbf{1}}_2...\rangle\!\langle...{\mathbf{0}}_2...|  &\longmapsto& 0,	
	\end{eqnarray*}
	and dynamics of the Hermitian conjugates follows from the Hermiticity-preservation of the dynamics (i.e., Hermitian conjugation of equations above). Here, we assumed the system of $N\geq3$ spins (see below for the discussion of finite-size effects). The above dynamics can be interpreted as \emph{quantum dynamics} with three types of jumps,
	\begin{eqnarray}\label{eq:C:1}
	J_0&=&\sqrt{\frac{\gamma}{2}}\, |\mathbf{1}_2 \mathbf{0}_2\rangle\! \langle \mathbf{1}_2 \mathbf{1}_2|,\\\nonumber
	J_1&=&\sqrt{\frac{\gamma}{3}} \,\left(|\mathbf{1}_2 \mathbf{0}_2\rangle\! \langle \mathbf{1}_2 \mathbf{0}_2|+|\mathbf{1}_2 \bar{\mathbf{1}}_2\rangle\! \langle  \mathbf{1}_2\bar{\mathbf{1}}_2|+|\mathbf{1}_2 \mathbf{1}_2\rangle\! \langle \mathbf{1}_2 \mathbf{1}_2|\right),\\\nonumber
	J_2&=&\sqrt{\frac{2\gamma}{3}}  \!\left(\! |\mathbf{1}_2 \mathbf{0}_2\rangle\! \langle \mathbf{1}_2 \mathbf{0}_2|+|\mathbf{1}_2 \bar{\mathbf{1}}_2\rangle\! \langle  \mathbf{1}_2\bar{\mathbf{1}}_2|+\frac{1}{2}|\mathbf{1}_2 \mathbf{1}_2\rangle\! \langle \mathbf{1}_2 \mathbf{1}_2|\!\right)\!.
	\end{eqnarray}	
	The jump operator $J_0$, corresponds to the decay of neighbouring excitations, while the jump operators $J_1$ and $J_2$, cause dephasing of states with different locations of excitations. In particular, the dephasing jumps $J_1$ and $J_2$ lead to decay of all coherences in the DFS.
	Therefore, the manifold of states stationary with respect to the first-order dynamics is \emph{classical}. Furthermore, due to the decay represented by $J_0$, the stationary states consist of isolated excitations followed by at least two empty sites, $|\mathbf{1}_3\rangle=|100\rangle$,
	\begin{equation}\label{eq:C:rho_ss1}
	\lVert...0...0...\rangle\!\rangle, \qquad \lVert...\mathbf{1}_3...0...\rangle\!\rangle,...\qquad \lVert...\mathbf{1}_3...\mathbf{1}_3...\rangle\!\rangle.
	\end{equation}

	{\bf \em Second-order dynamics}. In the second order, the dynamics due to jumps $J^+$ is \emph{classical} and features decay of the neighbouring excitations,
	\begin{eqnarray}\label{eq:C:2}
	\lVert...\mathbf{1}_3\mathbf{1}_3...\rangle\!\rangle &\longmapsto& \frac{2}{3}\frac{\gamma^2}{\kappa} \big(	\lVert...\mathbf{1}_3\mathbf{0}_3...\rangle\!\rangle -	\lVert...\mathbf{1}_3\mathbf{1}_3...\rangle\!\rangle\big),\qquad\\\nonumber
	\lVert...\mathbf{1}_4\mathbf{1}_3...\rangle\!\rangle &\longmapsto& \frac{1}{4}\frac{\gamma^2}{\kappa} \big(	\lVert...\mathbf{1}_4\mathbf{0}_3...\rangle\!\rangle -	\lVert...\mathbf{1}_4\mathbf{1}_3...\rangle\!\rangle\big),\qquad
	\\\nonumber
	\lVert...\mathbf{1}_5...\rangle\!\rangle &\longmapsto& 0,
	\end{eqnarray}
	where $|\mathbf{1}_4\rangle=|1000\rangle=|\mathbf{1}_2\mathbf{0}_2\rangle$ and $|\mathbf{1}_5\rangle=|10000\rangle$. Here, we assumed $N\geq5$ spins (see below for the discussion of finite-size effects).
	Therefore, the remaining stationary states are composed of empty sites and excitations followed by at least 4 empty sites,
	$|\mathbf{1}_5\rangle$,
	\begin{equation}\label{eq:C:rho_ss2}
	\lVert...0...0...\rangle\!\rangle, \qquad \lVert...\mathbf{1}_5...0...\rangle\!\rangle,...\qquad \lVert...\mathbf{1}_5...\mathbf{1}_5...\rangle\!\rangle.
	\end{equation}

	{\bf \em Third-order dynamics}. In the third order, we have two contributions to the dynamics of the states in Eq.~\eqref{eq:C:rho_ss2}: from the (third-order) perturbation by the temperature outside the dark space [Eq.~\eqref{eq:dark}], and the (second-order) perturbation with the effective dynamics [Eq.~\eqref{eq:C:2}] inside the classical space [Eq.~\eqref{eq:C:rho_ss1}]. We thus obtain the decay of the neighbouring excitations,
	\begin{eqnarray}\label{eq:C:3}
	\lVert...\mathbf{1}_5\mathbf{1}_5...\rangle\!\rangle &\longmapsto& \frac{4}{3}\frac{\gamma^3}{\kappa^2} \big(	\lVert...\mathbf{1}_5\mathbf{0}_5...\rangle\!\rangle -	\lVert...\mathbf{1}_5\mathbf{1}_5...\rangle\!\rangle\big),\qquad\\\nonumber
	\lVert...\mathbf{1}_6\mathbf{1}_5...\rangle\!\rangle &\longmapsto& \frac{2}{3}\frac{\gamma^3}{\kappa^2} \big(	\lVert...\mathbf{1}_6\mathbf{0}_5...\rangle\!\rangle -	\lVert...\mathbf{1}_6\mathbf{1}_5...\rangle\!\rangle\big),\qquad
	\\\nonumber
	\lVert...\mathbf{1}_7\mathbf{1}_5 ...\rangle\!\rangle &\longmapsto& \frac{4}{11}\frac{\gamma^3}{\kappa^2} \big(	\lVert...\mathbf{1}_7\mathbf{0}_5...\rangle\!\rangle -	\lVert...\mathbf{1}_7\mathbf{1}_5...\rangle\!\rangle\big),\\\nonumber
	\lVert...\mathbf{1}_8\mathbf{1}_5...\rangle\!\rangle &\longmapsto& \frac{1}{8}\frac{\gamma^3}{\kappa^2} \big(	\lVert...\mathbf{1}_8\mathbf{0}_5...\rangle\!\rangle -	\lVert...\mathbf{1}_8\mathbf{1}_5...\rangle\!\rangle\big),\\\nonumber
	\lVert...\mathbf{1}_9...\rangle\!\rangle &\longmapsto &0,
	\end{eqnarray}
	which leads to remaining stationary states composed of empty sites and excitations followed by at least 8 empty sites,
	\begin{equation}\label{eq:C:rho_ss3}
	\lVert...0...0...\rangle\!\rangle, \qquad \lVert...\mathbf{1}_9...0...\rangle\!\rangle,...\qquad \lVert...\mathbf{1}_9...\mathbf{1}_9...\rangle\!\rangle.
	\end{equation}
	We have assumed $N\geq9$ spins (see below for the discussion of finite-size effects). We note that the order or perturbation necessary for the decay of neighbouring excitation is not linear in the distance between excitations, but follows the \emph{logarithmic scaling} instead.
	
	Alternatively, we can consider the third-order dynamics in the set of states of Eq.~\eqref{eq:C:rho_ss1}, which will reintroduce neighbouring excitations as,
	\begin{eqnarray}\label{eq:C:3_2}
	\lVert...\mathbf{1}_5\mathbf{1}_5...\rr &\longmapsto& \frac{\gamma^3}{\kappa^2} \bigg( \frac{2}{3}	\lVert...\mathbf{1}_3\mathbf{1}_7...\rr + \frac{1}{4}	\lVert...\mathbf{1}_4\mathbf{1}_6...\rr\\\nonumber
	&&\qquad + \frac{5}{12}	\lVert...\mathbf{1}_5\mathbf{0}_5...\rr- \frac{4}{3}	\lVert...\mathbf{1}_5\mathbf{1}_5...\rr\bigg),\qquad\\\nonumber
	\lVert...\mathbf{1}_6\mathbf{1}_5...\rr &\longmapsto& \frac{\gamma^3}{\kappa^2} \bigg( \frac{2}{3}	\lVert...\mathbf{1}_3\mathbf{1}_3\mathbf{1}_5...\rr + \frac{1}{4}	\lVert...\mathbf{1}_4\mathbf{1}_7...\rr\\\nonumber
	&&\qquad + \frac{1}{12}	\lVert...\mathbf{1}_6\mathbf{0}_5...\rr- 	\lVert...\mathbf{1}_6\mathbf{1}_5...\rr\bigg),\qquad\\\nonumber
	\lVert...\mathbf{1}_7...\rr &\longmapsto& \frac{\gamma^3}{\kappa^2} \bigg( \frac{1}{2}	\lVert...\mathbf{1}_3\mathbf{1}_4...\rr + \frac{1}{4}	\lVert...\mathbf{1}_4\mathbf{1}_3...\rr\\\nonumber
	&&\qquad -\frac{3}{4}	\lVert...\mathbf{1}_7...\rr \bigg).\\\nonumber
	\end{eqnarray}
	Here, we omitted the modes decaying in the second order, as they will lead to higher order corrections, e.g., in the stationary state, and we assumed $N\geq7$ spins (see below for the discussion of finite-size effects).   \\
	
	{\bf \em Hierarchy of timescales and metastable manifolds}. In the discussion above, we obtained a hierarchy of timescales corresponding to the dynamics with different orders of perturbation in the temperature parameter $\gamma$. In particular, the structure of the modes invariant to the dynamics of a particular order [see Eqs.~\eqref{eq:C:rho_ss1},~\eqref{eq:C:rho_ss2}, and~\eqref{eq:C:rho_ss3}] determines the metastable manifold in the timescales until the contribution from the following-order becomes significant. Finally, we note that for the perturbation theory to hold, the parameter $\gamma$ must be small enough when multiplied by the system size $N$, due to the locality of the perturbative dynamics.
	We discuss the examples of finite system sizes in the Appendix~\ref{sec:finitesize}.
	
	\subsubsection{Corrections to state structure}
	
	Introduction of non-zero dynamics, not only changes the timescale of the dynamics, introducing decaying modes, but also changes their structure.\\
	
	{\bf \em First-order corrections}. We now consider first-order corrections to Eq.~\eqref{eq:C:rho_ss1}. Only the states with an excitation are corrected with
	\begin{eqnarray}\label{eq:rho_ssT2}
	&&\left(1-4\frac{\gamma}{\kappa}\right)\lVert {10000}...\rangle\!\rangle\\\nonumber
	&&+\frac{\gamma}{\kappa}\left(\lVert11000...\rr+\lVert10100...\rr+\lVert10010...\rr+\lVert10001...\rr\right),
	\end{eqnarray}
	where the first-order  corrections are due to the  first-order perturbation outside the dark DFS [Eq.~\eqref{eq:dark}], the second-order corrections inside the DFS, but beyond the invariant states in Eq.~\eqref{eq:C:rho_ss1}, and the third-order corrections inside this set, but beyond the states in Eq.~\eqref{eq:C:rho_ss2} (the last two terms), respectively; see~\cite{Macieszczak2016}. We therefore recover the structure of the stationary states in Eq.~\eqref{eq:rho_ssT}. \\
	
	{\bf \em Corrections acquired during dynamics}. Although all the corrections in Eq.~\eqref{eq:rho_ssT2} are of the first order, their origin is due to different-orders of perturbative dynamics. Therefore, for an initial state  $\lVert {10000}...\rangle\!\rangle$, the term $\lVert11000...\rr$ will be acquired after the first-order dynamics takes place, i.e., for times $t\gg \gamma^{-1}$, the term  $\lVert10100...\rr$ will be acquired  for times $t\gg \kappa \gamma^{-2}$, while the terms $\lVert10010...\rr+\lVert10001\rr$ for $t\gg \kappa^2 \gamma^{-3}$, etc. This directly corresponds to the fact that the state   $\lVert10000...\rr$ fulfils the constraint for the dynamics of the second spin, which takes place at the rate $\gamma$ leading to each stationary state (cf.~the first-order dynamics in Appendix~\ref{app:dynamicsT}). The presence of the excited state in the stationary state of second spin can further facilitate the dynamics of the third spin (see the second-order dynamics in Appendix~\ref{app:dynamicsT}), etc.

	\subsubsection{Finite size}\label{sec:finitesize}
	We now consider how the perturbative dynamics in the first, second, and third order is changed for $N=3,4,5,6$, which are system sizes relevant for the discussion in the main text.\\
	
	{\bf \em 3 spins}. There are 4 dark states of $N=3$ spins [cf.~Eq.~\eqref{eq:dark}]
	\begin{equation}
	|000\rangle,|100\rangle,|010\rangle,|001\rangle.
	\end{equation}
	As there is only at most a single excitation present, the first-order dynamics in Eq.~\eqref{eq:C:1} leads to dephasing of coherences between different states, 
	which gives the classical manifold,
	\begin{equation}
	\lVert 000\rr,\lVert100\rr,\lVert010\rr,\lVert001\rr.
	\end{equation}
	In the second order, the single excitation couples to itself via the periodic boundary [cf.~Eq.~\eqref{eq:C:2}],
	\begin{eqnarray}
	\lVert 100\rr&\longmapsto&\! \frac{\gamma^2}{\kappa}\!\left[\frac{1}{3}\! \left(\lVert 100\rr+\lVert 010\rr+\lVert 001\rr \right)\!-\lVert 100\rr \right]\!\!,\qquad\,\,\,
	\end{eqnarray}
	yielding the uniform stationary state $\rho_\text{ss}^{(1)}$ in Eq.~\eqref{eq:rho_ssT}.\\

	{\bf \em 4 spins}. There are 7 dark states of $N=4$ spins [cf.~Eq.~\eqref{eq:dark}]
	\begin{equation}
	|\mathbf{0}_4\rangle,|\mathbf{1}_4\rangle,...,|\mathbf{1}_2\mathbf{1}_2\rangle,|\bar{\mathbf{1}}_2\bar{\mathbf{1}}_2\rangle.
	\end{equation}
	with dots between the states denoting the three states obtained under the translation. The first-order dynamics in Eq.~\eqref{eq:C:1} leads to the classical stationary states with at most a single excitation present, without coherences  [cf.~Eq.~\eqref{eq:C:rho_ss1}]
	\begin{equation}
	\lVert\mathbf{0}_4\rr,\lVert\mathbf{1}_4\rr,....
	\end{equation}
	In the second order, similarly as in the case $N=3$, the single excitation, can couple to itself via the periodic boundary [cf.~Eq.~\eqref{eq:C:2}]
	\begin{eqnarray}
		\lVert \mathbf{1}_4\rr&\longmapsto& \frac{\gamma^2}{4\kappa} \left(\lVert  \mathbf{0}_2\mathbf{1}_2\rr -\lVert \mathbf{1}_4\rr \right).\qquad
	\end{eqnarray}
	Finally, in the third order, the two modes left invariant by the second-order dynamics, couple as
	\begin{eqnarray*}
		\frac{1}{2} \left(\lVert \mathbf{1}_4\rr+\lVert  \mathbf{0}_2\mathbf{1}_2\rr \right)&\longmapsto& \frac{\gamma^3}{2\kappa^2}\bigg[\frac{1}{2} \left(\lVert \bar{\mathbf{1}}_2\mathbf{0}_2\rr+\lVert  \mathbf{0}_2\bar{\mathbf{1}}_2\rr \right)\\
		&&\qquad-\frac{1}{2} \left(\lVert \mathbf{1}_4\rr+\lVert  \mathbf{0}_2\mathbf{1}_2\rr \right)\bigg],
	\end{eqnarray*}
	leading to the uniform stationary state $\rho_\text{ss}^{(1)}$ [cf.~Eq.~\eqref{eq:rho_ssT}].\\
	
	{\bf \em 5 spins}. There are 11 dark states of $N=5$ spins [cf.~Eq.~\eqref{eq:dark}]
	\begin{equation}
	|\mathbf{0}_5\rangle,|\mathbf{1}_5\rangle,...,|\mathbf{1}_2\mathbf{1}_3\rangle, ....
	\end{equation}
	The first-order dynamics in Eq.~\eqref{eq:C:1} leads to the classical stationary states with at most a single excitation present, without coherences  [cf.~Eq.~\eqref{eq:C:rho_ss1}]
	\begin{equation}
	\lVert\mathbf{0}_5\rr,\lVert\mathbf{1}_5\rr,....
	\end{equation}
	These states are also invariant to the second-order dynamics in Eq.~\eqref{eq:C:2}. In the third order, the remaining degeneracy is lifted, by coupling of the single excitation to itself, as follows
	\begin{eqnarray}\label{eq:C:3:N5}
	\lVert \mathbf{1}_5\rr&\longmapsto& \frac{4}{3}\frac{\gamma^3}{\kappa^2}\left[ \frac{1}{2}\left(\lVert \mathbf{0}_2\mathbf{1}_3\rr+\lVert  \mathbf{0}_3\mathbf{1}_2\rr\right)-\lVert \mathbf{1}_5\rr \right],\qquad
	\end{eqnarray}
	with other transformations following by the translation symmetry of the dynamics. Therefore, we recover two stationary states as [cf.~Eq.~\eqref{eq:rho_ssT}]
	\begin{equation}
	\lVert\mathbf{0}_5\rr,\frac{1}{5}(\lVert\mathbf{1}_5\rr+...).
	\end{equation}
	\\

	{\bf \em 6 spins}. There are 18 dark states at 0-temperature are [cf.~Eq.~\eqref{eq:dark}]
	\begin{equation}
	|\mathbf{0}_6\rangle,|\mathbf{1}_6\rangle,..., |\mathbf{1}_2\mathbf{1}_4\rangle,..., |\mathbf{1}_3\mathbf{1}_3\rangle,..., |\mathbf{1}_2\mathbf{1}_2\mathbf{1}_2\rangle,....\qquad
	\end{equation}
	The first-order dynamics [Eq.~\eqref{eq:C:1}] leads to only the following classical states being stable [cf.~Eq.~\eqref{eq:C:rho_ss1}]
	\begin{equation}\label{eq:C:rho_ss1:N6}
	\lVert\mathbf{0}_6\rangle\!\rangle,\lVert\mathbf{1}_6\rangle\!\rangle,...,  \lVert\mathbf{1}_3\mathbf{1}_3\rangle\!\rangle,...,
	\end{equation}
	while in the second order [Eq.~\eqref{eq:C:2}] the decay
	\begin{eqnarray}\label{eq:C:2:N6}
	\lVert \mathbf{1}_3\mathbf{1}_3\rr&\longmapsto& \frac{2}{3} \frac{\gamma^2}{\kappa}\bigg( \frac{\lVert \mathbf{1}_6\rr+\lVert  \mathbf{0}_3\mathbf{1}_3\rr}{2} -\lVert\mathbf{1}_3\mathbf{1}_3\rr \bigg)\qquad
	\end{eqnarray}
	leads only two types of states being stationary [cf.~Eq.~\eqref{eq:C:rho_ss2}]
	\begin{equation}
	\lVert\mathbf{0}_6\rangle\!\rangle,\lVert\mathbf{1}_6\rangle\!\rangle,...
	\end{equation}
	In the third order, due to coupling of a single excitation to itself via the boundary, we obtain
	\begin{eqnarray}\label{eq:C:3:N6}
	\lVert \mathbf{1}_6\rr&\longmapsto& \frac{2}{3} \frac{\gamma^3}{\kappa^2}\bigg[\frac{1}{2}\lVert  \mathbf{0}_3\mathbf{1}_3\rr+  \frac{1}{4}\left(\lVert  \mathbf{0}_2\mathbf{1}_4\rr-\lVert  \mathbf{0}_4\mathbf{1}_2\rr\right)\qquad \\\nonumber
	&&\qquad-  \lVert \mathbf{1}_6\rr \bigg],
	\end{eqnarray}
	which again recovers the two uniform stationary states of Eq.~\eqref{eq:rho_ssT}. Alternatively, we can consider their-order dynamics including  double excitations~\eqref{eq:C:rho_ss1:N6},
	\begin{eqnarray}\label{eq:C:3_2:N6}
	\lVert \mathbf{1}_6\rr&\longmapsto&  \frac{\gamma^3}{\kappa^2}\bigg[\frac{2}{3}\lVert  \mathbf{1}_3\mathbf{1}_3\rr+ \frac{1}{6} \left(\lVert  \mathbf{0}_2\mathbf{1}_4\rr+\lVert  \mathbf{0}_4\mathbf{1}_2\rr\right) \qquad\\\nonumber
	&&\qquad-\lVert \mathbf{1}_6\rr \bigg]
	\end{eqnarray}
	which dynamics together with the second-order dynamics, Eq.~\eqref{eq:C:2:N6}, obeys~\emph{detailed balance} (as a consequence of translation symmetry and at most a single excitation removed or injected at a time). Note that we neglected the third-order dynamics of  $\lVert \mathbf{1}_3\mathbf{1}_3\rr$, which already undergoes the second-order dynamics, as it will lead to second-order corrections in the stationary state.
	\\
	
	{\bf \em Directionality in the perturbative dynamics}. Above we discuss the perturbative dynamics with constraint from the spin to the left; see Eqs.~\eqref{eq:Jkappa} and~\eqref{eq:Jgamma}. This directionality is apparent in the notation used for the description of the dynamics in Appendix~\ref{app:dynamicsT}. For the system of a finite size, it can directly be observed in the first-order dynamics for sizes $N=5,6$. Indeed, in the first order, the decay of the states $|\mathbf{1}_2\mathbf{1}_3\rangle$, $|\mathbf{1}_2\mathbf{1}_4\rangle$, ... to $|\mathbf{1}_5\rangle$, $|\mathbf{1}_6\rangle$, ..., respectively, manifests interactions to the left. However, in the second and the third orders, dynamics would be the same when considering action of the constraint from the spin to the right.

	\subsection{Decay dynamics due to soft constraint}\label{app:soft}

	\subsubsection{Dynamics due to soft constraint at small  temperature and without coherent field}  \label{app:soft_small}
	
	We now consider changing the hard constraint $|1\rangle\!\langle 1|$ to $|1\rangle\!\langle 1|+\epsilon |0\rangle\!\langle 0|$ with $0<\epsilon\ll1$, i.e., changing the jump operator~\eqref{eq:Jkappa} to
	\begin{eqnarray} \label{eq:Jdelta}
	J^-+\epsilon\,\delta J^-&=&\sqrt{\kappa}  (|1\rangle\!\langle 1|+\epsilon |0\rangle\!\langle 0|)\otimes |0\rangle\!\langle 1|.
	\end{eqnarray}
    The stationary state of this dynamics is  \emph{unique} and equal to the non-interacting (tensor-product) state $\lVert 0\rr^{\otimes N}$  [cf.~Eq.~\eqref{eq:rhoss_pbc_soft}].

	At low temperature, $\gamma\ll\kappa$, the perturbation from $\delta J^-$, will compete with  the temperature itself, $J^+$ in Eq.~\eqref{eq:Jgamma}, leading to the interplay of dynamics with different timescales, as we explain below. The change in the constrain of $J^+$ will contribute in a higher order with $\delta J^+=\sqrt{\gamma}  (|1\rangle\!\langle 1|+\epsilon |0\rangle\!\langle 0|)\otimes |1\rangle\!\langle 0|$ [cf.~Eq.~\eqref{eq:Jdelta}], and since, as we explain below, two former contributions lead to a unique stationary state, it can be neglected in the dynamics.   \\

	{\bf \em First-order dynamics}. There are no first-order corrections in $\epsilon$, as we now explain. First, since the DFS in Eq.~\eqref{eq:dark} is dark to $J^-$, the action of the jump is 0 on any dark state, while the anti-commutator terms contain $(\delta J^-)^\dagger J^-=0=(J^-)^\dagger \delta J^-$ (due to orthogonality of the constraints $|0\rangle$ and $|1\rangle$). Therefore, there are no first-order corrections in dynamics of the structure of states due to $\delta J^-$, nor there are any-higher orders corrections. Similarly,  $\delta J^+$ will not contribute in the first order proportional to $\gamma\epsilon$. Indeed, $(\delta J^+)^\dagger J^+=0=(J^+)^\dagger\delta  J^+$, and thus only coherence $|...01...\rangle\!\langle...11...|$ can be created from $|...00...\rangle\!\langle...10...|$ at the rate $\gamma\epsilon$ (and the same for the Hermitian conjugates), which nevertheless decays to $0$. As we will see below, however, in the higher order, $J^+$ will crucially contribute to the corrections to the structure of the stationary state.
	\\

	{\bf \em Second-order dynamics}. Depending on the ratio between $\epsilon^2$ and $\gamma$, we need to consider the second-order dynamics induced by the soft constraint in different metastable manifolds; see Eqs.~\eqref{eq:C:rho_ss1},~\eqref{eq:C:rho_ss2}, and~\eqref{eq:C:rho_ss3}.
	
	\emph{Regime of $\gamma=O(\epsilon^2)$}. We consider the dissipative dynamics with $\delta J^-$ in the DFS of dark states [Eq.~\eqref{eq:dark}]. In this case, we obtain decay of isolated excitations,
	\begin{eqnarray*}
		|...\bar{\mathbf{1}}_2...\rangle\!\langle...\bar{\mathbf{1}}_2...|  &\longmapsto& \epsilon^2\,\kappa\, \big(|...\mathbf{0}_2 ...\rangle\!\langle...\mathbf{0}_2...| -|...\bar{\mathbf{1}}_2...\rangle\!\langle...\bar{\mathbf{1}}_2...|\big),\qquad
	\end{eqnarray*}
	which associated decay of coherences as
	\begin{eqnarray*}
		|...\bar{\mathbf{1}}_2...\rangle\!\langle...\mathbf{0}_2...|  &\longmapsto& - \frac{\epsilon^2}{2}\kappa\,	|...\bar{\mathbf{1}}_2...\rangle\!\langle...\mathbf{0}_2...| ,\\
		|...\bar{\mathbf{1}}_2...\rangle\!\langle...{\mathbf{1}}_2...|  &\longmapsto& - \frac{\epsilon^2}{2}\kappa\,	|...\bar{\mathbf{1}}_2...\rangle\!\langle...{\mathbf{1}}_2...| ,
	\end{eqnarray*}
	which dynamics exactly corresponds to the jump operator $\delta J^-$. Therefore, even in the presence of the competing first-order dynamics due to temperature [Eq.~\eqref{eq:C:1}], all the modes decay at timescales proportional to $\epsilon^2$ (plus $\gamma$), with the unique stationary state without any excitations,
	\begin{equation}\label{eq:rhoss_soft}
	\rhoss=\left(1-N\frac{\gamma}{\kappa}\right)\lVert 0\rr ^{\otimes N} +\frac{\gamma}{\kappa}\left(\lVert \mathbf{1}_N\rr+... \right),
	\end{equation}
	where ... denotes the translations.
	At $0$-temperature, $\gamma=0$, this is the stationary state to all orders in $\epsilon$ [cf.~Eq.~\eqref{eq:rho_ssT}], while for $\gamma>0$ the corrections are due to the second-order perturbation from $\delta J^+$, which can be equivalently understood as the result of the second-order dynamics with $\delta J^+$.

	\emph{Regime of $\gamma^2=O(\epsilon^2)$}. Here, the perturbations  act on the classical manifold in Eq.~\eqref{eq:C:rho_ss2} invariant to the first-order temperature dynamics.
	The second-order dynamics with $\delta J^-$ leads again to decay of excitations
	\begin{eqnarray*}
	\lVert...\mathbf{0}_2\mathbf{1}_3...\rr  &\longmapsto&  \epsilon^2\,\kappa\,\left(\lVert...\mathbf{0}_5...\rr-\lVert...\mathbf{0}_2\mathbf{1}_3...\rr\right).
	\end{eqnarray*}
	This will again lead to the stationary state without excitations in Eq.~\eqref{eq:rhoss_soft}. We note that the dynamics induced by the soft constrain, $\delta J^-$, preserves the structure of the classical manifold, and thus there are no higher order corrections to the dynamics. Here, the perturbation acting on the \emph{second} most left spin indicated in the state.
	
	In contrast,  the second-order dynamics introducing excitations due to $\delta J^+$ when projected on  the classical manifold in Eq.~\eqref{eq:C:rho_ss2}, would introduce excitations
	\begin{eqnarray}
	\lVert...\mathbf{0}_5...\rr  &\longmapsto&  \epsilon^2\,\gamma\,\left(\lVert...\mathbf{0}_2\mathbf{1}_3...\rr-\lVert...\mathbf{0}_5...\rr\right),
	\end{eqnarray}
	but also leads to  decay of neighbouring excitations and movement of excitations,
	\begin{eqnarray}
	\lVert...\mathbf{1}_3\mathbf{1}_3...\rr  &\longmapsto&  \epsilon^2\,\gamma\,\left(\lVert...\mathbf{1}_6...\rr-\lVert...\mathbf{1}_3\mathbf{1}_3...\rr\right),\quad\\\nonumber
	\lVert...\mathbf{1}_4\mathbf{1}_3...\rr  &\longmapsto&  \frac{\epsilon^2}{2}\,\gamma\,\left(\lVert...\mathbf{1}_7...\rr-\lVert...\mathbf{1}_4\mathbf{1}_3...\rr\right),\\\nonumber
	\lVert...\mathbf{0}_3\mathbf{1}_3...\rr  &\longmapsto&  \epsilon^2\,\gamma\,\left(\lVert...\mathbf{0}_2\mathbf{1}_4...\rr-\lVert...\mathbf{0}_3\mathbf{1}_3...\rr\right),\\\nonumber
	\lVert...\mathbf{0}_4\mathbf{1}_3...\rr  &\longmapsto&  \epsilon^2\,\gamma\,\left(\lVert...\mathbf{0}_2\mathbf{1}_5...\rr-\lVert...\mathbf{0}_4\mathbf{1}_3...\rr\right).
	\end{eqnarray}
	The dynamics described above is induced by the perturbation acting on the \emph{second} most left spin indicated in the state and takes place in the higher order $\epsilon^2\,\gamma=\mathcal{O}(\epsilon^4)$.

	\emph{Regime of $\gamma^{(n+1)}=O(\epsilon^2)$}. In general, the soft constraint will lead to decay of neighbouring excitations with rates proportional to $\kappa$ in the metastable states invariant to $n$-th order dynamics with $J^+$. In particular, when degeneracy is lifted by the temperature to only two states, the soft constraint will lead further to the unique stationary state. We discuss such dynamics for arbitrary values of the temperature and the coherent field in Appendix~\ref{app:soft_finite}. \\

	{\bf \em Finite-size example}. We consider regime $\gamma^2=O(\epsilon^2)$ and the system of $N=6$ spins with the stationary states of the first-order dynamics in $\gamma$ given by Eq.~\eqref{eq:C:rho_ss1:N6}. The soft constraint in jump operators yields the following dynamics,
	\begin{eqnarray}\nonumber
	\lVert\mathbf{1}_3\mathbf{1}_3\rr  &\longmapsto&  2\,\epsilon^2\,(\kappa+\gamma)\left(\frac{\lVert\mathbf{1}_6\rr+\lVert\mathbf{0}_3\mathbf{1}_3\rr}{2}-\lVert\mathbf{1}_3\mathbf{1}_3\rr\right),\\\nonumber
	\lVert\mathbf{1}_6\rr  &\longmapsto&  \epsilon^2\,\bigg[\kappa\left(\lVert\mathbf{0}_6\rr-\lVert\mathbf{1}_6\rr\right)\\\nonumber
	&&\quad+3\gamma\!\left(\!\frac{\lVert\mathbf{1}_3\mathbf{1}_3\rr+\lVert\mathbf{0}_4\mathbf{1}_2\rr+\lVert\bar{\mathbf{1}}_6\rr}{3}-\lVert\mathbf{1}_6\rr\!\right)\!\bigg],\\
	\lVert\mathbf{0}_6\rr  &\longmapsto&  6\,\epsilon^2\,\gamma\left(\frac{\lVert\mathbf{1}_6\rr+...}{6}-\lVert\mathbf{0}_6\rr\right),
	\end{eqnarray}
	where ... denotes possible translations and $\lVert\bar{\mathbf{1}}_6\rr=\lVert000001\rr$. This dynamics obeys \emph{detailed balance} as a consequence of translation symmetry and at most a single excitation removed or injected at a time (a ladder structure), which structure will also be left unchanged by the second-order perturbation theory with $\gamma$; see Eq.~\eqref{eq:C:3_2:N6}. In particular, the soft constraint leads to the unique stationary state
	\begin{eqnarray}
	\rhoss&=&\left(1-6\frac{\gamma}{\kappa}+33\frac{\gamma^2}{\kappa^2}\right)\lVert \mathbf{0}_6\rr \qquad \\\nonumber
	&&+\left(\frac{\gamma}{\kappa}-6\frac{\gamma^2}{\kappa^2}\right)\left(\lVert \mathbf{1}_6\rr+... \right)+\frac{\gamma^2}{\kappa^2}\left(\lVert \mathbf{1}_3\mathbf{1}_3\rr+... \right) ,
	\end{eqnarray}
	where we expanded in the two-lowest order of $\gamma$, but neglected the first-order corrections outside the metastable manifold in Eq.~\eqref{eq:C:rho_ss2} [cf.~Eq.~\eqref{eq:rho_ssT2}].

		\subsubsection{Dynamics due to soft constraint at finite temperature and coherent field} \label{app:soft_finite}
	
	For dynamics with a hard constraint at finite temperature, $\gamma>0$, and finite coherent field, $\Omega\neq 0$, there exist two stationary states given in~Eq.~\eqref{eq:rhoss_pbc}. The soft constraint in the limit $|\epsilon|\ll 1$ induces perturbative dynamics between those states, which in the second order is given by
	\begin{eqnarray*}
		&&\rho_\text{ss}^{(0)}\longmapsto \epsilon^2\, N \,\left(\kappa |e_0|^4 +\gamma  |e_1|^4 \right) \left(\rho_\text{ss}^{(1)}-\rho_\text{ss}^{(0)}\right), \\
		&&\rho_\text{ss}^{(1)}\longmapsto  \epsilon^2\,N \frac{\lambda_u^{N-1}}{1-\lambda_u^{N}} \lambda_e\left(\kappa |e_1|^4 +\gamma |e_0|^4    \right) \left(\rho_\text{ss}^{(0)}-\rho_\text{ss}^{(1)}\right), \\
	\end{eqnarray*}
	where we defined we defined $e_0= \langle 0| e\rangle $ and $e_1= \langle 1| e\rangle $, so that $|e_0|^2=1-|e_1|^2$.
	This leads to a single \emph{non-interacting} stationary state  in Eq.~\eqref{eq:rhoss_pbc_soft}
	as we have $\lambda_u(\kappa   |e_0|^4 +\gamma  |e_1|^4)=\lambda_e(\kappa |e_1|^4+\gamma |e_0|^4   )$.
	In particular, for a small field $\Omega$ and a low temperature $\gamma$, we obtain a \emph{separation of timescales} in the lifetime of two phases
	\begin{equation}\label{eq:tau_pbc}
	\frac{\tau_{1}}{\tau_{0}}=\frac{1-\lambda_u^{N}}{\lambda_u^{N}}=N\left(\frac{\gamma}{\kappa} +\frac{\Omega^4}{\kappa^4} \right)+...,
	\end{equation}
	so that the dark state should be prevalent in quantum trajectories.

	\subsection{Dephasing dynamics due to coherent field} \label{app:dynamicsOmega}
	
	We now consider influence of coherent dynamics due to non-zero coherent field $\Omega$, i.e., the Hamiltonian
	\begin{equation}\label{eq:HOmega}
	H(\Omega)=\Omega \, |e\rangle\!\langle e| \otimes \frac{1}{2}(|1\rangle\!\langle 0|+|0\rangle\!\langle 1|)=:\Omega H+\Omega^2 \delta H+....
	\end{equation}
	The new constraint also appears in jump operators,
	\begin{equation}\label{eq:JOmega}
	J^-(\Omega)=\sqrt{\kappa}\, |e\rangle\!\langle e| \otimes|1\rangle\!\langle 0|=:J^-+\Omega \delta J^-+\Omega^2 \delta^2 J^-,
	\end{equation}
	and is determined by the eigenbasis of a single-spin stationary state, which is rotated due to the presence of the coherent field $\Omega$,
	\begin{eqnarray}\label{eq:e_matrix2}
	|e\rangle\!\langle e|&=&\left(
	\begin{array}{cc}
		\frac{1}{2}-\frac{\kappa }{ 2\Delta} & -\frac{i \Omega }{\Delta} \\ \frac{i \Omega }{\Delta} &\frac{1}{2}+\frac{\kappa}{ 2\Delta}  \\
	\end{array}
	\right),
	\end{eqnarray}
	where $\Delta=\sqrt{\kappa^2+4 \Omega ^2}$ [cf.~Eq.~\eqref{eq:e_matrix}].
	
	\subsubsection{Stationary states}
	
	The stationary states of dynamics with~\eqref{eq:HOmega} and~\eqref{eq:JOmega}, while at the hard constraint ($\epsilon=0$) and zero temperature ($\gamma=0$), are given by [cf.~Eq.~\eqref{eq:rhoss_pbc}]
	\begin{eqnarray}
	\rho_\text{ss}^{(0)}&=&\lVert u\rr ^{\otimes N},\\
	\rho_\text{ss}^{(1)}&=&\left[\left(\lambda_u\lVert u\rr\!+\!\lambda_e\lVert e\rr\right)^{\otimes N} \!\!-\! \lambda_u^N \lVert u\rr ^{\otimes N}\right]\!/(1\!-\!\lambda_u^N),\qquad\quad
	\end{eqnarray}
	where $\lVert e\rr= |e\rangle\!\langle e|$ is defined in Eq.~\eqref{eq:e_matrix2} and
	\begin{eqnarray}\label{eq:u_matrix2}
	\lVert u\rr&=&\left(
	\begin{array}{cc}
		\frac{1}{2}+\frac{\kappa }{ 2\Delta} & \frac{i \Omega }{\Delta} \\ -\frac{i \Omega }{\Delta} &\frac{1}{2}-\frac{\kappa  }{ 2\Delta}  \\
	\end{array}\right),
	\end{eqnarray}
	[cf.~Eq.~\eqref{eq:u_matrix}], with the corresponding probabilities
	\begin{eqnarray}\label{eq:p_ue_Omega}
	\lambda_{u,e}=\frac{1}{2}\pm\frac{\kappa  \Delta}{\kappa^2+\Delta ^2}
	\end{eqnarray}
	[cf.~Eq.~\eqref{eq:p_ue}]. Therefore, in the limit of small field $\Omega$, we obtain
	\begin{eqnarray}\label{eq:rhoss_Omega}
	\rho_\text{ss}^{(0)}&=&\lVert u\rr ^{\otimes N},\\
	\rho_\text{ss}^{(1)}&=& \left[\!1-\frac{\Omega^4}{\kappa^4}(N-1)\right]\frac{1}{N}\!\sum_{j=1}^N\,  \lVert ...e_j...\rr\nonumber\\&&\quad+ \frac{\Omega^4}{\kappa^4}\frac{1}{N}\sum_{j=1}^N\sum_{j>k}^N \, \lVert ...e_j...e_k...\rr+...,\quad\,
	\end{eqnarray}
	where $\lVert...e_j...\rr=\lVert u\rr^{\otimes(j-1)} \otimes\lVert e\rr\otimes\lVert u\rr^{\otimes(N-j)} $ and $\lVert ...e_j...e_k...\rr=\lVert u\rr^{\otimes(j-1)} \otimes\lVert e\rr\otimes\lVert u\rr^{\otimes(k-j-1)}\otimes\lVert e\rr\otimes \lVert u\rr^{\otimes(N-k)}$. Comparing Eqs.~\eqref{eq:rhoss_Gamma} and~\eqref{eq:rhoss_Omega}, suggests that the field $\Omega$ could act in the fourth order as the temperature parameter $\gamma$, in the rotated basis formed by $ \lVert u\rr$ and  $\lVert e\rr$. Below, we recover metastable states as classical states with isolated excitations in the second-order perturbation theory, so that the next order corrections are of the fourth order, while in the main text we confirm numerically that the fourth-order and higher perturbations indeed correspond to the temperature.

	\subsubsection{Perturbative dynamics}\label{app:dynamicsOmega2}
	
	{\bf \em First-order dynamics}. There are no first-order corrections in $\Omega$. In the first order, we have perturbations from the Hamiltonian $H=\Omega \, |1\rangle\!\langle 1| \otimes (|1\rangle\!\langle 0|+|0\rangle\!\langle 1|)/2$, and anti-commutator with $[(J^-)^\dagger \delta J^-+(\delta J^-)^\dagger J^-]/2 = i\Omega  (|1\rangle\!\langle 0|-|0\rangle\!\langle 1|)/2\otimes  |1\rangle\!\langle 1|$, which introduce coherences of the dark states [Eq.~\eqref{eq:dark}]  to the states with double excitations, and thus such contributions decay to $0$.\\
	
	{\bf \em Second-order dynamics}. We have the following contributions to the second-order dynamics in $\Omega$.
	
	The dynamics due to the second-order perturbation in the Hamiltonian, $\delta H=i  (|1\rangle\!\langle 0|-|0\rangle\!\langle 1|)\otimes (|1\rangle\!\langle 0|+|0\rangle\!\langle 1|)/2$, leads to the unitary dynamics in the dark DFS corresponding to movement of excitations,
	\begin{eqnarray}\label{eq:deltaH}
	|...0100...\rangle\!\langle...|&\longmapsto&-\frac{\Omega^2}{2\kappa}|...0010...\rangle\!\langle...|,\\\nonumber
	|...0010...\rangle\!\langle...|&\longmapsto&\frac{\Omega^2}{2\kappa}|...0100...\rangle\!\langle...|,
	\end{eqnarray}
	where in the above expressions we considered the perturbation $\delta H$ acting on the second and third spins, on the ket only.
	
	The dynamics due to the second-order perturbation in the jump operator, $\delta^2 J^-=  \Omega^2 (|0\rangle\!\langle 0|-|1\rangle\!\langle 1|)/2\otimes  |0\rangle\!\langle 1|/\sqrt{\kappa}^{3}$, is analogous to the first-order dynamics with the soft constraint discussed in Sec.~\ref{app:soft_small}, and thus gives no contribution.
	
	The dissipative dynamics with the first-order perturbation to the jump operators, $\delta J^-=i\Omega  (|1\rangle\!\langle 0|-|0\rangle\!\langle 1|)\otimes  |0\rangle\!\langle 1|/\sqrt{\kappa}$, which similarly to $\delta H$ facilitate movement of interactions, in the DFS causes both  movements of excitations and decay of neighbouring excitations,
	\begin{eqnarray}\label{eq:Leff:Jdelta}
	|...0010...\rangle\!\langle...0010...|&\longmapsto&\frac{\Omega^2}{\kappa}\big(|...0100...\rangle\!\langle...0100...|\\\nonumber&&\qquad-|...0010...\rangle\!\langle...0010...|\big),\\\nonumber
	|...1010...\rangle\!\langle...1010...|&\longmapsto&\frac{\Omega^2}{\kappa}\big(|...1000...\rangle\!\langle...1000...|\\\nonumber&&\qquad-|...1010...\rangle\!\langle...1010...|\big),
	\end{eqnarray}
	with the corresponding decay of coherences
	\begin{eqnarray}\label{eq:Leff:Jdelta2}
	|...\cdot 010...\rangle\!\langle...|&\longmapsto&-\frac{\Omega^2}{2\kappa}|...\cdot 010...\rangle\!\langle...|,
	\end{eqnarray}
	where in the above expression $(\cdot)$ stands for either $0$ or $1$ and we considered the perturbation $\delta J^-$ acting on the second and third spins, and on  the ket only.
	
	The second-order correction due to $H$ taking the states outside the dark DFS also leads to completely positive trace-preserving dynamics, similar to the first-order corrections from the non-zero temperature in Eq.~\eqref{eq:C:1},
	\begin{eqnarray}
	|...1010...\rangle\!\langle...1010...|&\longmapsto&\frac{\Omega^2}{4\kappa}\big(|...1000...\rangle\!\langle...1000...|\\\nonumber&&\qquad-|...1010...\rangle\!\langle...1010...|\big),\\\nonumber
	|...100\cdot...\rangle\!\langle...100\cdot...|&\longmapsto&0,
	\end{eqnarray}
	with coherences decaying as
	\begin{eqnarray}
	|...1010...\rangle\!\langle...100\cdot...|&\longmapsto&-\frac{\Omega^2}{4\kappa}|...1010...\rangle\!\langle...100\cdot...|,\qquad\,\,\\\nonumber
	|...1010...\rangle\!\langle...010\cdot...|&\longmapsto&-\frac{\Omega^2}{4\kappa}|...1010...\rangle\!\langle...010\cdot...|,\\\nonumber
	|...1010...\rangle\!\langle...00\cdot\cdot...|&\longmapsto&-\frac{\Omega^2}{4\kappa}|...1010...\rangle\!\langle...00\cdot\cdot...|,\\\nonumber
	|...100...\rangle\!\langle...010...|&\longmapsto&-\frac{\Omega^2}{2\kappa}|...100...\rangle\!\langle...010...|,\\\nonumber
	|...100...\rangle\!\langle...00\cdot...|&\longmapsto&-\frac{\Omega^2}{2\kappa}|...100...\rangle\!\langle...00\cdot...|,
	\end{eqnarray}
	where we considered the perturbation $H$ acting on the first and second spins.
	
	The second-order dynamics due to the first-order perturbation in dissipation, which takes the states outside the dark DFS,  is
	\begin{eqnarray}
	|...0010...\rangle\!\langle...0010...|&\longmapsto&-\frac{\Omega^2}{\kappa}\big(|...0100...\rangle\!\langle...0100...|\qquad\,\,\,\\\nonumber&&\qquad-|...0010...\rangle\!\langle...0010...|\big),\\\nonumber
	|...1010...\rangle\!\langle...1010...|&\longmapsto&-\frac{3}{4}\frac{\Omega^2}{\kappa}\big(|...1000...\rangle\!\langle...1000...|\\\nonumber&&\qquad-|...1010...\rangle\!\langle...1010...|\big),
	\end{eqnarray}
	where the perturbations act on the second and third spins,
	while the coherences obey
	\begin{eqnarray}
	|...0010...\rangle\!\langle...|&\longmapsto&\frac{\Omega^2}{2\kappa}|...0010...\rangle\!\langle...|,\\\nonumber
	|...1010...\rangle\!\langle... |&\longmapsto&\frac{\Omega^2}{4\kappa}|...1010...\rangle\!\langle... |,
	\end{eqnarray}
	where the perturbation acts on the second and third spins, and on the ket only. Note that the generated dynamics is trace-preserving but not positive. Nevertheless, when added to Eqs.~\eqref{eq:Leff:Jdelta} and~\eqref{eq:Leff:Jdelta2}, it becomes completely positive.

	Finally, the mixed second-order contribution from the perturbation in dissipation and the Hamiltonian gives
	\begin{eqnarray}
	|...1010...\rangle\!\langle...1010...|&\longmapsto&-\frac{\Omega^2}{2\kappa}\big(|...1000...\rangle\!\langle...1000...|\\\nonumber&&\qquad-|...1010...\rangle\!\langle...1010...|\big),\qquad\\\nonumber
	\end{eqnarray}
	where the first Hamiltonian perturbation acts on the first and second spins, while the first dissipative perturbation on the second and third spins. This also leads to the dynamics of coherences as follows,
	\begin{eqnarray}
	|...100\cdot...\rangle\!\langle...1010...|&\longmapsto&\frac{\Omega^2}{2\kappa}|...100\cdot...\rangle\!\langle...1010...|,\qquad
	\end{eqnarray}
	Finally, we also obtain unitary dynamics [cf.~Eq.~\eqref{eq:deltaH}],
	\begin{eqnarray}
	|...0100...\rangle\!\langle...|&\longmapsto&\frac{\Omega^2}{2\kappa}|...0010...\rangle\!\langle...|,\\\nonumber
	|...0010...\rangle\!\langle...|&\longmapsto&-\frac{\Omega^2}{2\kappa}|...0100...\rangle\!\langle...|,
	\end{eqnarray}
	where the contribution corresponds to the second and third spins being perturbed.\\

	{\bf \em Total second-order dynamics}. Summing all the second-order contributions above we obtain no dynamics of occupations,
	\begin{eqnarray}
	|...1010...\rangle\!\langle...1010...|&\longmapsto&0,\\\nonumber
	|...0010...\rangle\!\langle...0010...|&\longmapsto&0,
	\end{eqnarray}
	while the coherences undergo \emph{dephasing},
	\begin{eqnarray}
	|...1010...\rangle\!\langle...|&\longmapsto&-\frac{\Omega^2}{2\kappa}|...1010...\rangle\!\langle...|,\\\nonumber
	|...100\cdot...\rangle\!\langle...|&\longmapsto&-\frac{\Omega^2}{4\kappa}|...100\cdot...\rangle\!\langle...|.
	\end{eqnarray}
	Therefore the manifold of states invariant to the second-order dynamics is \emph{classical} with isolated excitations followed at least by one empty site [cf.~Eq.~\eqref{eq:C:rho_ss1}]
	\begin{equation}
	\label{eq:Omega:rho_ss2}
	\lVert...0...0...\rangle\!\rangle, \qquad \lVert...010...0...\rangle\!\rangle,...\qquad \lVert...010...010...\rangle\!\rangle,....
	\end{equation}\\

		{\bf \em Third-order dynamics}. As stationary states of the second-order dynamics are classical,  it follows that  there are \emph{no third-order corrections in $\Omega$} to the perturbative dynamics. Indeed, the eigenvalues of the induced stochastic dynamics must be negative, while $\Omega^3$ can be both positive and negative.

	\subsubsection{Corrections to state structure}
	
	{\bf \em First-order corrections}. We now consider first-order corrections to~\eqref{eq:C:rho_ss1} for states with isolated excitation, which are stationary under the second-order dynamics in $\Omega$. We have that the perturbation by the field outside the dark DFS yields [cf.~Eq.~\eqref{eq:dark}],
	\begin{eqnarray}\nonumber
	&&|...{010}...\rangle\!\langle...{010}...|\\\nonumber
	&& -i\frac{\Omega}{\kappa}\left(|...0{11}...\rangle\!\langle...0{10}...|-|...{010}...\rangle\!\langle...0{11}...|\right)+...\qquad\\\nonumber
	&&-i\frac{\Omega}{\kappa}\left(|...{11}0...\rangle\!\langle...0{10}...|-|...0{10}...\rangle\!\langle...1{10}...|\right)+...,
	\end{eqnarray}
	where the corrections in the first line are due to the Hamiltonian $H$, while in the second line due to the anti-commutator with $[(J^-)^\dagger \delta J^-+(\delta J^-)^\dagger J^-]/2$ [cf.~Eqs.~\eqref{eq:HOmega} and~\eqref{eq:JOmega}]. These corrections directly correspond to the transformation of $|0\rangle$ and $|1\rangle$ into the rotated basis of $|u\rangle$ and $|e\rangle$ [cf.~Eqs.~\eqref{eq:e_matrix2} and~\eqref{eq:u_matrix2}].
	
	\section{Perturbation theory for open quantum East model with open boundary conditions}\label{app:obc}
	
	Here, we show that the dynamics at a very low softness ($\epsilon\ll1$), in the case of open boundary conditions,  leads to dynamical heterogeneity. For the case of periodic boundary conditions, see Appendix~\ref{app:soft_finite}.

	\subsection{Stationary states for hard constraint}

	We first consider dynamics at any temperature and with arbitrary coherent field, but with a hard constraint, i.e., the Hamiltonian and jump operators,
	\begin{eqnarray} \label{eq:HJJhard}
	&&H = \Omega\, |e\rangle\!\langle e|\otimes(|0\rangle\!\langle 1|+|1\rangle\!\langle 0|)/2,\\\nonumber
	&&J^-=\sqrt{\kappa}\,|e\rangle\!\langle e|\otimes |0\rangle\!\langle 1|,\\\nonumber
	&&J^+=\sqrt{\gamma}\,|e\rangle\!\langle e|\otimes|1\rangle\!\langle 0|,
	\end{eqnarray}
	for a neighbouring pair of spins.

	\subsubsection{Stationary states}
	There are $N+1$ orthogonal stationary states in the model with open boundary conditions and the hard constraint $(\epsilon=0)$ [cf.~Eq.~\eqref{eq:rhoss_pbc}]
	\begin{eqnarray}\label{eq:rhoss_obc}
	\rho_\text{ss}^{(0)}&=&\lVert u\rr^{\otimes N}, \\
	\rho_\text{ss}^{(j)}&=&\lVert u\rr^{\otimes (j-1)} \otimes \lVert e\rr \otimes \left(\lambda_u\lVert u\rr+\lambda_e\lVert e\rr\right) ^{\otimes (N-j)},\quad\nonumber
	\end{eqnarray}
	where $j=1,...,N$. The states $\lVert u\rr=| u\rangle\!\langle u|$ and $\lVert e\rr=| e\rangle\!\langle e|$ defined in Eqs.~\eqref{eq:u_matrix} and~\eqref{eq:e_matrix}, and the probabilities $\lambda_u$, $\lambda_e$ defined in Eq.~\eqref{eq:p_ue}, correspond to the stationary state of single-spin dynamics. Moreover, the coherences between the pure $\rho_\text{ss}^{(0)}$ and $\rho_\text{ss}^{(N)}$ (i.e., the coherences of $N$-th spin) are also stationary,
	\begin{eqnarray}\label{eq:Css_obc}
	C^{+}&=&\lVert u\rr^{\otimes (N-1)} \otimes | e\rangle\!\langle u|, \\
	C^{-}&=&\lVert u\rr^{\otimes (N-1)} \otimes | u\rangle\!\langle e| \nonumber,
	\end{eqnarray}
	which corresponds to the existence of a qubit DFS. This structure of stationary states is a consequence of the first spin undergoing no dynamics due to the absence of its neighbour to the left, but acting as a constraint, while the last spin undergoing dynamics only in the presence of the penultimate spin in the excited state. We note, however, that the coherence in the first spin, $| e\rangle\!\langle  u| \otimes \left(\lambda_u\lVert u\rr+\lambda_e\lVert e\rr\right) ^{\otimes (N-1)}$ (and similarly in other spins $\lVert u\rr^{\otimes (j-1)} \otimes | e\rangle\!\langle  u|\otimes \left(\lambda_u\lVert u\rr+\lambda_e\lVert e\rr\right) ^{\otimes (N-j)}$), is not conserved, but instead, due to the first spin acting as constraint, decays with the effective Hamiltonian $-iH_\text{eff}=-i \Omega (|0\rangle\!\langle 1|+|1\rangle\!\langle 0|)-\frac{\kappa}{2} |1\rangle\!\langle 1| -\frac{\gamma}{2} |0\rangle\!\langle 0| $, acting on the second spin ($j+1$-th spin) with the eigenvalues $-\frac{\kappa+\gamma}{4}\pm\sqrt{(\frac{\kappa-\gamma}{4})^2-\Omega^2}$ featuring negative real parts.

	Finally, we note that for small values of the temperature and the coherent field $\gamma, |\Omega|\ll \kappa$, we have  $\lambda_e\approx0$, so that the stationary states $\rho_\text{ss}^{(0)}$ and $\rho_\text{ss}^{(j)}$ in Eq.~\eqref{eq:rhoss_obc} can be viewed as states with none and a single excitation of $j$th spin, respectively. This is analogous to the structure of the latest metastable manifold in the model with periodic boundary conditions.

	\subsubsection{Conserved quantities}
	 The corresponding projections on the stationary states in Eq.~\eqref{eq:rhoss_obc}  are determined by their support,
	\begin{eqnarray}\label{eq:Pss_obc}
	&&\Pi_{0}=\LL u\rVert ^{\otimes N}, \\
	&&\Pi_j=\LL u\rVert ^{\otimes (j-1)} \otimes \LL e\rVert \otimes \mathds{1}_2^{\otimes (N-j)} ,\quad\nonumber\\\nonumber
	&&(C^{+})^\dagger\quad\text{and} \quad (C^{-})^\dagger,
	\end{eqnarray}
	and conserved by the dynamics.
	We have introduced $\LL \cdot\rVert  =\lVert \cdot\rr^\dagger$. The asymptotic state is determined as, $\lim_{t\rightarrow\infty}\rho_t= p_0 \,\rho_\text{ss}^{(0)}+ \sum_{j=1}^N p_j\,\rho_\text{ss}^{(j)}+c\, C^{+}+c^*\,C^{-}$, with probabilities $p_j= \Tr (\Pi_j \rho_0) $ and coefficients $c= \Tr\{ [C^{+}]^\dagger\rho_0\}$. In particular, we have that $\Pi_{0}+\sum_{j=1}^N \Pi_j=\mathds{1}$, which corresponds to the trace-preservation of the dynamics.

	\subsection{Dynamics due to soft constraint}
	
	We now consider dynamics due to soft constraint $0<\epsilon \ll 1$. The change of softness in the constraint leads to the following shifts in the  original Hamiltonian $H$ and jump operators $J^-$ and $J^+$ in Eq.~\eqref{eq:HJJhard} [cf.~Eq.~\eqref{eq:HJJ}],
	\begin{eqnarray}
	&& \delta H = \epsilon^2\,\Omega\, |u\rangle\!\langle u|\otimes(|0\rangle\!\langle 1|+|1\rangle\!\langle 0|)/2,\\\nonumber
	&&\delta J^-=\epsilon\,\sqrt{\kappa}\, |u\rangle\!\langle u|\otimes|0\rangle\!\langle 1|,\\\nonumber
	&&\delta J^+= \epsilon\,\sqrt{\gamma}\,|u\rangle\!\langle u|\otimes|1\rangle\!\langle 0|.
	\end{eqnarray}
	
	There are no first-order perturbations to the dynamics. This is due to orthogonality of the constraint in $J^\pm$ and $\delta J^\pm$, which gives $(J^\pm)^\dagger \delta J^\pm = 0=(\delta J^\pm)^\dagger  J^\pm$.

	We have two independent second-order contributions from the dissipative dynamics with the jumps operators $\delta J^+$, and $\delta J^-$, and from the unitary dynamics in the DFS induced by the Hamiltonian $\delta H$. There are no mixed contributions from $J^\pm$ and $\delta J^\pm$, again due to orthogonality of their constraints.
	
	\subsubsection{Classical dynamics}
	
	{\bf \em Effective dynamics}. In the absence of the field, $\Omega=0$, we obtain
	\begin{eqnarray*}
		&&\rho_\text{ss}^{(0)}\longmapsto\epsilon^2 \,\gamma \left[ \sum_{j=2}^N \rho_\text{ss}^{(j)}-(N-1) \rho_\text{ss}^{(0)}\right],\qquad\\
		&&\rho_\text{ss}^{(1)}\longmapsto 0, \\
		&&\rho_\text{ss}^{(j)}\longmapsto \epsilon^2 \,\gamma \left[\sum_{k=2}^{j-1} \rho_\text{ss}^{(k)}-(j-2) \rho_\text{ss}^{(j)}\right]\qquad \\
		&&\qquad\qquad\!\!+\epsilon^2 \kappa \left[\! \lambda_0^{N-j}\rho_\text{ss}^{(0)}+   \lambda_1\!\!\!\sum_{k=j+1}^{N} \! \!\lambda_0^{k-(j+1)}  \rho_\text{ss}^{(k)}- \rho_\text{ss}^{(j)}\!\right],\qquad\\
		&&\rho_\text{ss}^{(N)}\longmapsto \epsilon^2 \,\gamma \left[\sum_{k=2}^{N-1} \rho_\text{ss}^{(k)}-(N-2) \rho_\text{ss}^{(N)}\right]\qquad \\
		&&\qquad\qquad\!\!+\epsilon^2 \,\kappa\left[\rho_\text{ss}^{(0)}- \rho_\text{ss}^{(N)}\right],\qquad\\
		&&C^+\longmapsto -\frac{\epsilon^2}{2} \left(\kappa+\gamma\right)C^+,
	\end{eqnarray*}
	where $j=2,..., N-1$. Therefore, in the \emph{classical model} at small temperature, $\gamma\ll \kappa$, from Eq.~\eqref{eq:p_ue} we recover that, due to softening of the constraint, the coherences simply decay at the rate $\epsilon^2(\kappa+\gamma)/2$, while the inverse of the lifetime of the states $\rho_\text{ss}^{(0)}$ and $\rho_\text{ss}^{(j)}$,  $j=2,...,N$, is given, respectively, by
	\begin{eqnarray*}\label{eq:tau_obc_full0}
		&&\tau_{0}^{-1}= \epsilon^2 \,\gamma \,(N-1)+...,\qquad \\
		&&\tau_{j}^{-1}= \epsilon^2 \,\left[\kappa  +\gamma\,(j-2) \right]+...,
	\end{eqnarray*}
	while $\tau_{1}^{-1}=0$, as $\rho_\text{ss}^{(1)}$ is decoupled from the perturbative dynamics. \\

	{\bf \em Stationary states}. Indeed, there are two stationary states of the perturbative dynamics, given by two possible states of the unconstrained first spin with the rest of the system in a local stationary state,
	\begin{eqnarray}\label{eq:rhoss_obc_full0}
	&&\lVert e\rr \otimes \left(\lambda_0\lVert 0\rr+\lambda_1\lVert 1\rr\right) ^{\otimes (N-1)}=\rho_\text{ss}^{(1)},\quad\\\nonumber
	&&\lVert u\rr \otimes \left(\lambda_0\lVert 0\rr+\lambda_1\lVert 1\rr\right) ^{\otimes (N-1)}= \\\nonumber
	&&\qquad\qquad\qquad \lambda_0^{N-1}\rho_\text{ss}^{(0)}+  \lambda_1 \sum_{j=2}^{N}\lambda_0^{j-2} \rho_\text{ss}^{(j)} .
	\end{eqnarray}
	 The stationary states in Eq.~\eqref{eq:rhoss_obc_full0} are actually the solutions of the dynamics with open boundary conditions to all orders in $\epsilon$.\\
	
	 {\bf \em Dynamical heterogeneity}. Furthermore, we observe from Eq.~\eqref{eq:tau_obc_full0}, that  there exists a \emph{separation of timescales}, $\tau_{j}/\tau_{0}=(N-1)\gamma/\kappa+...$. Moreover, the excited state $\rho_\text{ss}^{(j)}$ is transformed into $\rho_\text{ss}^{(0)}$ with the overwhelming probability $1-(N-2)\gamma/\kappa+...$, rather than into $\rho_\text{ss}^{(k)}$, $k\neq j$. This together with the separation of timescales leads to dynamical heterogeneity for small enough $\epsilon$ and $\gamma/\kappa$.

	\subsubsection{Quantum dynamics}
	
	{\bf \em Effective dynamics}. In the presence of the field, $\Omega\neq0$, from the perturbation of jump operators we obtain the dissipative dynamics
	\begin{eqnarray*}
		&&\rho_\text{ss}^{(0)}\longmapsto\epsilon^2 \left(\kappa |e_0|^4+\gamma |e_1|^4 \right)\left[ \sum_{j=2}^N \rho_\text{ss}^{(j)}-(N-1) \rho_\text{ss}^{(0)}\right],\qquad\\
		&&+\frac{\epsilon^2}{2}  \left\{  e^{i\phi} e_0^* e_1^*\left[ \kappa  (1
		+2|e_0|^2)  -\gamma(1+2|e_1|^2)\right]C^+ \,+\,\text{h.c.}\right\}   \\
		&&\rho_\text{ss}^{(1)}\longmapsto 0, \\
		&&\rho_\text{ss}^{(j)}\longmapsto \epsilon^2 \left(\kappa |e_0|^4+\gamma |e_1|^4 \right)\left[\sum_{k=2}^{j-1} \rho_\text{ss}^{(k)}-(j-2) \rho_\text{ss}^{(j)}\right]\qquad \\
		&&+\epsilon^2 \left(\kappa |e_1|^4 +\gamma |e_0|^4 \right)\!\!\left[\! \lambda_u^{N-j}\rho_\text{ss}^{(0)}+   \lambda_e\!\!\!\sum_{k=j+1}^{N} \! \!\lambda_u^{k-(j+1)}  \rho_\text{ss}^{(k)}- \rho_\text{ss}^{(j)}\!\right],\qquad\\
		&&\rho_\text{ss}^{(N)}\longmapsto \epsilon^2 \left(\kappa |e_0|^4 +\gamma |e_1|^4 \right)\left[\sum_{k=2}^{N-1} \rho_\text{ss}^{(k)}-(N-2) \rho_\text{ss}^{(N)}\right]\qquad \\
		&&\qquad\!+\epsilon^2 \left(\kappa |e_1|^4 +\gamma |e_0|^4 \right)\left[\rho_\text{ss}^{(0)}- \rho_\text{ss}^{(N)}\right],\qquad\\
		&&+\frac{\epsilon^2}{2}  \left\{  e^{i\phi} e_0^* e_1^*\left[\kappa (1
		+2|e_1|^2)    -\gamma(1+2|e_0|^2)\right]C^+ \,+\,\text{h.c.}\right\} \\
		&&C^+\longmapsto \frac{\epsilon^2}{2} \left(\kappa -\gamma\right) e^{-i\phi} e_0 e_1  \left(1-2 |e_1|^2\right)\left[\rho_\text{ss}^{(0)}- \rho_\text{ss}^{(N)}\right]\\
		&&\qquad\qquad\!-\frac{\epsilon^2}{2}\left[\kappa|e_1|^2\left(1+2|e_0|^2\right) +\gamma|e_0|^2\left(1+2|e_1|^2\right) \right]C^+\\
		&&\qquad\qquad\!-\epsilon^2 \left(\kappa+\gamma\right)  e^{-i2\phi} e_0^{2}e_1^2\,C^-,
	\end{eqnarray*}
	where $j=2,..., N-1$ and we defined $e_0= \langle 0| e\rangle $, $e_1= \langle 1| e\rangle $, so that $|e_0|^2=1-|e_1|^2$ and $u_0= \langle 0| u\rangle =e^{i\phi} e_1^* $, $u_1=  \langle 1| u\rangle =-e^{i\phi} e_0^*$ with the relative phase $\phi\in\mathbb{R}$. The perturbation of constraint in the coherent field, $\delta H$,  yields the unitary dynamics in the DFS,
	\begin{eqnarray*}
			\rho_\text{ss}^{(0)}\,&\longmapsto& \frac{\epsilon^2}{2} \, \Omega \left[ i e^{i\phi}\left(e_0^{*2}-e_1^{*2}\right)C^+ +\text{h.c.}\right],\\
		\rho_\text{ss}^{(j)}\,&\longmapsto& 0, \\
		\rho_\text{ss}^{(N)}\,&\longmapsto&  -\frac{\epsilon^2}{2} \, \Omega \left[ i e^{i\phi}\left(e_0^{*2}-e_1^{*2}\right)C^+ +\text{h.c.}\right],\\
		C^+\,&\longmapsto&   \frac{\epsilon^2}{2} \, \Omega i e^{-i \phi}\left(e_0^2-e_1^2\right) \left[\rho_\text{ss}^{(0)} -\rho_\text{ss}^{(N)}\right] \\
		&&-i \epsilon^2\,\Omega \left( e_0  e_1^*+e_0^*e_1\right) C^+,
	\end{eqnarray*}
	where $j=1,..., N-1$. \\

	{\bf \em Stationary states}. Although the dynamics of coherences is more complex than the simple decay present in the classical dynamics, we note the dissipative dynamics of $N$-th spin in the DFS (i.e., in the coupling between $\rho_\text{ss}^{(N)}$, $\rho_\text{ss}^{(0)}$, $C^+$ and $C^-$) features the stationary state given by $	\rho_\text{ss}^{(0,N)}=	\lambda_u	\rho_\text{ss}^{(0)}+	\lambda_e	\rho_\text{ss}^{(N)}=\lVert u\rr^{\otimes (N-1)} \otimes \left(\lambda_u\lVert u\rr+\lambda_e\lVert e\rr\right)$, without any stationary coherences.  
	Indeed, this follows from $\lambda_u(\kappa   |e_0|^4 +\gamma  |e_1|^4)=\lambda_e(\kappa |e_1|^4+\gamma |e_0|^4   )$ and 
$\lambda_u  e_0^* e_1^*[ \kappa  (1
+2|e_0|^2)  -\gamma(1+2|e_1|^2)]
		+ \lambda_e  e_0^* e_1^*[\kappa (1
		+2|e_1|^2)    -\gamma(1+2|e_0|^2)]
	+ (\lambda_u-\lambda_e) \Omega  i (e_0^{*2}-e_1^{*2})=0$
	[see~Eqs.~\eqref{eq:p_ue}-\eqref{eq:e_matrix} and consider $e_0\geq 0$, so that $e_0^*e_1^*=-i\Omega/\Delta$ and $e_0^{*2}-e_1^{*2}=1$]. The effective dynamics involving the rest of spins can then be represented as 
	\begin{eqnarray*}
	&&	\rho_\text{ss}^{(0,N)}\longmapsto\epsilon^2 \left(\kappa |e_0|^4+\gamma |e_1|^4 \right)\left[ \,\sum_{j=2}^N \rho_\text{ss}^{(j)}-(N-2) \rho_\text{ss}^{(0)}\right],\qquad\quad\\
	&&	\rho_\text{ss}^{(1)}\longmapsto 0, \\
	&&	\rho_\text{ss}^{(j)}\longmapsto \epsilon^2 \left(\kappa |e_0|^4+\gamma |e_1|^4 \right)\left[\,\sum_{k=2}^{j-1} \rho_\text{ss}^{(k)}-(j-2) \rho_\text{ss}^{(j)}\right]\qquad \\
		&&\qquad\quad+\epsilon^2 \left(\kappa |e_1|^4 +\gamma |e_0|^4 \right)\\
		&&\qquad\qquad\times\!\!\left[ \lambda_u^{N-(j+1)}\rho_\text{ss}^{(0,N)}+   \lambda_e\!\!\!\sum_{k=j+1}^{N-1} \! \!\lambda_u^{k-(j+1)}  \rho_\text{ss}^{(k)}- \rho_\text{ss}^{(j)}\right]\!\!,
	\end{eqnarray*}
    which leads ultimately to two stationary states,
	\begin{eqnarray}\label{eq:rhoss_obc_full}
	&&\lVert e\rr \otimes \left(\lambda_u\lVert u\rr+\lambda_e\lVert e\rr\right) ^{\otimes (N-1)}=\rho_\text{ss}^{(1)},\quad\\\nonumber
	&&\lVert u\rr \otimes \left(\lambda_u\lVert u\rr+\lambda_e\lVert e\rr\right) ^{\otimes (N-1)}= \\\nonumber
	&&\qquad\qquad\qquad \lambda_u^{N-2}\rho_\text{ss}^{(0,N)}+ \lambda_e\sum_{j=2}^{N-1}   \lambda_u^{j-2} \rho_\text{ss}^{(j)} ,
	\end{eqnarray}
	which directly correspond to two possible states of the unconstrained first spin, with the rest of the system equilibrated.
	Note that $\rho_\text{ss}^{(1)}$ is again disconnected from the perturbative dynamics. Moreover, the stationary states in Eq.~\eqref{eq:rhoss_obc_full} are actually the solutions of the dynamics with open boundary conditions to all orders in $\epsilon$.\\

	{\bf \em Dynamical heterogeneity}.
	In the presence of the small field and at the small temperature, the inverses of the lifetime of $\rho_\text{ss}^{(0,N)}$ and $\rho_\text{ss}^{(j)}$ are given by
	\begin{eqnarray*}
		&&\tau_{0,N}^{-1}= \epsilon^2 \left(\frac{\Omega^4}{\kappa^3}  +\gamma\right)(N-2)+...,\\
		&&\tau_{j}^{-1}= \epsilon^2 \left[\kappa-2\frac{\Omega^2}{\kappa}  +\left(\frac{\Omega^4}{\kappa^3}  +\gamma\right)(j-2) \right]+...,\qquad 
	\end{eqnarray*}
	where $j=2,...,N-1$, which again corresponds to a \emph{separation of timescales},
	\begin{equation}\label{eq:tau_obc}
		\frac{\tau_{j}}{\tau_{0,N}}=\left(\frac{\Omega^4}{\kappa^4}  +\frac{\gamma}{\kappa}\right)(N-2)+...,
	\end{equation}
	where $j=2,...,N-1$. Furthermore,  the excited state $\rho_\text{ss}^{(j)}$, $j=2,..,N-1$ is transformed into $\rho_\text{ss}^{(0,N)}$ again with the overwhelming probability equal $1-(N-3)[\Omega^4/\kappa^4+\gamma/\kappa]+...$. This, together with the separation of timescales, leads to dynamical heterogeneity for small enough $\epsilon$, $\gamma/\kappa$ and $\Omega^4/\kappa^4$.

	While coherences contribute non-trivially to dynamics in the DFS of  $\rho_\text{ss}^{(0)}$ and $\rho_\text{ss}^{(N)}$, in the limit of small temperature and the field, the dynamics of $\rho_\text{ss}^{(N)}$ is dominated by the decay to $\rho_\text{ss}^{(0)}$ with rate $\epsilon^2\kappa+...$. Similarly, $\rho_\text{ss}^{(0)}$ is excited to $\rho_\text{ss}^{(1)}$ directly with rate $\epsilon^2(\Omega^4/\kappa^3  +\gamma)+...$, as the second-order contribution in softness constraint via coherences, which should be proportional to $\epsilon^4 \Omega^2/\kappa$,  cancels out.

\section{Aspects of classical dynamics}	\label{app:Wall}

\subsection{Classical stochastic dynamics}\label{app:W}

For a classical system with configurations labelled by $l=1,...,m$, the stochastic dynamics of corresponding probabilities $p_l$ (with $0\leq p_l\leq 1$ and $\sum_{l=1}^m p_l =1$),
\begin{equation}
	\frac{d}{dt}\,p_l(t)=\sum_{k=1}^m (\W)_{lk} p_k(t),
\end{equation}
 is governed  by the generator which fulfils
 \begin{eqnarray}
 	\sum_{k=1}^m (\W)_{kl}=0,
 \end{eqnarray}
so that total probability is conserved,
and
\begin{eqnarray}
	(\W)_{ll}\leq 0, \quad (\W)_{kl}\geq 0 \,\,\text{for} \,\, k\neq l,
\end{eqnarray}
so that probabilities remain positive.

\subsection{Distance to classical dynamics}\label{app:Wdistance}

For a probability-conserving operator $\Wt$, the distance to the set of classical generators measured in the operator norm induced by $L1$ vector norm [that is, $\Vert \Wt\rVert_1=\max_{1\leq  l\leq m} \sum_{k=1}^m | (\Wt)_{kl}|$] is given by
\begin{eqnarray}\label{eq:Delta0}
	\min_{\W} \Vert \Wt-\W\rVert_1=2\underset{1\leq  l\leq m} {\max}\sum_{k\neq l} \big|\min[(\Wt)_{kl},0]\big|.\quad
\end{eqnarray}
The normalised distance [that is, $\Vert \Wt-\W\rVert_1/(\Vert \Wt\rVert_1+\Vert \W\rVert_1)$ for $\W$ being the closest classical generator] is
\begin{eqnarray}\label{eq:Delta}
	\Delta_+
	&=&\frac{\underset{1\leq  l\leq m} {\max}\sum_{k\neq l} \big|\min[(\Wt)_{kl},0]\big|}{ \underset{1\leq  l\leq m} {\max} \Big|\Wt_{ll}+\sum_{k\neq l} \min[(\Wt)_{kl},0]\Big|}.
\end{eqnarray}
This normalised distance is shown in Fig.~\ref{fig:EffDyn}\textcolor{blue}{(b)}.
 For derivations of Eqs.~\eqref{eq:Delta0} and~\eqref{eq:Delta}, see Ref.~\cite{Macieszczak2020}.

\subsection{Distance to detailed balance}\label{app:detailed-balance}

\subsubsection{Detailed balance}

 The stationary current of probability from $k$th to $l$th configuration is given by
\begin{equation}\label{eq:current}
	j_{kl}= (\W)_{kl} (\pss)_l -(\W)_{lk} (\pss)_k,
\end{equation}
where $k,l=1,...,m$ and $\pss$ denotes the stationary probability distribution for $\W$. 	Detailed balance takes place when there are no stationary currents, i.e.,
\begin{equation}\label{eq:Wdb}
	(\W)_{kl} (\pss)_l =(\W)_{lk} (\pss)_k.
\end{equation}
In this case, the generator $\W$ becomes symmetric under a similarity transformation
\begin{equation}
	(\W')_{kl}=(\pss)_k^{-\frac{1}{2}} (\W)_{kl}(\pss)_l^{\frac{1}{2}}.
\end{equation}

\subsubsection{Breaking of detailed balance}

For a classical generator $\W$, breaking of detailed balance can be quantified with respect to its stationary distribution $\pss$ as
\begin{equation}\label{eq:WminusWdb}
\min_{\W_\text{db}} \sum_ {k,l=1}^m|(\W)_{kl}-(\W_\text{db})_{kl}| (\pss)_l ,
\end{equation}
where $\W_\text{db}$ features detailed balance and the stationary distribution identical to $\pss$ [cf.~Eq.~\eqref{eq:Wdb}]. Note that in Eq.~\eqref{eq:WminusWdb} we consider an entry-wise matrix norm weighted with respect to $\pss$, which is in general smaller than $\lVert \W-\W_\text{db} \rVert_1=\max_{1\leq l\leq m}\sum_{k=1}^m|(\W)_{kl}-(\W_\text{db})_{kl}|$. Below, we show that it is bounded by twice the total stationary current [cf.~Eq.~\eqref{eq:current}]
\begin{equation}\label{eq:J}
J= \frac{1}{2}\sum_{k,l=1}^m |j_{kl}|.
\end{equation}
Furthermore, the normalised distance $\Delta_\text{db}$ corresponds to Eq.~\eqref{eq:WminusWdb} divided by  twice the sum of the total activity $K$ in $\W$,
\begin{equation}\label{eq:K}
K=\frac{1}{2}\sum_ {k,l=1}^m|(\W)_{kl}|(\pss)_l=\sum_{l=1}^m|(\W)_{ll}|(\pss)_l,
\end{equation}
and in the optimal $\W_\text{db}$ (which equals $\frac{1}{2}\sum_{k,l=1}^m q_{kl}\geq 0$), so that
\begin{equation}
\Delta_\text{db} \leq \frac{J}{K}.
\end{equation}\\

\emph{Proof}. We now prove that Eq.~\eqref{eq:WminusWdb} is bounded by $2J$ in Eq.~\eqref{eq:J}. The sum on the left-hand side of Eq.~\eqref{eq:WminusWdb} corresponds to
\begin{eqnarray*}\nonumber
	&&\frac{1}{2}\sum_{\substack{l=1}}^m \Bigg\{ \sum_{\substack{k=1\\k\neq l}}^m\big[|(\W)_{kl}(\pss)_l-q_{kl}|+|(\W)_{lk}(\pss)_k-q_{kl}|\big]\\
	&&+ \Bigg|\sum_{\substack{k=1\\k\neq l}}^m [ (\W)_{kl}(\pss)_l- q_{kl}]\Bigg|+\Bigg|\sum_{\substack{k=1\\k\neq l}}^m [ (\W)_{lk}(\pss)_k- q_{kl}]\Bigg|\Bigg\},
\end{eqnarray*}
where in the first line we introduced $q_{kl}=(\W_\text{db})_{kl}(\pss)_l =q_{lk} $ and the second line corresponds to $|(\W)_{ll}(\pss)_l-(\W_\text{db})_{ll}(\pss)_l|$ from the probability-conservation of $\W$ and $\W_\text{db}$.   The minimum of the first line equals total current [cf.~Eq.~\eqref{eq:J}] and is achieved for $\min[(\W)_{kl}(\pss)_l,(\W)_{lk}(\pss)_k]\leq q_{lk}\leq \max[(\W)_{kl}(\pss)_l,(\W)_{lk}(\pss)_k]$ [as the minimisation can be considered separately for each $q_{kl}$ (where $k>l$)]. By the triangle inequality the second line is bounded by the first line and thus the minimum is not larger than $2J$.

\end{appendix}

\end{document}